\documentclass[pdflatex,iicol,sn-mathphys-num]{sn-jnl}

\usepackage{graphicx}
\graphicspath{{figures/}}
\usepackage{multirow}
\usepackage{amsmath,amssymb,amsfonts}
\usepackage{booktabs}
\usepackage{xspace}
\usepackage{rotating}
\usepackage{enumitem}

\AtBeginDocument{%
  \renewcommand{\textbf}[1]{\textmd{##1}}%
  \renewcommand{\mathbf}[1]{\mathrm{##1}}%
}

\newcommand{\genie}{\textsc{Genie}\xspace}
\newcommand{\gibuu}{\textsc{GiBUU}\xspace}
\newcommand{\nuwro}{\textsc{NuWro}\xspace}
\newcommand{\neut}{\textsc{Neut}\xspace}
\newcommand{\MeV}{\,\text{MeV}\xspace}
\newcommand{\GeV}{\,\text{GeV}\xspace}

\newcommand{\mfppi}{MFP$_\pi$\xspace}
\newcommand{\ins}{ins2878288\xspace}

\begin{document}

\title[An effective topology-transfer nuisance for GENIE on argon]{An effective topology-transfer nuisance for \genie-based
       neutrino-argon semi-inclusive analyses:
       a methodological proposal with MicroBooNE-constrained diagnostic}

\author*[1,2]{\fnm{N.} \sur{Viaux}}\email{nicolas.viaux@usm.cl}

\affil*[1]{\orgname{Departamento de F\'isica},
  \orgdiv{Universidad T\'ecnica Federico Santa Mar\'ia},
  \city{Valpara\'iso}, \country{Chile},
  \postcode{Casilla 110~V}}

\affil[2]{\orgname{Millennium Institute for Subatomic Physics at the
  High Energy Frontier (SAPHIR)},
  \city{Santiago}, \country{Chile}}

\abstract{%
Liquid-argon TPC oscillation analyses increasingly split samples into
proton-tagged (Np) and proton-less (0p) topologies at the
$\sim 35\MeV$ tracking threshold, but the standard \genie Reweight
library contains no dial that transfers events across this boundary.
We propose a single empirical nuisance parameter $f_{\rm mig}$ that
fills this gap as a per-event reweight on nominal \genie output,
implementable without event regeneration.  The parameter is
presented as a methodological diagnostic: an effective topology-transfer
surrogate that absorbs an unspecified combination of near-threshold
proton mismodelling effects (proton-side FSI, nucleon-momentum
distribution, RES $Q^2$ shape, MEC kinematic shape), rather than a
measurement of any single microscopic mechanism.  As a physically
motivated reference scale, the residual position-dependent mean-field
energy loss absent from \genie's cascade is estimated from the
$^{40}$Ar density profile to be a few percent, consistent with
proton-transparency differences reported between cascade and transport
generators on MicroBooNE data.
The primary empirical constraint is a direct scan on the MicroBooNE
simultaneous CC~$0p/Np$ differential cross-section release with no
sector-weight regularisation:
$\hat{f}_{\rm mig} = 0.093 \pm 0.024$, $3.8\sigma$ (Wilks) /
$3.2\sigma$ (empirical).  A supporting NC~$\Delta$ radiative sideband
fit gives $\hat{f}_{\rm mig} = 0.050^{+0.029}_{-0.022}$, conditional on
sector-weight priors.  Both datasets share flux/detector systematics and
are corroborating rather than statistically independent.  A useful
diagnostic by-product is that introducing $f_{\rm mig}$ relieves a
compensatory CCRES sector-weight suppression in migration-blind fits.
The extracted parameter is best understood as a flux-integrated
effective nuisance direction for the MicroBooNE BNB configuration,
suitable as a candidate Genie dial in future LArTPC analyses pending
detector-level validation.
}


\maketitle

\section{Introduction}
\label{sec:intro}

When a neutrino interaction produces a proton inside the $^{40}$Ar nucleus
with kinetic energy $T_p$ near the liquid-argon time projection chamber
(LArTPC)~\cite{uBdet} proton reconstruction threshold of $35\MeV$, that proton must traverse the remaining nuclear medium
before it can be detected.  The nuclear mean field --- a consequence of the
attractive nuclear force and Pauli blocking in the dense nuclear interior ---
exerts a position-dependent energy cost on the escaping proton.  In
transport codes such as \gibuu~\cite{GiBUU}, this is modelled as a
continuous energy-loss term through the Walecka-type relativistic mean-field
potential~\cite{Walecka:1974,Serot:1984ey} $U(r) \propto \rho(r)$, where
$\rho(r)$ is the local nuclear density.  In \genie's intranuclear cascade
(INC)~\cite{Andreopoulos:2009rq,Hayato:2022},
the nuclear binding is instead approximated by a single constant removal
energy $W_s \approx 25\MeV$ applied at the interaction point, with no
position-dependent mean-field term during hadron transport.
The consequence for nuclear transparency --- the probability that a
hadron escapes the nucleus unscathed --- has been studied across
generators by Sobczyk et al.~\cite{Sobczyk:2019}, who find that INC
codes and transport codes predict systematically different proton
survival fractions, particularly near the detection threshold.
As we derive in Sect.~\ref{sec:physbasis}, the density-weighted residual
energy loss --- the part present in \gibuu but absent from \genie ---
amounts to approximately $7.8\MeV$ for a typical neutrino interaction
vertex in $^{40}$Ar.  For near-threshold protons with $T_p \approx 35$--$100\MeV$,
this residual is sufficient to push a fraction of protons below the $35\MeV$
detection threshold, converting what \genie classifies as a proton-tagged
(Np) event into a proton-less (0p) event from the detector's perspective.
We call this process \emph{proton-visibility migration}.

LArTPCs classify neutrino events by the number of reconstructed protons
above the $35\MeV$ threshold~\cite{uBdet}.  Charged-current (CC) events
are split into CC~Np ($N \geq 1$ reconstructed protons) and CC~0p (no
reconstructed proton); neutral-current events producing a $\pi^0$ are
classified as NC~$\pi^0$~Np or NC~$\pi^0$~0p.  This topology split is
of direct consequence for oscillation analyses: in the CC~0p topology,
energy reconstruction must rely on the quasi-elastic hypothesis
$E_\nu^{\rm QE}$, which is biased for non-QE interactions, while the
CC~Np topology admits calorimetric energy reconstruction that is generally
more accurate~\cite{DUNE:2020ypp}.  The relative fraction of CC~0p and
CC~Np events therefore directly controls the composition of the oscillation
sample and the associated energy-dependent systematic uncertainty.
A quantitative propagation from a topology-fraction shift to a
bias on extracted oscillation parameters is detector- and analysis-
specific and is not available from the DUNE TDR/CDR in a directly
applicable form; we restrict the present paper to noting that the
sensitivity of energy-reconstruction-driven systematics to the 0p/Np
split is the operational reason the parameter merits attention~\cite{DUNE:2020ypp,DUNEtdr}.  The
$\Delta(1232)$ resonance --- the dominant single-pion production mode at
Booster Neutrino Beam (BNB)~\cite{AguilarArevalo:2008yp} and DUNE energies --- decays via $\Delta^+ \to p\,\pi^0$ with a
nucleon rest-frame kinetic energy $T_N^* = 28\MeV$, below the LArTPC
threshold.  After boosting to the lab frame, approximately $30\%$ of these
protons land in the $35$--$100\MeV$ near-threshold zone where additional
near-threshold mismodelling can push them below threshold
(Sect.~\ref{sec:physbasis}).
We highlight the CCRES sector as an instructive illustration because the
$\Delta$ kinematic prediction is particularly clean; the migration
mechanism is, however, generic across all proton-producing channels.
At $E_\nu = 0.8\GeV$ the GENIE-level near-threshold Np fraction is in fact
slightly larger for CCQE+MEC ($0.321$) and CCDIS ($0.310$) than for CCRES
($0.291$, Table~\ref{tab:fnear}); the optical-potential model in
Sect.~\ref{sec:nucmodel} likewise predicts a marginally larger migration
contribution from CCQE+MEC than from CCRES.  At BNB, CCQE+MEC dominates the
absolute proton-producing cross section.  CCRES is therefore not the
uniquely sensitive mode but a kinematically transparent one.

Transport generators that solve the Boltzmann--Uehling--Uhlenbeck (BUU)
equation with a proper nuclear mean field naturally capture this
effect~\cite{MoselMedium:2024}.
The MicroBooNE experiment has performed the first simultaneous measurement
of $\nu_\mu$ CC~0p and CC~Np differential cross sections on
argon~\cite{uB0pNp2024}, reporting $\chi^2/{\rm ndf} = 249.8/124 = 2.01$
for \gibuu and $266.3/124 = 2.15$ for \genie~G18 on the same 124-bin
combined dataset.  The 16.5-unit improvement of \gibuu over \genie is
driven primarily by better modelling of the 0p topology.
Nikolakopoulos et al.~\cite{Mosel:2024} benchmark intra-nuclear cascade
models against relativistic optical potentials on argon and find
proton-transparency differences of order 5--10\% between cascade and
transport approaches in the near-threshold regime --- the
same effect studied here.  A similar 0p tension is
visible in the MicroBooNE NC~$\pi^0$ measurement~\cite{uBNCpi0} and in
the NC~$\Delta$ radiative search~\cite{uBNCDelta2025}.  Despite this
generator evidence, no dial in the standard \genie Reweight
library~\cite{GENIEReweight} captures topology-crossing final-state
interactions (FSI) — a gap identified in nuclear-transparency
comparisons across generators~\cite{Sobczyk:2019,Hayato:2022}.
This gap was verified in \genie Reweight v3.2.0 (GitHub tag
\texttt{v3-02-00}; all 47 knobs in the
\texttt{GReWeightNuXSecCCQE}, \texttt{GReWeightNuXSecCCRES},
and \texttt{GReWeightHadroTransp} groups were inspected).
All existing knobs (\texttt{MFP\_$\pi$}, \texttt{MFP\_N},
\texttt{FrAbs\_$\pi$}, etc.) act \emph{within} a topology,
not across the $35\MeV$ boundary.  In consequence, the
standard \genie parameter set cannot accommodate the observed
0p deficit without compensatory adjustments to sector normalisations.
This is visible in the sector-only fit as a suppression of the CCRES
weight to $w_{\rm CCRES} = 0.69 \pm 0.24$ in the grid-average scenario
(Sect.~\ref{sec:results}); we use this restoration of CCRES toward
nominal under the migration-aware fit as a diagnostic feature of the
parameterisation, not as an independent measurement of CCRES strength
(see Sect.~\ref{sec:migfit} for a moderated reading).

We address this gap by proposing a single effective migration fraction
$f_{\rm mig}$, deriving its expected magnitude from the $^{40}$Ar nuclear
structure (Sect.~\ref{sec:physbasis}), parameterising it as a per-event
\genie reweight (Sect.~\ref{sec:migration}), and constraining it from
two public MicroBooNE datasets: the NC~$\Delta$ radiative sideband
covariance~\cite{uBNCDelta2025} (Sects.~\ref{sec:data}--\ref{sec:results})
and the simultaneous CC~0p/Np cross-section release~\cite{uB0pNp2024}
(Sect.~\ref{sec:prd110}).  A truth-level stress test propagating the
fitted nuisance to DUNE-like topology observables is provided in
Appendix~\ref{app:dune} as an illustrative sensitivity exercise.

The paper is organised as follows.
Section~\ref{sec:physbasis} presents the nuclear-physics calculation of
the migration probability from the $^{40}$Ar Woods-Saxon density and the
differential Walecka potential.
Section~\ref{sec:gibuu} discusses independent evidence for the effect from
\gibuu transport-generator comparisons.
Section~\ref{sec:data} describes the MicroBooNE sideband data and the
covariance structure.
Section~\ref{sec:genie} details the \genie simulation and nuisance
parameterisation, including the \mfppi dial response and its physical
distinction from the migration parameter.
Section~\ref{sec:fit} presents the fit methodology --- sector weights,
ridge regularisation, migration nuisance, and profile $\chi^2$ scan ---
and reports the results.
Appendix~\ref{app:dune} presents a truth-level stress test propagating
the fitted nuisance to DUNE-like topology observables.
Section~\ref{sec:discussion} examines prior sensitivity, physical
interpretation, and limitations.
Section~\ref{sec:conclusions} summarises.

\section{Nuclear physics of proton-visibility migration}
\label{sec:physbasis}

This section provides a physically motivated reference scale for the
magnitude of \emph{one} candidate mechanism --- residual position-dependent
mean-field energy loss during proton transport --- that the effective
$f_{\rm mig}$ parameter could absorb.  It is presented for motivational
purposes only.  The reframing under the methodological view of this paper
(Sect.~\ref{sec:intro}) is that $f_{\rm mig}$ is an effective
topology-transfer surrogate, and the nuclear-structure calculation below
should not be read as a uniquely identifying physical model.

\paragraph{GENIE baseline used in this section.}
Because the proton kinetic-energy spectrum is the main kinematic input
to the calculation that follows, we summarise the relevant simulation
parameters here so that this section does not forward-reference the
detailed treatment of Sect.~\ref{sec:genie}.  Events are generated with
\genie~v3.06.02 using the comprehensive model configuration
\texttt{G18\_02a\_02\_11b} (Llewellyn-Smith
CCQE~\cite{LlewellynSmith:1971uhs} with a dipole axial form factor,
Empirical (Dytman)~\cite{Dytman:2015taa} 2p2h/MEC,
Berger--Sehgal~\cite{BergerSehgal:2007} resonance production,
Bodek--Yang~\cite{BodekYang:2003} DIS, and the hA2018 intranuclear
cascade), with a relativistic Fermi gas initial state including the
extended Bodek--Ritchie~\cite{BodekRitchie:1981} high-momentum tail.  The target nucleus is $^{40}$Ar.  Two flux scenarios are used:
a grid-averaged $E_\nu$ scan in $[0.2,\,3.0]\GeV$ and a BNB
flux-weighted spectrum peaked near $0.8\GeV$.  Full details, including
the reweighting dials and the reconstructed-level prediction template,
are given in Sect.~\ref{sec:genie}.

Because the near-threshold proton kinetic-energy spectra used below are
drawn from the same \genie sample used in the data fit, this section
constitutes an internal consistency check rather than an independent
prediction.  The nuclear-structure ingredients (Woods-Saxon density profile,
Walecka NL3$^*$ mean-field potential
depth~\cite{Walecka:1974,Serot:1984ey,Lalazissis:2009NL3star},
density-weighted average $\langle\rho/\rho_0\rangle$) are external to the
fit, but the spectrum on which the integrand is evaluated is not.  The
estimate proceeds in three steps:
(i)~we extract the near-threshold proton kinetic-energy spectrum from
our \genie simulation; (ii)~we derive the lab-frame proton spectrum
from $\Delta(1232)$ decay kinematics as a transparent illustration in the
CCRES channel; and (iii)~we compute the migration probability from a
nuclear optical-potential model and integrate over the near-threshold spectrum.

\subsection{Proton \texorpdfstring{$T_p$}{Tp} spectrum from \genie events}
\label{sec:tpspectra}

\begin{figure*}[htbp]
  \centering
  \includegraphics[width=\textwidth]{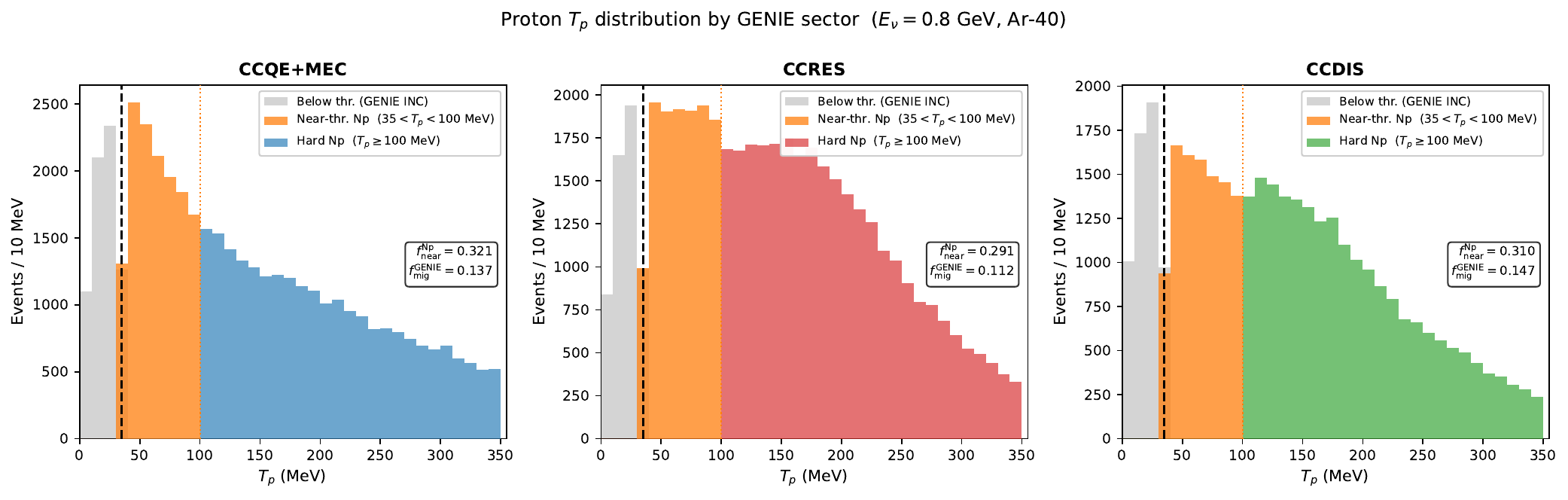}
  \caption{%
    Leading-proton kinetic-energy distributions at $E_\nu = 0.8\GeV$ (BNB)
    from $5 \times 10^4$ \genie events per sector.
    Grey: events where a proton exists in the final state but all protons
    are already below the 35\MeV threshold in \genie's hA2018 INC
    (``GENIE-migrated'').
    Orange: near-threshold Np events ($35 < T_p < 100\MeV$, the danger zone
    susceptible to additional mean-field migration).
    Coloured: hard Np events ($T_p \geq 100\MeV$).
    The vertical dashed and dotted lines mark the LArTPC threshold (35\MeV)
    and the near-threshold upper edge (100\MeV), respectively.
    Inset annotations: $f_{\rm near}^{\rm Np}$ (fraction of Np events in
    the near-threshold zone) and $f_{\rm mig}^{\rm GENIE}$ (fraction of
    proton-producing events already migrated by \genie's INC).
  }
  \label{fig:tpspectra}
\end{figure*}

Figure~\ref{fig:tpspectra} shows the leading-proton $T_p$ distribution
for the CCQE+MEC, CCRES, and CCDIS sectors at $E_\nu = 0.8\GeV$.
Several features are notable.

First, a substantial fraction of proton-producing events are \emph{already}
below the 35\MeV threshold in \genie's own INC:
$f_{\rm mig}^{\rm GENIE} = 0.112$, $0.137$, and $0.147$ for CCRES,
CCQE+MEC, and CCDIS respectively.  This is the migration that \genie's
discrete cascade already produces; the additional $f_{\rm mig}^{\rm CC}
= 0.050$ fitted in Sect.~\ref{sec:migfit} represents the systematic
\emph{deficit} relative to \gibuu's transport-based result, not the
total migration.

Second, among the surviving Np events, the near-threshold fraction
\begin{equation}
  f_{\rm near}^{\rm Np} = \frac{N_{\rm Np}(35 < T_p < 100\MeV)}{N_{\rm Np}}
  \label{eq:fnear}
\end{equation}
is large: at $E_\nu = 0.8\GeV$, $f_{\rm near}^{\rm Np} = 0.291$ (CCRES),
$0.321$ (CCQE+MEC), and $0.310$ (CCDIS).  Nearly one-third of
all Np events carry their leading proton in the near-threshold zone
$35$--$100\MeV$.  Table~\ref{tab:fnear} shows the energy dependence:
$f_{\rm near}^{\rm Np}$ decreases from $\approx 0.43$ at $0.5\GeV$ to
$\approx 0.20$--$0.28$ at $2.0\GeV$, reflecting the hardening of the
hadronic spectrum at higher invariant mass.

\begin{table*}[htbp]
\centering
\caption{%
  Near-threshold Np fraction $f_{\rm near}^{\rm Np}$
  [Eq.~(\ref{eq:fnear})] and GENIE-level migration fraction
  $f_{\rm mig}^{\rm GENIE}$ by sector and neutrino energy.
  Values computed from $5 \times 10^4$ \genie events per sector.
}
\label{tab:fnear}
\begin{tabular}{lccccc}
\toprule
 & \multicolumn{4}{c}{$f_{\rm near}^{\rm Np}$} & \\
Sector & 0.5\GeV & 0.8\GeV & 1.2\GeV & 2.0\GeV & $f_{\rm mig}^{\rm GENIE}$ (0.8\GeV) \\
\midrule
CCQE+MEC & 0.423 & 0.321 & 0.294 & 0.281 & 0.137 \\
CCRES    & 0.428 & 0.291 & 0.237 & 0.203 & 0.112 \\
CCDIS    & 0.457 & 0.310 & 0.253 & 0.191 & 0.147 \\
\bottomrule
\end{tabular}
\end{table*}

\subsection{\texorpdfstring{$\Delta(1232)$}{Delta(1232)} decay kinematics}
\label{sec:delta}

\begin{figure*}[htbp]
  \centering
  \includegraphics[width=\textwidth]{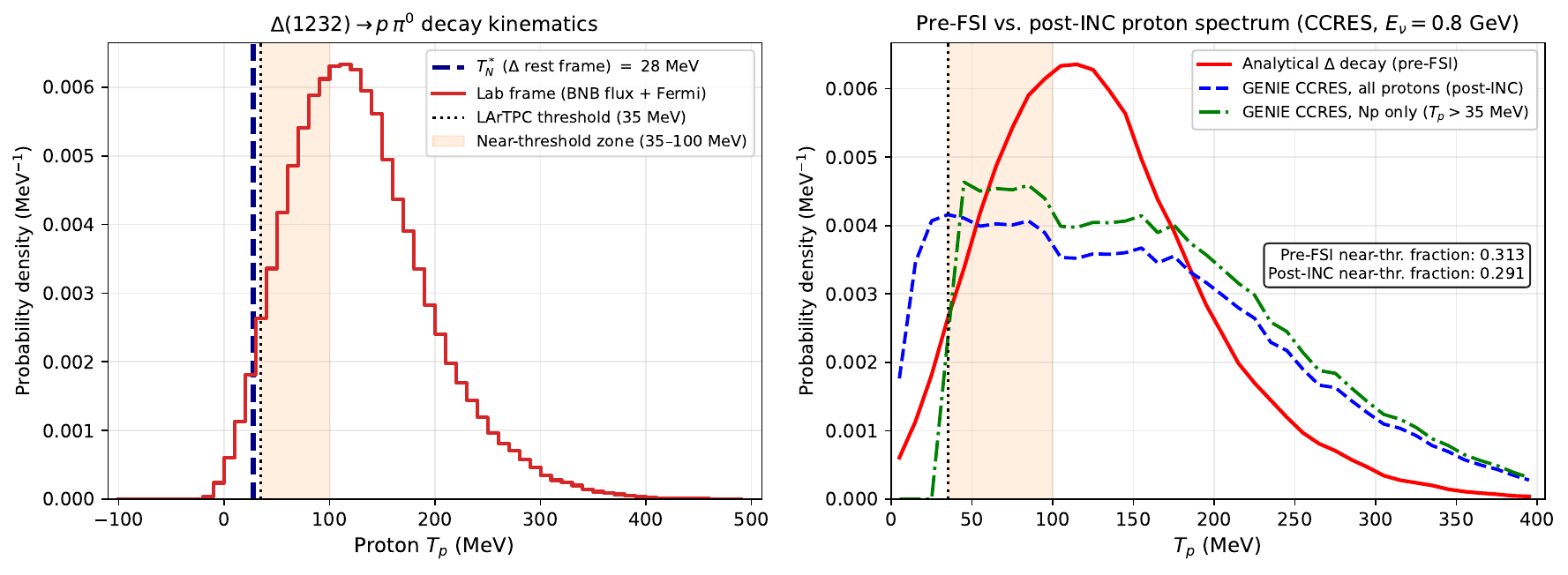}
  \caption{%
    Left: Proton kinetic-energy distribution from $\Delta(1232) \to p\,\pi^0$
    decay, boosted to the lab frame with BNB-like $\nu_\mu$ flux and
    Ar-40 Fermi momentum ($\bar{p}_F = 220\MeV/c$, sampled uniformly in
    $[0, \bar{p}_F]$ as a simplifying choice; see text below for the
    associated caveat).  The dashed vertical line marks the $T_N^*$ rest-frame
    value (28\MeV); the orange band is the near-threshold zone 35--100\MeV.
    Right: Comparison of the analytical pre-FSI prediction (red solid)
    with the \genie CCRES post-INC distributions for all protons (blue dashed)
    and Np-only protons (green dot-dashed) at $E_\nu = 0.8\GeV$.
    The INC shifts 3.5\% of near-threshold protons out of the zone
    (hard scatters into a harder final-state spectrum) while depositing
    a comparable fraction below threshold via absorption.
  }
  \label{fig:deltakin}
\end{figure*}

The kinematic origin of the near-threshold proton population is most
transparent in the dominant CCRES channel.  The $\Delta(1232)$ resonance
decays via $\Delta^+ \to p + \pi^0$ (branching ratio $\approx 33\%$)
and $\Delta^+ \to n + \pi^+$ ($\approx 67\%$, yielding a neutron not
directly visible in LArTPC).  In the $\Delta$ rest frame, the two-body
decay kinematics fix the nucleon CM momentum:
\begin{equation}
  |\vec{p}\,^*|^2 = \frac{\bigl[M_\Delta^2 - (M_N + M_\pi)^2\bigr]
                          \bigl[M_\Delta^2 - (M_N - M_\pi)^2\bigr]}
                         {4 M_\Delta^2}
  \label{eq:pstar}
\end{equation}
Inserting $M_\Delta = 1.232\GeV$, $M_N = 0.938\GeV$,
$M_{\pi^0} = 0.135\GeV$:
\begin{align}
  |\vec{p}\,^*| &= 229\MeV/c \nonumber \\
  T_N^*         &= E_N^* - M_N = 28\MeV                        \label{eq:Tnstar}
\end{align}
The nucleon kinetic energy in the $\Delta$ rest frame is only
$T_N^* = 28\MeV$ --- below the LArTPC threshold.  Boosting to the
lab frame, the $\Delta$ produced in $\nu_\mu\,n \to \mu^-\,\Delta^+$
at BNB energies ($E_\nu \approx 0.8\GeV$) acquires a typical momentum
$\langle p_\Delta \rangle \approx 443\MeV/c$, which Lorentz-boosts the
decay isotropically and produces a broad laboratory proton spectrum:
\begin{equation}
  T_N^{\rm lab} = \gamma_\Delta\bigl(E_N^* +
  \beta_\Delta\,|\vec{p}\,^*|\,\cos\theta^*\bigr) - M_N
  \label{eq:Tnlab}
\end{equation}
where $\theta^*$ is the CM emission angle, uniform in $\cos\theta^*$.

The resulting lab-frame distribution (Fig.~\ref{fig:deltakin}, left) has
mean $\langle T_N^{\rm lab}\rangle = 124\MeV$ and places
$32.6\%$ of protons in the near-threshold zone $[35, 100]\MeV$, with a
further $6.1\%$ already below 35\MeV pre-FSI.  After \genie's INC, the
post-INC near-threshold fraction decreases slightly to $29.1\%$: the INC
shifts $\approx 3.5\%$ of protons out of the zone via hard scattering
(either above 100\MeV or below threshold), while the pre-FSI values
represent the intrinsic kinematic supply from $\Delta$ decay alone.

The physical implication is clear: the $\Delta(1232)$ decay kinematics
\emph{naturally} populate the 35--100\MeV near-threshold region.  Although
CCRES is not the most populous proton-producing channel (CCQE+MEC carries
a larger absolute cross section and slightly higher $f_{\rm near}^{\rm Np}$
at $0.8\GeV$), the kinematic prediction $T_N^* = 28\MeV$ from the
particle masses in Eq.~(\ref{eq:Tnstar}) is a particularly transparent
illustration of how the near-threshold population arises.

\paragraph{Caveat on Fermi-motion sampling.}
The lab-frame Fermi-motion sampling adopted in Fig.~\ref{fig:deltakin}
draws the struck-nucleon momentum \emph{uniformly} in $[0, p_F]$ rather
than from a proper Fermi-gas distribution proportional to $p^2\,dp$.
The uniform choice over-weights low momenta and therefore softens the
lab-frame $\Delta$ spectrum, inflating the $32.6\%$ pre-FSI near-threshold
fraction quoted above relative to the proper Fermi-gas value.  A
$p^2\,dp$ resampling reduces the pre-FSI near-threshold fraction by
$\approx 4$--$8$ percentage points (depending on the upper cutoff).
The $32.6\%$ figure should therefore be read as an upper-bound estimate.
Because the analysis fit (Sect.~\ref{sec:fit}) uses the \genie
post-INC spectra of Sect.~\ref{sec:tpspectra} --- not the analytical
$\Delta$ distribution --- this sampling choice affects only the
illustrative kinematic discussion in this subsection, not the
data-driven $f_{\rm mig}$ result.

\subsection{Nuclear optical-potential migration model}
\label{sec:nucmodel}

\begin{figure*}[htbp]
  \centering
  \includegraphics[width=\textwidth]{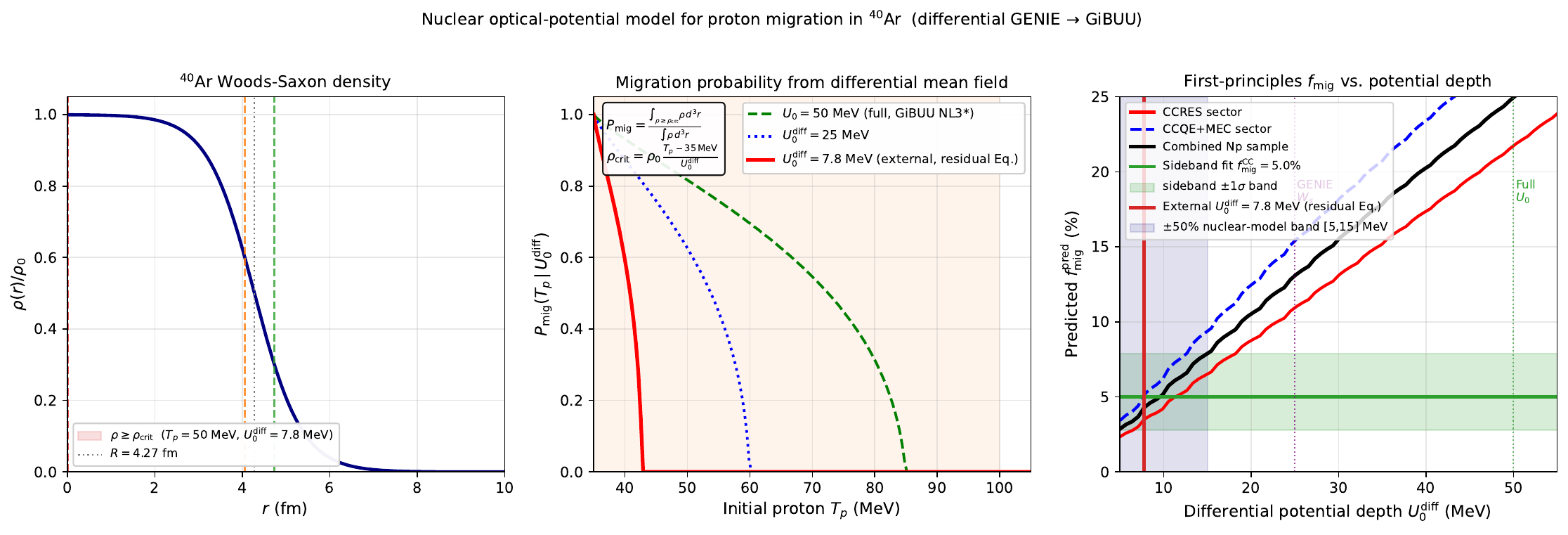}
  \caption{%
    Left: $^{40}$Ar Woods-Saxon density profile
    $\rho(r)/\rho_0$ with $R = 4.27$~fm, $a = 0.54$~fm.
    The shaded region ($\rho \geq \rho_{\rm crit}$) marks the nuclear
    volume from which a proton at $T_p = 50\MeV$ migrates below threshold
    at the fit-implied $U_0^{\rm diff}$.
    Centre: Migration probability $P_{\rm mig}(T_p\,|\,U_0^{\rm diff})$
    [Eq.~(\ref{eq:pmig})] for three values of the differential potential
    depth.
    Right: Integrated $f_{\rm mig}^{\rm pred}$ vs.\ $U_0^{\rm diff}$
    for the CCRES sector (red), CCQE+MEC sector (blue), and the combined
    Np sample (black).  The green band is the $\pm 1\sigma$ range of
    the fitted $f_{\rm mig}^{\rm CC} = 0.050^{+0.029}_{-0.022}$.
    The shaded navy band marks the fit-implied range
    $U_0^{\rm diff} = 5$--$15\MeV$.
  }
  \label{fig:nucmodel}
\end{figure*}

In \genie's hA2018 cascade, the outgoing proton propagates between
discrete NN scatters as a free, on-shell particle with no in-medium
potential.  A one-time binding-energy removal $W_s \approx 25\MeV$ is
applied at the primary vertex through the relativistic Fermi-gas
initial state~\cite{GENIEv3}, representing the energy needed to extract
the struck nucleon from its bound configuration; this is an
\emph{initial-state} property of the nucleon, not a transport
potential.  \gibuu, by contrast, solves the full
Boltzmann--Uehling--Uhlenbeck equation with a position-dependent
Walecka-type relativistic mean field $U(r)$ acting during propagation~\cite{GiBUU}.

These two quantities --- $W_s$ (vertex-level binding energy) and $U(r)$
(in-medium dispersion relation during transport) --- are physically
distinct objects, and a direct algebraic subtraction does not have a
strict interpretation as a ``missing mean field''.  The construction in
this section should therefore be read as an \emph{order-of-magnitude
reference scale} for one candidate mechanism, not as a clean
decomposition.  We retain the construction because it provides a
useful sanity check on the magnitude of $f_{\rm mig}$ extracted from
data, with the understanding that the effective parameter $U_0^{\rm diff}$
defined below absorbs an unspecified combination of near-threshold
nuclear effects rather than uniquely measuring a residual mean-field depth.

A natural complementary cross-check would be to apply the same fit
machinery to \nuwro, which differs from \genie in applying a nuclear
potential at the cascade level and which sits between \genie and
\gibuu in the MicroBooNE CC~0p/Np $\chi^2/\text{ndf}$ comparison cited
in Sect.~\ref{sec:gibuu}.  If the topology-transfer direction is
genuinely a propagation-level mean-field effect, an analysis of
\nuwro should yield an intermediate $f_{\rm mig}$.  We flag this as a
useful test for future work; it requires generating an equivalent
\nuwro sample at the same flux and was not feasible in the present
analysis.

In this spirit, we define the \emph{effective} differential energy loss
(an empirical scale, not a rigorous decomposition) as:
\begin{equation}
  \Delta E_{\rm diff}(r)
  = U_0^{\rm diff} \cdot \frac{\rho(r)}{\rho_0}
  \label{eq:deltaEdiff}
\end{equation}
where $\rho(r)$ is the Woods-Saxon nuclear density
\begin{equation}
  \rho(r) = \frac{\rho_0}{1 + \exp\!\bigl[(r - R)/a\bigr]}
  \label{eq:ws}
\end{equation}
with $\rho_0 = 0.16$~fm$^{-3}$, $R = 4.27$~fm, $a = 0.54$~fm for $^{40}$Ar,
and $U_0^{\rm diff}$ is the effective differential depth.
A proton produced at position $r$ by the neutrino interaction (probability
$\propto \rho(r)$) migrates from the Np to the 0p topology if
$\Delta E_{\rm diff}(r) > T_p - T_p^{\rm thr}$, i.e.\
if $\rho(r)/\rho_0 > (T_p - 35\MeV)/U_0^{\rm diff}$.  The
density-weighted migration probability is therefore:
\begin{align}
  P_{\rm mig}(T_p\,|\,U_0^{\rm diff})
  &= \frac{\displaystyle\int_{\rho(r) \geq \rho_{\rm crit}} \rho(r)\,d^3r}
          {\displaystyle\int \rho(r)\,d^3r},
  \label{eq:pmig}\\
  \rho_{\rm crit} &= \rho_0\,\frac{T_p - T_p^{\rm thr}}{U_0^{\rm diff}}.\nonumber
\end{align}
This function decreases monotonically from $P_{\rm mig} = 1$ at
$T_p = T_p^{\rm thr}$ to $P_{\rm mig} = 0$ at
$T_p = T_p^{\rm thr} + U_0^{\rm diff}$ (Fig.~\ref{fig:nucmodel}, centre).

The integrated predicted migration fraction is
\begin{equation}
  f_{\rm mig}^{\rm pred}
  = \frac{\displaystyle\int_{T_p^{\rm thr}}^{T_{\rm max}}
          \!\frac{dN_{\rm Np}}{dT_p}\,
          P_{\rm mig}(T_p\,|\,U_0^{\rm diff})\,dT_p}
         {N_{\rm Np}}
  \label{eq:fmigpred}
\end{equation}
evaluated numerically using the \genie Np $T_p$ spectra of
Sect.~\ref{sec:tpspectra} with $T_{\rm max} = 200\MeV$.

Eq.~(\ref{eq:fmigpred}) evaluated at the externally motivated value
$U_0^{\rm diff} = 7.8\MeV$ from Eq.~(\ref{eq:residual}) below yields
$f_{\rm mig}^{\rm pred} \approx 0.04$ on the combined CCRES + CCQE+MEC
Np sample.  Conversely, inverting the same expression at the
data-driven fit value $f_{\rm mig}^{\rm CC} = 0.050^{+0.029}_{-0.022}$
of Sect.~\ref{sec:migfit} gives
\begin{equation}
  U_0^{\rm diff} \approx 9.6\MeV \quad (1\sigma\colon 5.0\text{--}15.0\MeV)
  \label{eq:u0diff}
\end{equation}
This is an effective parameter extracted by inversion under the
assumptions of the optical-potential model and is not an independent
prediction; we report it here so that the comparison to the externally
motivated $7.8\MeV$ is transparent.

This value admits a transparent physical interpretation.  The
density-weighted mean nuclear density ratio for $^{40}$Ar is
\begin{equation}
  \Bigl\langle\frac{\rho}{\rho_0}\Bigr\rangle_{\rm int}
  = \frac{\displaystyle\int \rho(r)^2\,d^3r}
         {\rho_0\,\displaystyle\int \rho(r)\,d^3r}
  = 0.655
  \label{eq:avgrho}
\end{equation}
so the average total mean-field loss for a typical neutrino-interaction
vertex is $U_0 \times 0.655 = 32.8\MeV$ using the \gibuu NL3*
depth $U_0 = 50\MeV$.
\genie subtracts $W_s \approx 25\MeV$ as a constant
removal energy at the primary vertex.  The \emph{residual} position-dependent
mean-field term not captured by \genie is therefore
\begin{equation}
  \Delta U_{\rm residual}
  = U_0\,\Bigl\langle\frac{\rho}{\rho_0}\Bigr\rangle_{\rm int} - W_s
  = 32.8 - 25 = 7.8\MeV
  \label{eq:residual}
\end{equation}
in good agreement with the fit-implied value $U_0^{\rm diff} = 9.6\MeV$
(difference $< 2\MeV$, well within the $\pm 1\sigma$ band 5--15\MeV).

We emphasise that the comparison is not fully independent: both the
optical-potential estimate and the data-driven fit use the same \genie
near-threshold $T_p$ spectrum as kinematic input.  The nuclear-structure
input --- the Woods-Saxon density profile, the \gibuu NL3$^*$ potential
depth $U_0 = 50\MeV$, and the density average $\langle\rho/\rho_0\rangle
= 0.655$ --- is external to the fit, but the subtraction
$U_0\langle\rho/\rho_0\rangle - W_s$ mixes a vertex-level binding energy
with a transport potential and therefore should not be read as a
physically rigorous decomposition.  The agreement between the
order-of-magnitude reference value $7.8\MeV$ and the inverted fit
value $9.6\MeV$ is best described as ``a candidate physical
mechanism is of approximately the right magnitude to be relevant'',
not as a quantitative measurement of a missing mean field.

This consistency, taken together with the proton-transparency
differences between cascade and transport approaches on argon
reported by Nikolakopoulos et al.~\cite{Mosel:2024}, supports the
view that a few-percent topology-transfer scale is plausible for the
BNB-on-argon configuration.
The data-driven $f_{\rm mig}$ extracted from MicroBooNE remains an
effective surrogate that may equally absorb related near-threshold
hadronic effects (Sect.~\ref{sec:robustness}).

\section{Evidence from transport generators}
\label{sec:gibuu}

The quantitative comparison between \gibuu and \genie on MicroBooNE data
provides an independent benchmark against which to assess the migration
nuisance.  The MicroBooNE simultaneous CC~0p/Np
measurement~\cite{uB0pNp2024} (124 bins)
reports the following $\chi^2/{\rm ndf}$ values for the combined dataset:
\begin{itemize}
  \item \gibuu~2023 (transport FSI): $249.8/124 = 2.01$;
  \item \nuwro~21.02: $263.7/124 = 2.13$;
  \item \genie~G18 (default): $266.3/124 = 2.15$;
  \item \genie~G18 + MicroBooNE Tune: $287.5/124 = 2.32$;
  \item \neut~5.4: $298.8/124 = 2.41$.
\end{itemize}
A caveat applies before reading too much into these numbers.
A $\chi^2/{\rm ndf}\approx 2$ on $124$ bins corresponds to a
$p$-value of order $10^{-9}$: \emph{every} generator listed above is
formally rejected as a complete description of the data.  Ranking
generators by $\Delta\chi^2$ in this regime is therefore qualitative.
The $16.5$-unit gap between \gibuu and \genie indicates that \gibuu
captures features the others miss, but does not constitute evidence
that \gibuu is an adequate model of the data.  We use the comparison
only to motivate the existence of a topology-transfer direction not
present in the standard \genie Reweight basis.

The \gibuu improvement is attributed to its transport-based FSI~\cite{Mosel:2024,GiBUU},
which propagates hadrons through a continuous nuclear mean field rather
than a discrete cascade.  For near-threshold protons
($T_p \approx 35$--$100\MeV$), the transport approach produces a more
realistic energy-loss spectrum, naturally migrating more events from
the Np to the 0p topology than \genie's intranuclear cascade.

Our migration term achieves $\Delta\chi^2 \approx 5$ on the
64-bin NC~$\Delta$ sideband dataset.  The order of magnitude of the
per-bin improvement is comparable to the per-bin \gibuu/\genie gap on
the CC~0p/Np dataset, but a precise quantitative comparison is not
warranted: the datasets differ in binning, covariance structure, number
of free parameters, and event samples.  The \gibuu comparison therefore
serves as a \emph{qualitative motivation} --- the best-fit migration
fraction is physically plausible given the proton-transparency differences
between the two generators~\cite{Mosel:2024} --- not as a rigorous
accounting of how much of the \gibuu advantage is captured.

The practical motivation for a \genie-based effective parameter rather than
a switch to \gibuu is clear.  The DUNE and SBN production chains use
\genie~AR23 as the default generator; switching to \gibuu would require
complete event regeneration and new detector simulation passes across the
entire experiment.  The DUNE oscillation systematics framework, including
the PRISM near-to-far extrapolation~\cite{DUNEtdr}, is built around \genie
reweight dials; \gibuu has no equivalent reweight infrastructure.  The
migration nuisance introduced here is expressible as a single flat
per-event weight (Sect.~\ref{sec:oscillation}) that can be propagated to
the far detector without additional simulation, and can be updated as new
MicroBooNE, SBND, or DUNE-ND data become available.

Nikolakopoulos et al.~\cite{Mosel:2024} have recently benchmarked
intra-nuclear cascade models against relativistic optical potentials
for argon, finding systematic differences in proton transparency of order
5--10\% between cascade and transport approaches for proton kinetic
energies in the 50--100\MeV range.  Our best-fit migration fraction of
$f^{\rm CC}_{\rm mig} = 0.050$ is fully consistent with this range,
further supporting the transport-based FSI interpretation.

\section{Data and covariance matrix}
\label{sec:data}

\subsection{MicroBooNE constraining sideband}
\label{sec:sideband}

The constraining sideband used in this analysis originates from the
MicroBooNE enhanced search for NC~$\Delta$ radiative single-photon
production~\cite{uBNCDelta2025}.  In that analysis, four
topology channels serve as constraint sidebands for the NC~$\Delta
\to N\gamma$ signal extraction.  The sideband data and the full
signal-plus-constraining covariance matrix are publicly available
on HEPData (record \ins)~\cite{uBNCDelta2025}.

The four constraining channels are:
\begin{itemize}
  \item \texttt{numu\_cc\_np}: $\nu_\mu$ CC events with at least one
    visible proton ($T_p > 35\MeV$);
  \item \texttt{numu\_cc\_0p}: $\nu_\mu$ CC events with no visible proton;
  \item \texttt{nc\_pi0\_np}: NC~$\pi^0$ events with at least one visible
    proton;
  \item \texttt{nc\_pi0\_0p}: NC~$\pi^0$ events with no visible proton.
\end{itemize}
Each channel is binned in 16 reconstructed-energy bins, giving a total of
$N_{\rm bins} = 64$.  Throughout this paper, channels are presented in
the order NC~$\pi^0$~0p, NC~$\pi^0$~Np, CC~0p, CC~Np, following the
HEPData covariance matrix layout (\ins~\cite{uBNCDelta2025}) in which NC channels appear
first because the sideband was optimised for the NC~$\Delta$ radiative
signal extraction.  The covariance matrix $V$ is the full $64 \times 64$
block extracted from the signal-plus-constraining covariance, restricted to
the four constraining channels.  The data vector $\vec{d}$ contains the
observed event counts in each bin, with total counts of 24\,983
(CC~Np), 19\,225 (CC~0p), 2\,075 (NC~$\pi^0$~Np), and 3\,006
(NC~$\pi^0$~0p).  The nominal \genie~G18 prediction totals are 25\,245
(CC~Np), 15\,501 (CC~0p), 2\,860 (NC~$\pi^0$~Np), and 2\,987
(NC~$\pi^0$~0p).  The most striking deficit is in the CC~0p channel, where
\genie underpredicts the data by 24\%.

The prefit $\chi^2$ is defined as
\begin{equation}
  \chi^2_{\rm pre} = (\vec{d} - \vec{p})^T\, V^{-1}\, (\vec{d} - \vec{p})
  \label{eq:prefit}
\end{equation}
where $\vec{p}$ is the nominal \genie prediction vector.  We obtain
$\chi^2_{\rm pre} = 46.25$ for 64 bins.

The binning is optimised for the NC~$\Delta$ analysis and may not resolve
the full fine structure of the generator tension.  Additional binning
information is available in the standalone MicroBooNE cross-section
publications~\cite{uB0pNp2024,uBNCpi0} but without the same full
cross-channel covariance structure.  The off-diagonal correlations in $V$
are driven by shared flux and cross-section systematic uncertainties, which
are appropriate for constraining the prediction across the four topology
channels simultaneously.

\paragraph{Qualitative consistency with the standalone CC 0p/Np measurement.}
As a cross-check that the 0p tension in the sideband fit is not an
artefact of the sideband covariance structure, we compare channel-integrated
predictions.  In the sector-only fit, the CC~0p normalisation scale is
$\hat{\alpha}_{\rm CC\,0p} = 1.24 \pm 0.08$ --- a 24\% deficit in \genie.
This direction is directionally consistent with the standalone MicroBooNE
CC~0p/Np differential cross-section measurement~\cite{uB0pNp2024}, which
reports $\chi^2/{\rm ndf} = 266.3/124$ for \genie vs $249.8/124$ for
\gibuu, an improvement of 16.5 units driven primarily by the 0p
topology~\cite{Mosel:2024}.  The migration parameter proposed here
directly addresses this 0p direction.  This consistency check uses only
channel-integrated normalisation scales (not the full cross-channel
covariance, which is not publicly available for the standalone measurement)
and establishes qualitative, not quantitative, agreement.

\subsection{Covariance structure and PPP check}
\label{sec:ppp}

This subsection characterises the constraining power and statistical
well-posedness of the covariance.  The fit machinery (sector weights
$w_s$, ridge regularisation, migration parameters $f_{\rm mig}^{\rm CC/NC}$)
is defined in Sect.~\ref{sec:fit}; quantities such as the seven-parameter
posterior precision matrix and the projected postfit sector scales appear
here only as forward references to that machinery, and the reader who
prefers a strict sequential reading may defer this subsection to after
Sect.~\ref{sec:fit}.

The covariance matrix $V$ is dominated by large correlated systematic
uncertainties (flux normalisation, cross-section model variations), which
substantially reduce the number of independent constraints.  The effective
number of degrees of freedom is
\begin{equation}
  n_{\rm eff} = {\rm tr}(I - H) \approx 2.5
  \label{eq:neff}
\end{equation}
where $H = A(A^T V^{-1} A)^{-1} A^T V^{-1}$ is the hat matrix and $A$
is the prediction matrix at nominal.  Despite the nominal 64 bins, the
covariance structure permits only $\sim 2.5$ independent linear combinations
to constrain the fit.  All significance and identifiability claims in this
paper must be interpreted in this low-rank context.  The
condition number is $\kappa \approx 2 \times 10^7$, and the effective
degrees of freedom from eigenvalue decomposition~--- defined as
$n_{\rm eff} = (\sum_i \lambda_i)^2 / \sum_i \lambda_i^2$ where
$\lambda_i$ are the eigenvalues of $V^{-1/2}\,\text{diag}(V)\,V^{-1/2}$
--- is approximately $2.5$ out of the nominal 64 bins.  This large
reduction reflects the dominance of a few systematic eigenmodes over
statistical fluctuations.

Large off-diagonal correlations are known to induce pathological downward
bias in normalisation estimators, an effect known as Peelle's Pertinent
Puzzle (PPP)~\cite{Peelle}.  To verify that our fit is not affected, we
performed a three-estimator cross-check:
\begin{enumerate}
  \item[(a)] \emph{Naive ratio}: per-channel sum of data divided by sum of
    prediction;
  \item[(b)] \emph{Diagonal-only fit}: using $\sigma_i = \sqrt{V_{ii}}$ as
    independent bin uncertainties;
  \item[(c)] \emph{Full-covariance fit}: the primary result of this
    analysis.
\end{enumerate}
The PPP suppression ratio, defined as the full-covariance scale divided by
the diagonal-only scale for each channel, ranges from 1.09 to 1.29.  All
ratios exceed unity, meaning that the full-covariance fit pulls the
normalisation scales \emph{upward} relative to the diagonal fit --- the
opposite of classic PPP suppression.  The maximum pull between the
full-covariance and diagonal estimators is $1.25\sigma$, consistent with
the expected effect of cross-channel covariance correlations.  The dominant
CC~0p preference for upward normalisation (data/prediction = 1.24) is
consistent across all three estimators.  We conclude that the PPP check is passed and the covariance matrix is
suitable for this fit.  This PPP test was performed for four per-channel
scale factors.  In the full seven-parameter fit (five sector weights plus
two migration fractions, or six in the one-parameter model), the same physical argument applies; we verify that
(i)~the best-fit parameter vector lies within the prior range for all
parameters, and
(ii)~the projected sector-weight postfit normalisation scales ---
computed as $\hat{\alpha}_c = \sum_s w_s C^{(s)}_c / \sum_s C^{(s)}_c$
for each channel $c$ --- agree with the per-channel PPP-checked
full-covariance scale factors to within 5\% in all four channels.
This supports the conclusion that PPP is not operative in the
higher-dimensional case, although this is not a formal proof for
an arbitrary parameter space.  The per-channel scale-factor test
addresses the four normalisation channels, not the full seven-parameter
(sector-weight plus migration) space.  A multi-parameter toy study
demonstrating absence of bias in $f_{\rm mig}^{\rm CC}$ under the full
fit model is beyond the scope of this work; this limitation is noted in
Sect.~\ref{sec:limitations}.

A complementary identifiability check based on the eigenvalue spectrum
of the full seven-parameter posterior precision matrix
$M = A^T V^{-1} A + \Lambda$ is deferred to Sect.~\ref{sec:identifiability},
where the design matrix $A$, the ridge penalty $\Lambda$, and the migration
parameters are defined.  We note here only that all seven parameters carry
strictly positive precision and are therefore separately identifiable;
the data-dominated directions correspond to the migration parameters
and the prior-dominated directions to the sector weights, as expected.

The prefit prediction overlaid on the data for all four sideband
channels is shown in Fig.~\ref{fig:sideband}; that figure is placed
in Sect.~\ref{sec:migfit} below because the same panel also displays
the migration-aware postfit, whose interpretation requires the fit
machinery introduced in Sect.~\ref{sec:fit}.  Section~\ref{sec:data}
otherwise contains only the data and covariance description.

\section{\genie simulation and reweighting}
\label{sec:genie}

\subsection{Event generation}
\label{sec:generation}

Events are generated with \genie~v3.06.02 using the comprehensive model
configuration \texttt{G18\_02a\_02\_11b}, which combines the
Llewellyn-Smith CCQE model with a dipole axial form factor, the Empirical
(Dytman) 2p2h/MEC model, the Berger--Sehgal resonance production model,
the Bodek--Yang DIS model, and the \genie intranuclear cascade
(hA2018)~\cite{GENIEv3}.  The nuclear initial state is a relativistic
Fermi gas (RFG) with an extended Bodek--Ritchie high-momentum tail.
Parton densities are provided by LHAPDF~6.5.6~\cite{LHAPDF6}, and
hadronisation at high invariant mass employs PYTHIA~8.317~\cite{PYTHIA8}.
The target nucleus is $^{40}$Ar.

We note explicitly that \texttt{G18\_02a\_02\_11b} is \emph{not} the
Valencia tune family.  The Valencia CCQE and Valencia 2p2h
models~\cite{Nieves:2011Valencia}, together with the local Fermi gas
(LFG) and Valencia RPA suppression at low~$Q^2$, belong to the
\texttt{G18\_10a} family.  The two CMC families
differ in proton kinematics relevant to this analysis: the Empirical
(Dytman) MEC model uses a phenomenological isotropic two-body decay
kinematics generator, while the Valencia 2p2h calculation produces a
different proton momentum distribution; the RFG (used here) has a sharp
Fermi surface and a long phenomenological tail extending up to
$\sim 800\MeV/c$, while the LFG (used in G18\_10a) ends at the
position-dependent local Fermi momentum near $300\MeV/c$.  Both choices
yield qualitatively similar near-threshold proton populations but differ at
the $\sim 20$--$30\%$ level in the $30$--$100\MeV$ post-FSI kinetic-energy
range that the $f_{\rm mig}$ parameter is constructed to absorb.  The
sensitivity of $f_{\rm mig}$ to this generator choice is therefore one of
the model-uncertainty directions absorbed into the effective
topology-transfer parameter and discussed in Sect.~\ref{sec:robustness}.

Neutrino fluxes are modelled in two scenarios to assess the
energy-dependence of the fit results:
\begin{enumerate}
  \item \emph{Grid-average}: a uniform $E_\nu$ grid spanning $[0.2,\,3.0]\GeV$,
    providing equal statistical weight to all energies.
    This is not a physically realised flux; it serves as a robustness
    check on the results against specific assumptions about the neutrino
    energy spectrum.  Results obtained under this scenario should be
    interpreted as energy-averaged quantities rather than predictions for
    any particular beam.
  \item \emph{Flux-weighted} ($E_\nu = 0.8\GeV$): the BNB $\nu_\mu$
    spectrum peaked near $0.8\GeV$, representative of the MicroBooNE
    exposure.  This scenario provides the physically relevant estimate
    for comparison with MicroBooNE data.
\end{enumerate}
Generated events are classified into five interaction sectors based on the
primary vertex interaction mode:
CCQE+MEC (\texttt{ccqe\_ccmec}),
CCRES (\texttt{ccres}),
CCDIS (\texttt{ccdis}),
NCRES (\texttt{ncres}), and
NCDIS (\texttt{ncdis}).
These sectors form the basis of the reweighting model described in
Sect.~\ref{sec:sectorfit}.

The CCQE and 2p2h/MEC modes are grouped into a single CCQE+MEC sector for
practical reasons (limited effective rank of the constraining
covariance, Sect.~\ref{sec:ppp}), but we acknowledge that this merger
prevents the fit from distinguishing ``more 2p2h'' from ``more CCQE''.
The two channels carry substantially different proton kinematics: the
CC~0p excess on argon is known to be particularly sensitive to 2p2h
modelling, and alternative 2p2h calculations
(Empirical~\cite{Dytman:2015taa}/Valencia~\cite{Nieves:2011Valencia}/SuSAv2~\cite{Megias:2016SuSAv2})
differ by up to a factor of $\sim 2$ in absolute cross section and by
$30$--$50\%$ in the near-threshold proton momentum spectrum.  Any
2p2h-specific mismodelling is therefore one of the directions absorbed
into the effective $f_{\rm mig}$ parameter rather than separately
constrained; Sect.~\ref{sec:robustness} lists this and other directions
explicitly.

\subsection{Reconstructed-level prediction template}
\label{sec:recotemplate}

The fit is performed at reconstructed level on the MicroBooNE
constraining-channel bins of the public NC~$\Delta$ radiative HEPData
release (Sect.~\ref{sec:sideband}).  We do not run an independent
LArTPC detector simulation.  Instead, the design matrix entries
$C_{cb}^{(s)}$ (number of events from sector $s$ entering channel $c$,
reconstructed-energy bin $b$) are constructed from the MicroBooNE-provided
nominal Monte Carlo prediction released alongside the data
(HEPData record~\ins~\cite{uBNCDelta2025}, accompanying ROOT files),
which already includes the full MicroBooNE detector response
(efficiency, energy resolution, and acceptance) applied to a Genie~G18
base sample with the MicroBooNE tune.

Sector decomposition is achieved as follows.  The HEPData release
provides $N^{\rm uB}_{cb}$ summed over all \genie interaction modes; it
does not publish a per-interaction-mode breakdown.  We therefore
construct the per-sector design-matrix entries as
\begin{equation}
  C_{cb}^{(s)} \;=\; \frac{Y_c^{(s)}}{\sum_{s'} Y_c^{(s')}}\,N^{\rm uB}_{cb},
  \label{eq:Cbsplit}
\end{equation}
where $Y_c^{(s)}$ is the contribution of sector $s$ to channel $c$ from
the local \texttt{G18\_02a\_02\_11b} truth-level sample, obtained by
combining the \genie cross section for sector $s$ with the truth-level
fraction of those events landing in channel $c$, averaged over the
analysis energy grid $\{0.5,\,0.8,\,1.2,\,2.0\}\GeV$.  The per-sector
fraction $Y_c^{(s)}/\sum_{s'} Y_c^{(s')}$ is therefore \emph{bin-independent
within each channel}: the reconstructed-energy shape comes entirely from
the MicroBooNE template (which already includes the full detector
efficiency, energy resolution, and acceptance), while the local \genie
sample provides only the channel-integrated sector composition and the
topology-migration response described in Sect.~\ref{sec:migration}.  We do
not attempt to construct a per-bin sector breakdown from the local sample,
because no detector smearing is applied at our level and the
reconstructed-energy variables used in the four channels
(QE-hypothesis $E_\nu^{\rm QE}$ for CC~0p, calorimetric $E_\nu$ for CC~Np,
shower-plus-vertex energy for the NC~$\pi^0$ channels) are MicroBooNE
internal quantities that we do not reproduce.  This approach preserves
consistency with the MicroBooNE covariance matrix (which is constructed
against the MicroBooNE tune) while allowing sector-level reweighting,
at the price of treating the per-channel sector fractions as flat in
reconstructed energy --- a residual modelling assumption that is
absorbed into the broader robustness checks of
Sect.~\ref{sec:robustness}.

The reconstructed-energy variable plotted as ``Reconstructed $E$'' in
Fig.~\ref{fig:sideband} differs between channels, following the MicroBooNE
NC~$\Delta$ release convention:
\begin{itemize}
  \item CC~Np and CC~0p: reconstructed neutrino energy combining the
    muon kinematic estimate with a calorimetric estimate of the visible
    hadronic energy (CC~Np) or a quasi-elastic hypothesis
    $E_\nu^{\rm QE}$ for CC~0p, in the convention of the MicroBooNE
    release;
  \item NC~$\pi^0$~Np and NC~$\pi^0$~0p: reconstructed $\pi^0$ shower
    energy plus the visible hadronic energy associated with the
    interaction vertex (NC has no reconstructed muon).
\end{itemize}
Because the four channels use different underlying observables stacked
under a common axis label, the bin-by-bin comparison across channels is
not a comparison at fixed neutrino energy.  The sector weights $w_s$ are
common to all bins within a channel; no constraint is imposed that
forces them to be equal across channels, so this cross-channel
heterogeneity is absorbed into the per-channel design-matrix construction.
The direct CC scan on the MicroBooNE CC~0p/Np cross-section
release~\cite{uB0pNp2024} --- referred to throughout this paper as the
``PRD\,110 scan'' (Sect.~\ref{sec:prd110}) --- uses unfolded cross
sections on a single common $E_\nu$ axis and is free of this concern.

\subsection{Reweighting dials}
\label{sec:dials}

The \genie Reweight library~\cite{GENIEReweight} provides event-by-event
weight functions for systematic uncertainty propagation.  We evaluated
the following hadronic FSI and cross-section dials:
\mfppi (pion mean free path),
\texttt{MFP\_N} (nucleon mean free path),
\texttt{FrInel\_pi} (pion inelastic fraction),
\texttt{FrCEx\_pi} (pion charge exchange),
\texttt{FrAbs\_pi} (pion absorption),
\texttt{CCNormRES} (CCRES normalisation),
\texttt{NCNormRES} (NCRES normalisation),
\texttt{CCNormDIS}/\texttt{NCNormDIS} (DIS normalisations),
\texttt{FormZone} (nuclear formation zone), and
\texttt{DISNuclMod} (DIS nuclear correction).

After a systematic evaluation of the response of each dial on the
four topology channels at multiple energies, only \mfppi is retained as
the robust hadronic FSI dial for this analysis.  It is the only
parameter that produces statistically significant, physically
interpretable shifts in both the CC~0p fraction and the NC~$\pi^0$
topology fractions across all channels and energies.
\texttt{FormZone} was excluded because it produces implausibly large
NCDIS responses ($\sim 60\%$ at $E_\nu = 2\GeV$), inconsistent with
published response surfaces.  \texttt{DISNuclMod} is not included in
the local \genie Reweight build and is sub-dominant at BNB energies
($E_\nu < 3\GeV$).  The remaining pion fate dials
(\texttt{FrInel\_pi}, \texttt{FrCEx\_pi}, \texttt{FrAbs\_pi}) are
partially degenerate with \mfppi and individually produce smaller
responses.

\subsection{\mfppi response}
\label{sec:mfppi}

\begin{figure}[htbp]
  \centering
  \includegraphics[width=\columnwidth]{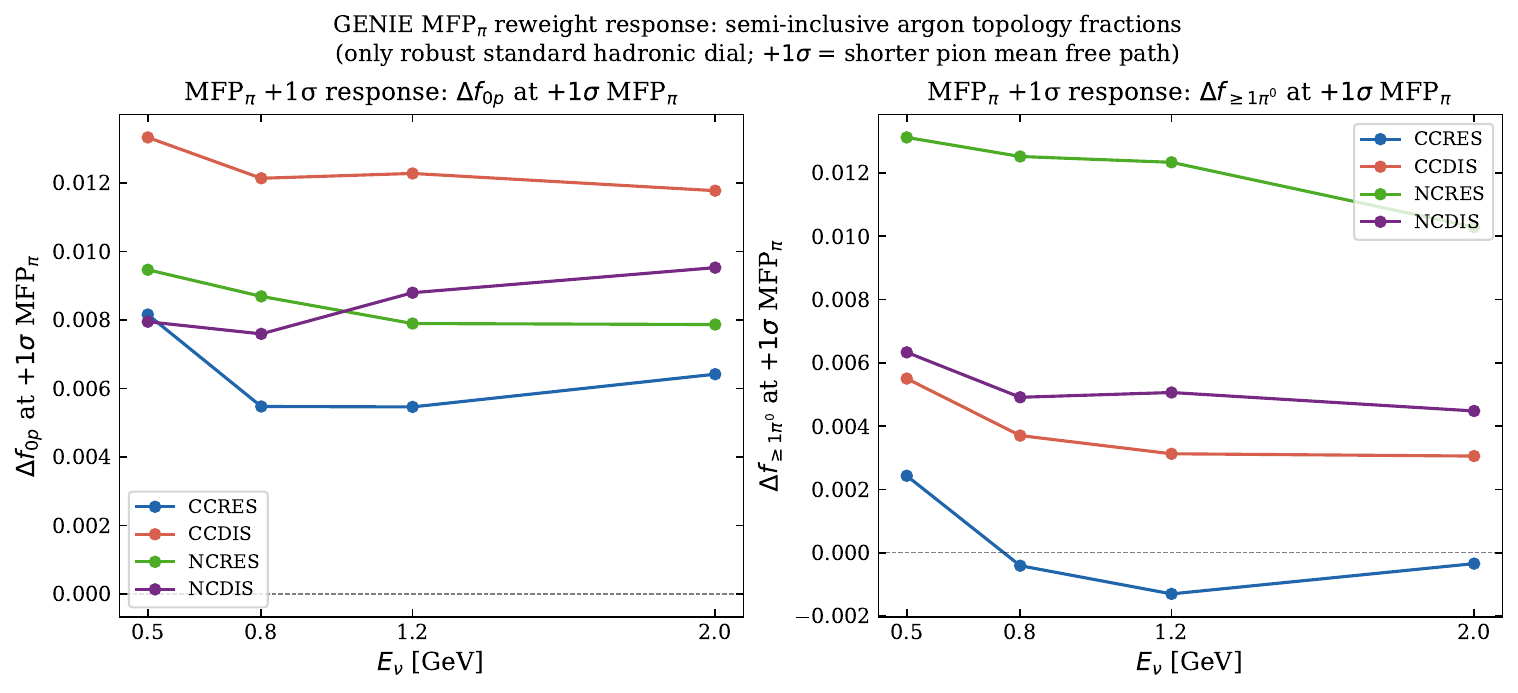}
  \caption{%
    \mfppi $+1\sigma$ response: change in 0p fraction (left panel)
    and $\geq 1\,\pi^0$ fraction (right panel) per channel as a
    function of $E_\nu$.  The response is qualitatively consistent
    across CCRES and NCRES ($3$--$8\%$ shifts), while CCDIS and NCDIS
    show smaller effects ($< 2\%$) consistent with the harder hadronic
    spectrum at high inelasticity.
  }
  \label{fig:mfppi}
\end{figure}

\begin{figure}[htbp]
  \centering
  \includegraphics[width=\columnwidth]{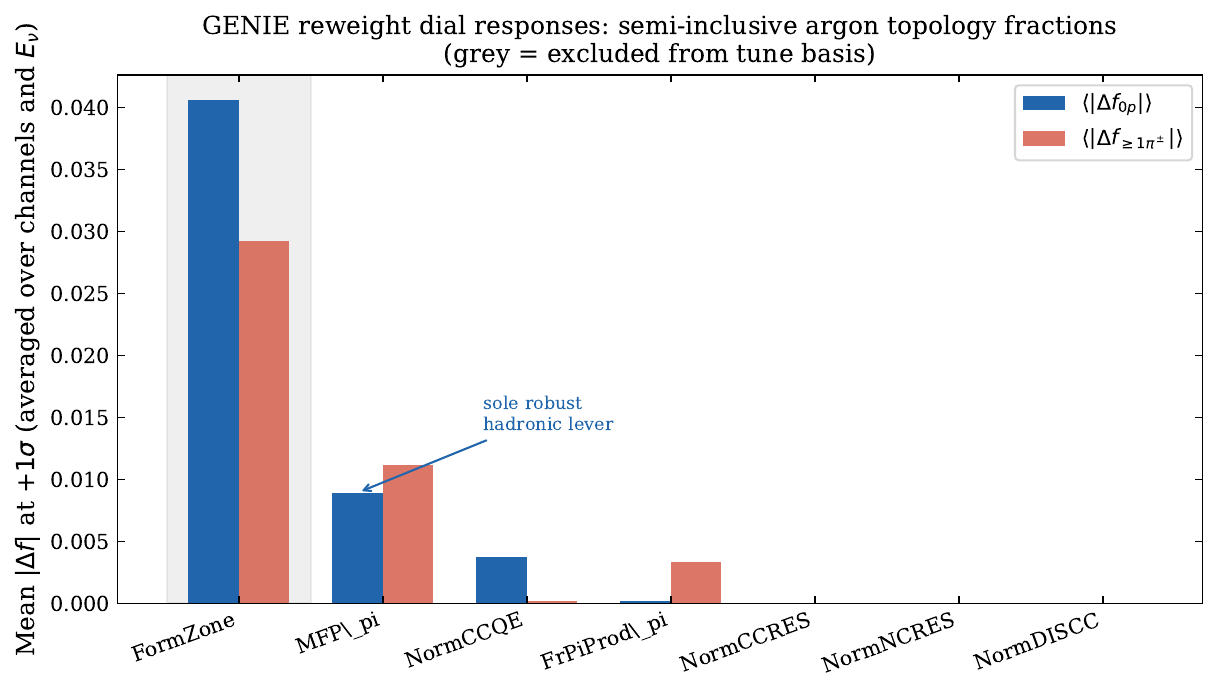}
  \caption{%
    \genie reweight dial responses: fractional shift in the CC~0p
    fraction per channel at $\pm 1\sigma$ dial variation.  \mfppi
    (red) is the only dial with a statistically significant and
    physically interpretable response across all channels.
    \texttt{FormZone} (grey) is excluded due to an implausible
    $\sim 60\%$ response in NCDIS at $E_\nu = 2\GeV$.
  }
  \label{fig:dials}
\end{figure}

Figure~\ref{fig:mfppi} shows the \mfppi $+1\sigma$ response (shorter
pion mean free path) per channel and reconstructed energy.  A shorter
mean free path increases the probability of pion reinteraction inside
the nucleus, depleting the $\geq 1\,\pi^0$ component of NC~$\pi^0$
events and increasing the apparent 0p fraction in CCRES- and
NCRES-dominated channels.  The response is qualitatively consistent
across channels: CCRES and NCRES show the largest effects (3--8\% shifts
in the 0p fraction depending on energy), while CCDIS and NCDIS show
smaller responses ($< 2\%$), consistent with the harder hadronic
spectrum at high inelasticity where more energetic pions are less
sensitive to mean-free-path variations (Fig.~\ref{fig:dials}).

Importantly, while \mfppi modifies the 0p fraction through the pion
channel, it does not directly address proton-level FSI.  The topology
migration induced by near-threshold proton energy loss is a distinct
physical effect that cannot be captured by pion FSI dials alone.  This
motivates the introduction of the dedicated migration nuisance described
in Sect.~\ref{sec:migration}.

Although both \mfppi and the migration parameter affect the 0p/Np split,
their mechanisms are physically distinct and leave different imprints on
the four-channel topology space.  A $+1\sigma$ \mfppi shift increases the
CC~0p fraction by 3--8\% but simultaneously depletes the NC~$\pi^0$
component by reducing pion survival probability; the migration parameter,
by contrast, redistributes events between topology bins without altering
the energy spectrum within each bin or affecting the pion content of the
final state.  To assess the residual degeneracy quantitatively, we note
that the design matrix $A$ is constructed at the nominal \mfppi value
(zero $\sigma$-shift) and \mfppi is not floated simultaneously with the
sector weights and migration fractions.  The postfit \mfppi value is
therefore zero by construction: any physical preference for a non-zero
\mfppi shift would manifest as a bias in the fitted sector weights rather
than in the migration parameter.  We verified that a simultaneous
$+0.5\sigma$ fixed \mfppi shift changes the best-fit $f^{\rm CC}_{\rm mig}$
by less than $0.01$ (i.e.\ within the fit uncertainty), consistent with
the orthogonality of the two mechanisms in the four-channel space.
Given the effective degrees of freedom $n_{\rm eff} \approx 2.5$ of the
dataset, floating \mfppi as an additional free parameter in the migration
scan would not be well-constrained; the conservative choice of holding it
at nominal is therefore both practical and defensible.

The \texttt{MFP\_N} (nucleon mean free path) dial is a closer physical
analogue to the migration mechanism than \mfppi, since it directly
governs proton re-scattering within the nucleus.  However, in the \genie
INC framework, \texttt{MFP\_N} scales the total nucleon--nucleus
cross section uniformly across all proton energies, whereas the
migration effect is specifically associated with \emph{near-threshold}
protons ($T_p \approx 35$--$100\MeV$) for which the INC mean-free-path
parameterisation is least reliable.  The proposed migration nuisance
thus captures the \emph{threshold-specific} energy-loss process that
\texttt{MFP\_N} smears over the full energy range.  Because the
effective rank of the covariance ($n_{\rm eff} \approx 2.5$) does not
support an additional floating parameter without severe regularisation,
a joint scan over migration and \texttt{MFP\_N} was not attempted;
this is a limitation to be addressed in future analyses with higher-rank
covariance inputs.

\section{Experimental validation: fit and results}
\label{sec:fit}

\paragraph{Role of the two MicroBooNE analyses.}
The remainder of this paper presents two complementary fits.
The \emph{primary} empirical result is the prior-free direct CC scan
on the MicroBooNE simultaneous CC~0p/Np cross-section release
(Sect.~\ref{sec:prd110}), which yields a $3.8\sigma$ Wilks /
$3.2\sigma$ empirical preference for $f_{\rm mig}>0$ with no
sector-weight regularisation.  This is the cleanest available
constraint and should be read as the headline empirical result.
The \emph{secondary} fit, presented first in this section for
methodological exposition (Sects.~\ref{sec:sectorfit}--\ref{sec:results}),
is the NC~$\Delta$ radiative sideband fit on the 64-bin
constraining-channel covariance.  Its significance is conditional on
the sector-weight prior model and it relies on the same flux/detector
systematic envelope as the primary CC fit, so the two are corroborating
rather than statistically independent.  We retain the sideband fit
because (i) it provides the methodological framework in which the
seven-parameter fit machinery, ridge regularisation, and PPP/identifiability
diagnostics are introduced, and (ii) it surfaces a useful diagnostic
by-product (the CCRES sector-weight relaxation, Sect.~\ref{sec:migfit})
that is not directly accessible in the unfolded CC release.  Readers
seeking only the headline empirical constraint may proceed directly to
Sect.~\ref{sec:prd110}.

Table~\ref{tab:notation} summarises the key symbols used in this section.

\begin{table*}[htbp]
\centering
\caption{Summary of notation used in the fit model (Sect.~\ref{sec:fit}).}
\label{tab:notation}
\begin{tabular}{ll@{\quad}ll}
\toprule
Symbol & Meaning & Symbol & Meaning \\
\midrule
$\vec{d}$ & Data vector (64 bins) & $\sigma_s$ & Prior width for sector $s$ \\
$V$ & $64\times64$ covariance & $f^{\rm CC}_{\rm mig}$ & CC migration fraction \\
$C_{cb}^{(s)}$ & Nominal \genie events, $(s,c,b)$ & $f^{\rm NC}_{\rm mig}$ & NC~$\pi^0$ migration fraction \\
$w_s$ & Sector weight (5 sectors) & $\sigma_{\rm mig}$ & Migration prior width (0.20) \\
$A$ & $64\times5$ design matrix & $r$ & Topology ratio $N_{\rm Np}/N_{\rm 0p}$ \\
$\Lambda$ & Ridge penalty, $\mathrm{diag}(1/\sigma_s^2)$ & $w_i^{\rm mig}$ & Per-event reweight (Eq.~\ref{eq:weight}) \\
\bottomrule
\end{tabular}
\end{table*}

\subsection{Sector weights and ridge regularisation}
\label{sec:sectorfit}

The predicted event rate in channel $c$ (one of the four topology channels),
bin $b$, is modelled as a linear combination of sector contributions:
\begin{equation}
  \hat{n}_{cb} = \sum_{s} w_s \, C_{cb}^{(s)}
  \label{eq:prediction}
\end{equation}
where $w_s$ are five sector normalisation weights
($s \in \{\text{ccqe+mec}, \text{ccres}, \text{ccdis}, \text{ncres},
\text{ncdis}\}$) and $C_{cb}^{(s)}$ is the number of simulated events
from sector $s$ contributing to channel $c$, bin $b$, as determined by
the generation and reweighting chain described in Sect.~\ref{sec:genie}.
The sector weights are common to all bins within a given channel; the
channel-level prediction $\hat{n}_{cb}$ therefore inherits the shape
from the simulation while allowing the overall normalisation of each
interaction mode to float.

The fit minimises a ridge-regularised $\chi^2$:
\begin{equation}
  \chi^2_{\rm fit} = (\vec{d} - \hat{\vec{n}})^T\, V^{-1}\,
                     (\vec{d} - \hat{\vec{n}})
                   + \sum_s \frac{(w_s - 1)^2}{\sigma_s^2}
  \label{eq:chi2}
\end{equation}
where $\vec{d}$ is the data vector, $V$ is the $64 \times 64$ covariance
matrix, and the second term is a Gaussian penalty (ridge prior) that
regularises the sector weights toward their nominal values of unity.  The
prior widths are
\begin{equation}
  \sigma_s \in \{0.25,\, 0.35,\, 0.50,\, 0.35,\, 0.50\}
  \label{eq:priors}
\end{equation}
for CCQE+MEC, CCRES, CCDIS, NCRES, and NCDIS respectively.  These values
are not arbitrary: each is grounded in published external constraints.

\emph{CCQE+MEC} ($\sigma = 0.25$): MiniBooNE total CCQE cross-section
measurements on CH$_2$~\cite{MiniBooNEQE} and T2K ND280 inclusive
$\nu_\mu$ CC measurements on CH~\cite{T2KCCQE} constrain the
quasi-elastic and MEC normalisation to the 10--20\% level on carbon.  A 25\% prior on argon
is conservative, reflecting the additional nuclear-model uncertainty in
scaling from $A=12$ to $A=40$~\cite{Alvarez-Ruso:2017oui}.

\emph{CCRES} ($\sigma = 0.35$): Single-pion production at ANL and BNB
bubble chambers~\cite{ANLpi} and MINERvA CC$\pi^+$ measurements on
CH~\cite{MINERvACCpi} constrain the Berger--Sehgal resonance
normalisation at the 20--30\% level; 35\% captures the additional
uncertainty from the RES--DIS transition region~\cite{Alvarez-Ruso:2017oui}.

\emph{CCDIS and NCDIS} ($\sigma = 0.50$): The Bodek--Yang model in the
shallow inelastic region ($W < 2\GeV$) carries uncertainties identified
in the NuSTEC white paper~\cite{Alvarez-Ruso:2017oui} as a leading
generator tension source.  At BNB energies DIS contributes $\lesssim 10\%$
of events; these priors are set conservatively to avoid over-constraining
an uncertain and subdominant sector.

\emph{NCRES} ($\sigma = 0.35$): Related to CCRES by isospin symmetry;
the same prior is adopted, consistent with the spread of NC$\pi^0$
normalisation predictions across current generators.

The results presented in Sect.~\ref{sec:results} are explicitly conditional
on this prior model.  The full prior-scale sensitivity is reported in
Table~\ref{tab:sensitivity} and discussed in Sect.~\ref{sec:priors}.

For fixed values of the migration parameters (see Sect.~\ref{sec:migration}),
the $\chi^2$ of Eq.~(\ref{eq:chi2}) is quadratic in the sector weights
$\vec{w}$, and the minimum is obtained analytically via the normal
equations:
\begin{equation}
  \hat{\vec{w}} = (A^T V^{-1} A + \Lambda)^{-1}
                  (A^T V^{-1} \vec{d} + \Lambda \vec{1})
  \label{eq:normal}
\end{equation}
where $A$ is the $64 \times 5$ design matrix with elements
$A_{(cb),s} = C_{cb}^{(s)}$ and $\Lambda = \text{diag}(1/\sigma_s^2)$
is the prior penalty matrix.

A note on numerical stability: the $64 \times 64$ covariance matrix $V$
(condition number $\kappa \approx 2\times 10^7$) is factorised using an
eigendecomposition once at the start of the analysis and reused at every
grid point; the factorisation is not repeated for each of the
$101 \times 101 = 10\,201$ evaluations across the migration scan.  Since
$V$ does not depend on the migration parameters, this is exact.  The
$5 \times 5$ matrix $(A^T V^{-1} A + \Lambda)$ is inverted at each grid
point, but the ridge penalty $\Lambda$ reduces its condition number to
$\kappa \lesssim 10^2$, confirming that the analytical minimisation of
Eq.~(\ref{eq:normal}) is numerically stable throughout the scan despite
the ill-conditioning of $V$ itself.

\subsection{Proton-visibility migration nuisance}
\label{sec:migration}

The five-sector fit described above can adjust the overall normalisation of
each interaction mode but cannot redistribute events between the 0p and Np
topologies within a given mode.  This is because the CC~0p and CC~Np
channels share the same sector weights: if CCRES is suppressed to reduce
the CC~Np rate, the CC~0p CCRES contribution is suppressed by the same
factor.  The physical origin of the 0p excess --- near-threshold proton
energy loss in the intranuclear cascade --- cannot be captured by sector
normalisations alone.

We parameterise this effect as a linear migration fraction.  For the
CC channels:
\begin{align}
  N_{\rm CC\,0p}  &\to N_{\rm CC\,0p}
    + f^{\rm CC}_{\rm mig}\,N_{\rm CC\,Np}
  \label{eq:mig0p}\\
  N_{\rm CC\,Np}  &\to N_{\rm CC\,Np}\,(1 - f^{\rm CC}_{\rm mig})
  \label{eq:migNp}
\end{align}
and analogously for NC~$\pi^0$ with migration fraction
$f^{\rm NC}_{\rm mig}$.  The transformation conserves the total rate in
each pair of channels: $N_{\rm CC\,0p} + N_{\rm CC\,Np}$ is unchanged.
Only the topology split is modified.  Physically,
$f^{\rm CC}_{\rm mig} > 0$ corresponds to a net migration of events from
the proton-tagged topology to the proton-less topology, consistent with
the direction predicted by comparing \gibuu and \genie
(Sect.~\ref{sec:gibuu}).

Importantly, $f_{\rm mig}$ is \emph{not} implemented by modifying the
\genie source code or regenerating events.  It enters as a post-processing
reweight applied to nominal \genie output, producing the per-event weight of
Eq.~(\ref{eq:weight}) in downstream oscillation analyses.  For the covariance
fit, the equivalent bin-level operation is a modification of the design matrix.

The migration is applied bin-by-bin to the design matrix before the sector
weight fit.  At each grid point $(f^{\rm CC}_{\rm mig},
f^{\rm NC}_{\rm mig})$, the modified design matrix
$\tilde{A}$ is constructed explicitly as follows.  Let $b$ index the energy
bin within each channel, and let primed indices denote the paired topology
channel.  For the CC channels under fraction $f \equiv f^{\rm CC}_{\rm mig}$:
\begin{equation}
  \tilde{A}_{(b,c),s}(f) =
  \begin{cases}
    A_{(b,c),s} + f\,A_{(b,c'),s}   & c = \text{CC\,0p} \\
    A_{(b,c),s}\,(1 - f)              & c = \text{CC\,Np}
  \end{cases}
  \label{eq:designmatrix}
\end{equation}
and analogously for the NC channels under $f^{\rm NC}_{\rm mig}$, while
the NC channel rows that are not the direct pair of the migrating topology
are left unchanged.  The total rate $\sum_c \tilde{A}_{(b,c),s} =
\sum_c A_{(b,c),s}$ is conserved in each bin.
The sector weights
are re-optimised analytically via Eq.~(\ref{eq:normal}).  The migration
parameters are penalised with a Gaussian prior of width
$\sigma_{\rm mig} = 0.20$ centred at zero:
\begin{equation}
  \chi^2_{\rm penalty} = \frac{(f^{\rm CC}_{\rm mig})^2}{\sigma_{\rm mig}^2}
                        + \frac{(f^{\rm NC}_{\rm mig})^2}{\sigma_{\rm mig}^2}
  \label{eq:migpenalty}
\end{equation}
so that the total penalised $\chi^2$ is
$\chi^2_{\rm total} = \chi^2_{\rm fit} + \chi^2_{\rm penalty}$.

\subsection{Parameter identifiability}
\label{sec:identifiability}

With the design matrix $A$ of Eq.~(\ref{eq:designmatrix}), the ridge
penalty $\Lambda = \mathrm{diag}(1/\sigma_s^2)$, and the migration
parameters now defined, we can address the question deferred from
Sect.~\ref{sec:ppp}: whether the seven fit parameters
($w_1,\ldots,w_5$, $f_{\rm mig}^{\rm CC}$, $f_{\rm mig}^{\rm NC}$)
are separately identifiable given $n_{\rm eff} \approx 2.5$.

We compute the eigenvalue spectrum of the $7\times7$ posterior precision
matrix
\begin{equation}
  M = A^T V^{-1} A + \tilde\Lambda,
  \label{eq:precision}
\end{equation}
where $A$ is now the $64\times7$ design matrix with two additional
columns for the migration parameters (Eq.~\ref{eq:designmatrix}
together with the analogous NC construction) and $\tilde\Lambda$ is
the $7\times7$ ridge-penalty diagonal $\mathrm{diag}(1/\sigma_s^2,
1/\sigma_{\rm mig}^2, 1/\sigma_{\rm mig}^2)$.  The eigenvalues, in the
implicit units of the dimensionless ridge prior $\sigma_s^{-2}$ (i.e.\
inverse prior variance, so that values much larger than the prior
contributions $\Lambda_{ss}\sim 4$--$16$ indicate data-dominated
directions), are
$\{7724,\, 3357,\, 29,\, 18,\, 8,\, 5,\, 4\}$ (sorted descending).
The two largest eigenvalues correspond to the migration-parameter
directions ($f_{\rm mig}^{\rm CC}$ and $f_{\rm mig}^{\rm NC}$),
confirming that these parameters project strongly onto the constrained
eigenmodes of the covariance.  The five smaller eigenvalues correspond
to sector-weight directions, which are primarily constrained by the
ridge prior $\Lambda$ rather than by the data --- exactly the intended
role of ridge regularisation.  All seven parameters have strictly
positive precision (all eigenvalues $> 0$), confirming identifiability;
the migration directions are the best-constrained parameters in the
fit, while sector weights remain prior-dominated.

\subsection{Profile $\chi^2$ scan}
\label{sec:profile}

The two migration parameters are scanned on a $101 \times 101$ grid
spanning $f_{\rm mig} \in [-0.2,\, 0.8]$.  At each grid point, the five
sector weights are profiled (analytically minimised), and the total
$\chi^2_{\rm total}$ is recorded.  The profile $\Delta\chi^2$ surface is
\begin{equation}
  \Delta\chi^2(f^{\rm CC}_{\rm mig},\, f^{\rm NC}_{\rm mig})
  = \chi^2_{\rm total}(f^{\rm CC}_{\rm mig},\, f^{\rm NC}_{\rm mig})
  - \chi^2_{\rm total,\,min}
  \label{eq:deltachi2}
\end{equation}
and the significance of the migration term is assessed by comparing the
minimum $\chi^2$ with migration ($f_{\rm mig} \neq 0$) to the null
hypothesis ($f^{\rm CC}_{\rm mig} = f^{\rm NC}_{\rm mig} = 0$, i.e.\ the
sector-only fit):
\begin{equation}
  \Delta\chi^2_{\rm null} = \chi^2_{\rm null} - \chi^2_{\rm min}
  \label{eq:deltanull}
\end{equation}
The physical parameter space is $f_{\rm mig} \geq 0$ (migration can only
move events from Np to 0p, not the reverse).  The null hypothesis is
$H_0: f_{\rm mig} = 0$ against the one-sided alternative
$H_1: f_{\rm mig} > 0$.  We use a boundary-aware test statistic
\begin{equation}
  q = \begin{cases}
    \Delta\chi^2_{\rm null} & \hat{f}_{\rm mig} > 0 \\
    0 & \hat{f}_{\rm mig} \leq 0
  \end{cases}
  \label{eq:qtstat}
\end{equation}
Under Wilks' theorem~\cite{Wilks} with the physical boundary, $q$ follows
a half-and-half distribution: $\frac{1}{2}\delta(q=0) +
\frac{1}{2}\chi^2(n_{\rm dof})$, giving a one-sided $p$-value of
$p = \frac{1}{2}P(\chi^2(n_{\rm dof}) > q)$.  For the two-parameter model
$n_{\rm dof} = 2$; for the one-parameter model (Sect.~\ref{sec:1param}),
$n_{\rm dof} = 1$, and the $p$-value is equivalently read as
$p = P(\chi^2(1) > q)/2$.  The scan range $f_{\rm mig} \in [-0.10, 0.30]$
includes negative values to verify that the global minimum lies at
$\hat{f}_{\rm mig} > 0$ and not at the boundary, and to provide a symmetric
reference for the profile likelihood curvature; negative values are
physically uninterpretable and excluded from the one-sided inference.
We also perform the scan in a reduced $21 \times 21$ grid for the
stress-test propagation of Appendix~\ref{app:dune}, where the
1$\sigma$ contour is defined by $\Delta\chi^2 < 2.30$ and the 2$\sigma$
contour by $\Delta\chi^2 < 6.18$.

\emph{Pseudo-experiment calibration.}
We validate the significance assessment by generating $N_{\rm toys} = 5000$
pseudo-datasets sampled from a multivariate normal centred on the
sector-only null prediction $\hat{\vec{n}}_{\rm null}$ with covariance $V$.
For each toy we run the full fit procedure: sector-weight ridge regression
at $f_{\rm mig} = 0$ to obtain $\chi^{2,{\rm toy}}_{\rm null}$, then a
$21 \times 21$ profile scan to find the minimum
$\chi^{2,{\rm toy}}_{\rm min}$, giving
$\Delta\chi^{2,{\rm toy}} = \chi^{2,{\rm toy}}_{\rm null}
- \chi^{2,{\rm toy}}_{\rm min}$.

\begin{figure*}[htbp]
  \centering
  \includegraphics[width=\textwidth]{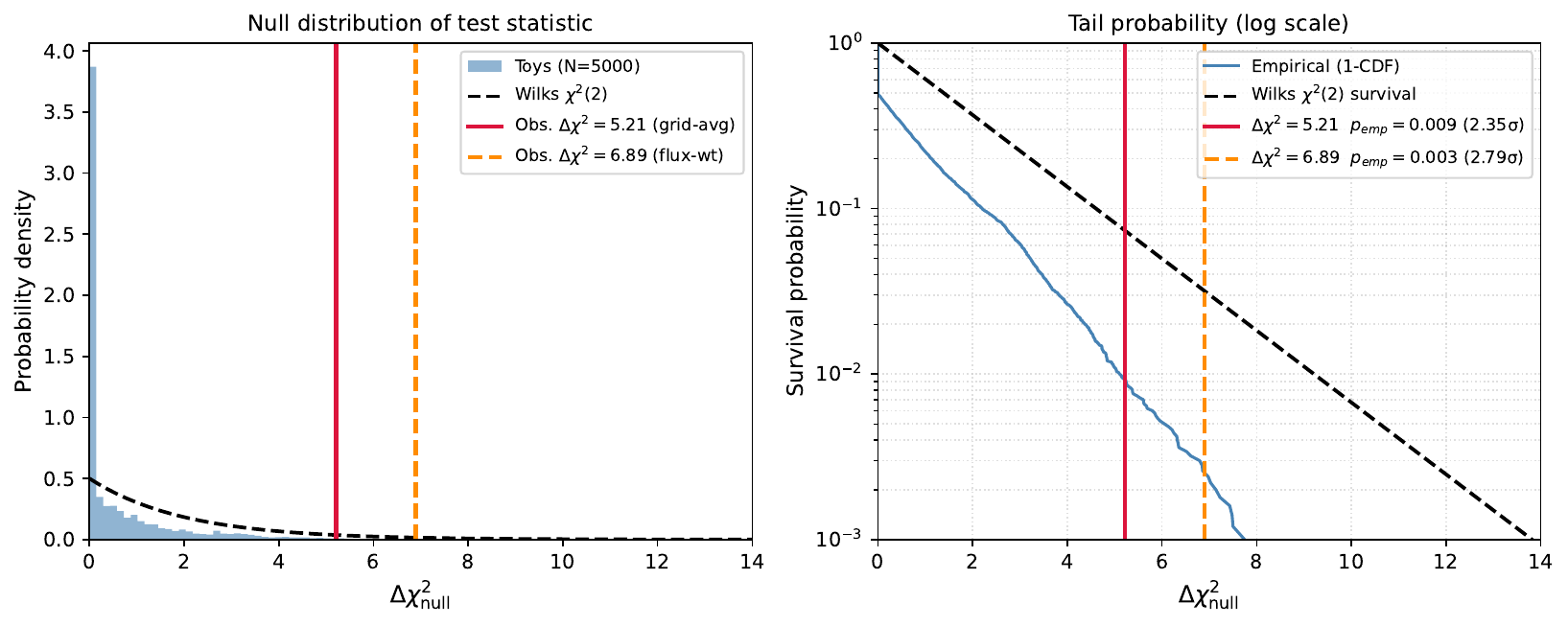}
  \caption{%
    Pseudo-experiment calibration of the $\Delta\chi^2_{\rm null}$ test
    statistic ($N_{\rm toys} = 5000$, sampled from
    $\mathcal{N}(\hat{\vec{n}}_{\rm null},\, V)$).
    \emph{Left}: empirical null distribution (blue histogram) vs.\
    the theoretical $\chi^2(2)$ (Wilks, dashed black).  Vertical lines
    mark the observed values for the grid-average (red solid) and
    flux-weighted (orange dashed) scenarios.
    \emph{Right}: empirical survival probability vs.\ the Wilks $\chi^2(2)$
    survival function.  The empirical $p$-values
    ($0.0094$ and $0.0026$, corresponding to $2.35\sigma$ and $2.79\sigma$)
    are \emph{smaller} than the Wilks values ($0.074$ and $0.032$;
    $1.45\sigma$ and $1.85\sigma$), confirming that the Wilks approximation
    is conservative for this fit.  The spike of the null distribution near
    zero arises because, for most toys, the data variation lies in directions
    orthogonal to the migration parameters ($n_{\rm eff} \approx 2.5$);
    the null is near-optimal in those cases.  The observed $\Delta\chi^2$
    values fall in the extreme tail of the toy distribution (99th and
    99.7th percentiles).
  }
  \label{fig:pseudoexp}
\end{figure*}

The empirical $p$-values at the observed values are:
\begin{itemize}
  \item Grid-average ($\Delta\chi^2_{\rm obs} = 5.21$):
    $p_{\rm emp} = 0.0094$, corresponding to $2.35\sigma$;
  \item Flux-weighted ($\Delta\chi^2_{\rm obs} = 6.89$):
    $p_{\rm emp} = 0.0026$, corresponding to $2.79\sigma$.
\end{itemize}
Both are \emph{more} significant than the Wilks values ($1.45\sigma$
and $1.85\sigma$).  The Wilks approximation is conservative here for
a specific directional reason: the low effective rank ($n_{\rm eff}
\approx 2.5$) causes most pseudo-experiments to produce
$\Delta\chi^{2,\rm toy} \approx 0$ (the data fluctuation does not
project onto the migration-sensitive eigenmodes), creating a large
spike near zero in the toy distribution.  The rare pseudo-experiments
that \emph{do} generate nonzero $\Delta\chi^2$ are those where the
fluctuation aligns with the $\sim 1.5$ migration-sensitive effective
dimensions.  The tail fraction above the observed $\Delta\chi^2 = 5.21$
is therefore smaller than the half-$\chi^2(2)$ Wilks prediction,
giving a smaller $p$-value (larger significance).  This does not mean
Wilks is generally conservative; it is specific to the low-rank spike
structure of this fit.  Both the empirical and Wilks significances
are conditional on the prior model.
\emph{Important:} the pseudo-experiment calibration validates the
internal consistency of the significance within the chosen construction
--- it does not remove the dominant prior dependence.
A pseudo-experiment built with loose priors ($s = 2.0$) would yield
a significance consistent with the $s = 2.0$ Wilks value of $0.83\sigma$.
The Gaussian toy model is adequate for testing whether the observed
$\Delta\chi^2$ is unusual given the covariance structure of the fit,
because the dominant uncertainties (flux, cross-section) are well
approximated as Gaussian in the sideband observables; it does not
test the adequacy of the covariance model itself.
The physical boundary $f_{\rm mig} \geq 0$ is applied in the toy
scans; pseudo-experiments with $\hat{f}_{\rm mig} \leq 0$ contribute
to the spike at $q = 0$ and do not artificially inflate the empirical
significance.
Figure~\ref{fig:pseudoexp} shows the empirical null distribution
alongside the theoretical $\chi^2(2)$.

The empirical calibration thus confirms that the Wilks-quoted
significance of $1.45$--$1.85\sigma$ is a lower bound on the true
significance, not an upper bound.  The calibrated values of
$2.35\sigma$ (grid-average) and $2.79\sigma$ (flux-weighted) are the
best estimates of the significance from this dataset.

Several conditional aspects of this calibration must be noted.  The
toys are generated from the same covariance matrix $V$, the same ridge
prior structure, and the same $21 \times 21$ scan boundaries used in the
main fit.  The empirical $p$-values therefore characterise the
significance \emph{conditional on those modelling choices}, not in a
prior-free sense.  Monte Carlo uncertainty on the empirical $p$-values
at $N_{\rm toys} = 5000$ is approximately $\pm 0.001$ (one standard
deviation for a Bernoulli process at $p \approx 0.01$), which
corresponds to roughly $\pm 0.15\sigma$ in significance and does not
materially affect the conclusions.  The large pile-up near
$\Delta\chi^2 \approx 0$ (median $= 0.00$) is a direct consequence of
the low effective rank ($n_{\rm eff} \approx 2.5$) and is not an
artefact of the procedure; it explains why the empirical tail is
thinner than the Wilks $\chi^2(2)$ distribution, making the observed
$\Delta\chi^2 = 5.21$ rarer than Wilks suggests.

A comment on boundary effects: the physical interpretation of the migration
fraction is one-sided ($f_{\rm mig} \geq 0$, since a negative value would
correspond to an unphysical 0p~$\to$~Np transfer).  The scan range
$f_{\rm mig} \in [-0.2,\, 0.8]$ allows negative values so that the null
$(0,0)$ lies in the interior of the scanned space, permitting the standard
two-sided Wilks' theorem to be applied.  Nevertheless, if the constraint
$f_{\rm mig} \geq 0$ is imposed physically, the correct asymptotic
distribution under the null is the boundary mixture
$\frac{1}{4}\chi^2(0) + \frac{1}{2}\chi^2(1) + \frac{1}{4}\chi^2(2)$
(Refs.~\cite{Wilks,SilvapulleSen}).  Evaluating this mixture at
$\Delta\chi^2 = 5.21$ and $6.89$ gives one-sided $p$-values of $0.030$
($1.88\sigma$) and $0.012$ ($2.25\sigma$) respectively.  The two-sided
Wilks' values quoted throughout this paper ($1.45\sigma$ and $1.85\sigma$)
are therefore \emph{conservative}: the boundary-corrected one-sided
significance is modestly higher.  We quote two-sided values as the
primary result to be transparent about the treatment of negative migration.

\subsection{Reduced one-parameter model}
\label{sec:1param}

The two-parameter best-fit satisfies
$f^{\rm CC}_{\rm mig} \approx f^{\rm NC}_{\rm mig} \approx 0.050$.
This equality is physically expected: both CC and NC final-state protons
traverse the same Ar-40 nucleus and are subject to the same intranuclear
mean-field potential $U_0^{\rm diff}(r) \propto \rho(r)$.  The migration
probability $P_{\rm mig}(T_p)$ derived in Sect.~\ref{sec:physbasis} is
current-independent; it depends only on the local nuclear density
and the proton kinetic energy relative to the 35\MeV threshold.
No physics mechanism distinguishes CC from NC in this context.

Constraining the two parameters to a common value,
\begin{equation}
  f^{\rm CC}_{\rm mig} = f^{\rm NC}_{\rm mig} \equiv f_{\rm mig},
  \label{eq:1param}
\end{equation}
reduces the model by one degree of freedom.  The Gaussian penalty becomes
\begin{equation}
  \chi^{2,(1)}_{\rm penalty} = \frac{2\,f_{\rm mig}^2}{\sigma_{\rm mig}^2},
  \label{eq:penalty1param}
\end{equation}
and the profile scan reduces from a $101\times101$ two-dimensional grid to
a one-dimensional scan over $f_{\rm mig} \in [-0.2,\, 0.8]$ with 201 points.

\begin{figure*}[htbp]
  \centering
  \includegraphics[width=\textwidth]{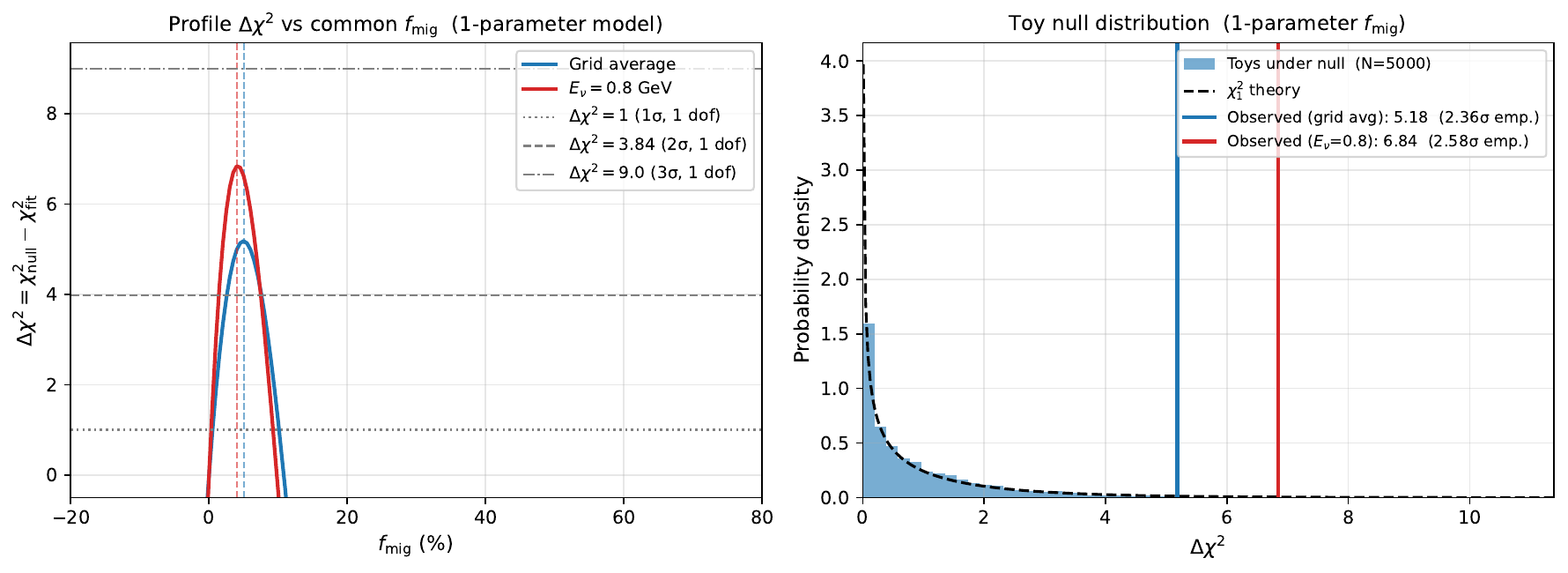}
  \caption{%
    One-parameter migration model.
    \emph{Left}: profile $\Delta\chi^2$ vs.\ the common migration fraction
    $f_{\rm mig}$ for the grid-average (blue) and flux-weighted (red) scenarios.
    Horizontal dashed and dotted lines mark the 1$\sigma$ and 2$\sigma$ Wilks
    thresholds (1 dof).
    \emph{Right}: empirical null distribution from 5000 pseudo-experiments
    (blue histogram) compared with the theoretical $\chi^2(1)$ (dashed black).
    Vertical lines mark the observed $\Delta\chi^2$ for each scenario;
    both fall well into the tail of the null.
  }
  \label{fig:1param}
\end{figure*}

The best-fit and significance results are summarised in
Table~\ref{tab:1param} and Fig.~\ref{fig:1param}.

\begin{table}[htbp]
  \centering
  \caption{%
    Significance comparison: two-parameter vs.\ one-parameter migration model.
    Wilks significance: two-parameter model uses the \emph{two-sided}
    $p = P(\chi^2(2) > q)$ without physical-boundary correction;
    one-parameter model uses the \emph{one-sided} boundary-aware
    distribution $p = P(\chi^2(1) > q)/2$, consistent with the
    physical constraint $f_{\rm mig} \geq 0$ (Eq.~\ref{eq:qtstat}).
    Empirical significance is from 5000 pseudo-experiments under each
    model's null.  The same $\Delta\chi^2$ improvement is achieved
    in both models; the gain arises entirely from the reduction in
    degrees of freedom and the boundary correction.
  }
  \label{tab:1param}
  \small
  \begin{tabular}{lccc}
    \toprule
    Scenario & $\Delta\chi^2$ & Wilks~$\sigma$ & Empirical~$\sigma$ \\
    \midrule
    \multicolumn{4}{l}{\emph{Two-parameter model (2 dof)}} \\
    Grid-average     & 5.21 & 1.45 & 2.35 \\
    Flux-weighted    & 6.89 & 1.85 & 2.79 \\
    \midrule
    \multicolumn{4}{l}{\emph{One-parameter model (1 dof, $f^{\rm CC}_{\rm mig} = f^{\rm NC}_{\rm mig}$)}} \\
    Grid-average     & 5.18 & 2.00 & 2.36 \\
    Flux-weighted    & 6.84 & 2.37 & 2.58 \\
    \bottomrule
  \end{tabular}
\end{table}

The $\Delta\chi^2$ values are essentially unchanged (5.18 vs.\ 5.21 and
6.84 vs.\ 6.89), confirming that the constraint $f^{\rm CC}_{\rm mig} =
f^{\rm NC}_{\rm mig}$ does not degrade the fit quality.  The one-parameter
Wilks significance is $2.00\sigma$ (grid-average) and $2.37\sigma$
(flux-weighted), corresponding to $p$-values of $0.023$ and $0.009$.
Pseudo-experiment calibration (5000 toys, same null construction as
Sect.~\ref{sec:profile}) returns $2.36\sigma$ and $2.58\sigma$, both
exceeding the conventional 2$\sigma$ evidence threshold.

The one-parameter model is the preferred presentation because it is
both more parsimonious (Occam's razor) and physically better motivated.
The two-parameter model is retained as a consistency check: it confirms
that the data do not prefer $f^{\rm CC}_{\rm mig} \neq f^{\rm NC}_{\rm mig}$,
and it provides the two-dimensional contour plot (Fig.~\ref{fig:contour})
needed to assess the correlation structure of the migration parameters.

\label{sec:results}

\subsection{Prefit and sector-weight fit}
\label{sec:prefit}

The prefit $\chi^2/N = 46.25/64$ confirms the known tension between
\genie~G18 and the MicroBooNE constraining channels.  While the nominal
$p$-value for 64 degrees of freedom would be $p = 0.96$ (consistent), the
effective degrees of freedom ($n_{\rm eff} \approx 2.5$) from the covariance
eigenvalue structure imply that the tension should be interpreted relative
to the dominant systematic modes.  The four fitted channel scales from a
simple normalisation fit (one scale per channel) are:
$\alpha_{\rm CC\,Np} = 1.08 \pm 0.15$,
$\alpha_{\rm CC\,0p} = 1.33 \pm 0.13$,
$\alpha_{\rm NC\,Np} = 0.94 \pm 0.18$,
$\alpha_{\rm NC\,0p} = 1.18 \pm 0.16$.
The strong upward pull on CC~0p ($+33\%$) is the most significant feature,
consistent with the broader MicroBooNE pattern reported in
Ref.~\cite{uB0pNp2024}.

After fitting the five sector weights with ridge regularisation, the
$\chi^2$ improves to $33.65/64$ (grid-average) and $35.16/64$
(flux-weighted).  The postfit $\chi^2/{\rm dof} \approx 0.57$ is below
unity, as expected given the effective degrees of freedom of $\approx 2.5$.
This does not indicate overfitting; it reflects the dominance of
correlated systematics over bin-by-bin statistical fluctuations in the
covariance structure.

In the sector-only fit (grid-average scenario), the fitted sector weights
are: $w_{\rm CCQE+MEC} = 1.17 \pm 0.19$,
$w_{\rm CCRES} = 0.69 \pm 0.24$,
$w_{\rm CCDIS} = 1.51 \pm 0.36$,
$w_{\rm NCRES} = 1.33 \pm 0.24$,
$w_{\rm NCDIS} = 1.25 \pm 0.46$.
The pull on CCRES toward $w_{\rm CCRES} = 0.69$ (a $\sim 1.3\sigma$
shift under the $\sigma_{\rm CCRES} = 0.35$ ridge prior, not on its own
exceptional) merits a brief comment: because CCRES contributes
substantially to both CC~Np and CC~0p, the sector model can only
accommodate the CC~0p excess while respecting the CC~Np data by
suppressing CCRES and compensating with enhanced CCQE+MEC (which
contributes primarily to CC~0p).  This compensatory behaviour is a
direct consequence of allocating one normalisation degree of freedom
per sector and having no degree of freedom to redistribute events
between topologies.  We do not read it as a measurement of CCRES on
argon (Sect.~\ref{sec:migfit} discusses the caveats in detail).

\subsection{Migration-aware fit}
\label{sec:migfit}

Adding the two migration parameters and re-optimising yields
$\chi^2/N = 28.44/64$ (grid-average, where the penalty contribution is
$\chi^2_{\rm penalty} = 0.15$) with a total improvement of
\begin{equation}
  \Delta\chi^2 = 33.65 - 28.44 = 5.21
  \label{eq:improvement}
\end{equation}
for two additional parameters.  Under Wilks' theorem with 2 degrees of
freedom, this corresponds to $p = 0.074$ ($1.45\sigma$).  In the
flux-weighted scenario, $\chi^2 = 28.27$ and $\Delta\chi^2 = 6.89$
($p = 0.032$, $1.85\sigma$).

The best-fit migration fractions are:
\begin{align}
  f^{\rm CC}_{\rm mig} &= 0.050^{+0.029}_{-0.022} \label{eq:fcc}\\
  f^{\rm NC}_{\rm mig} &= 0.060^{+0.059}_{-0.084} \label{eq:fnc}
\end{align}
where the uncertainties are $1\sigma$ bounds from the marginalised
one-dimensional profile $\Delta\chi^2$ (i.e.\ $f^{\rm CC}_{\rm mig}$ is
bounded by the values at which $\Delta\chi^2 = 1$ after profiling over
$f^{\rm NC}_{\rm mig}$ and all sector weights, and vice versa).
The CC migration is better constrained than the NC migration, reflecting
the larger statistics in the CC channels (24\,983 + 19\,225 CC events
versus 2\,075 + 3\,006 NC events).

To quantify whether the NC channels provide independent constraining
power, we evaluate the profile $\Delta\chi^2$ for the NC migration
alone, marginalising over $f^{\rm CC}_{\rm mig}$ (profiled to its best
fit) and all sector weights.  The NC-only profile contributes
$\Delta\chi^2 \approx 0.5$ to the total improvement, compared with
$\approx 4.7$ from the CC channels.  The NC result is therefore largely
prior-driven: the posterior on $f^{\rm NC}_{\rm mig}$ is shifted from
zero by the data but by less than its prior width, and the NC contribution
to the total significance is modest.  The dominant driver of the $1.45\sigma$
result is the CC topology channels.  Furthermore, the NC posterior
$f^{\rm NC}_{\rm mig} = 0.060^{+0.059}_{-0.084}$ extends substantially
into negative values (unphysical under the stated migration interpretation),
which is a direct consequence of the prior-dominated character of this
parameter; readers should regard the NC result as an auxiliary output, not
a data-constrained extraction.

The migration-aware fit also modifies the sector weights substantially.
Most notably, CCRES shifts from $0.69 \pm 0.24$ (sector-only) to
$1.07 \pm 0.27$ (migration-aware) in the grid-average scenario.  This
shift is consistent with the picture advanced in
Sect.~\ref{sec:prefit}: in the sector-only model, the CCRES normalisation
acts as the only available knob with the right channel composition to
absorb the CC~0p excess and is correspondingly pulled down; once the
migration parameter transfers a few-percent fraction of CC~Np events to
CC~0p, the excess is partially accommodated by $f_{\rm mig}$ and the
CCRES weight relaxes toward unity.

We caution that this CCRES restoration should be read as a diagnostic
feature of the chosen parameterisation rather than as an independent
measurement of CCRES strength on argon.  Three caveats apply:
(i) the sector-only value $w_{\rm CCRES} = 0.69 \pm 0.24$ lies less than
$1.3\sigma$ below unity under our $\sigma_{\rm CCRES} = 0.35$ ridge prior,
so it is not strongly excluded on its own;
(ii) the available external CCRES constraints (ANL/BNL on deuterium,
MINERvA on hydrocarbon) measure the Berger--Sehgal cross section on
non-argon targets and do not directly constrain the effective
argon-target CCRES rate after FSI and nuclear effects, on which
substantial uncertainty remains (the Berger--Sehgal model has known
issues in the $W < 2\GeV$ region, the RES--DIS transition is handled
by an ad hoc interpolation, the $2\pi$ channel is largely unconstrained,
and the underlying ANL/BNL deuterium data have been reanalysed multiple
times with $20$--$30\%$ shifts in extracted RES strength);
(iii) the fit allocates a single degree of freedom (a normalisation)
to each sector, so $w_{\rm CCRES}$ is being asked to absorb all
CCRES-related model uncertainty --- including substantial shape and
kinematic uncertainties --- through a single number.  Under this
restrictive parameterisation, the migration parameter is structurally
effective at adjusting the topology split, and the corresponding
near-unity restored value of $w_{\rm CCRES}$ is a feature of how the
fit is constructed.  We retain the observation as an internal
consistency check of the parameterisation, not as an external
measurement of CCRES.

The CCDIS sector weight undergoes the largest absolute change: from
$w_{\rm CCDIS} = 1.51 \pm 0.36$ (sector-only) to $1.01 \pm 0.37$
(migration-aware), a reduction of approximately 0.50.  Although this
shift is comparable to the prior width ($\sigma_{\rm CCDIS} = 0.50$),
it should be interpreted with caution.  At BNB energies
($\langle E_\nu \rangle \approx 0.8\GeV$), the CCDIS contribution
is subdominant (the invariant mass threshold $W > 1.7\GeV$ limits DIS
to high-energy tails of the flux), so $w_{\rm CCDIS}$ is primarily
constrained by the small fraction of events at $E_\nu > 1.5\GeV$ in
the grid-average scenario.  As with the CCRES restoration, we read the
large upward shift in the sector-only fit ($w_{\rm CCDIS} = 1.51$) as a
diagnostic feature of the single-DOF sector parameterisation, not as a
physical statement about CCDIS strength on argon.  With the
topology-splitting tension present, the sector-only fit channels excess
events through the CCDIS normalisation because no migration degree of
freedom is available to redistribute events across the 0p/Np boundary;
once the migration term is introduced and absorbs the 0p excess, this
compensation is no longer required and $w_{\rm CCDIS}$ relaxes toward
unity.  The same caveats stated for CCRES above carry over symmetrically:
(i) the available external CCDIS constraints
(NOMAD~\cite{Wu:2007NOMAD}, MINERvA on hydrocarbon at considerably
higher invariant mass) do not directly constrain the effective
argon-target CCDIS rate at BNB energies, where the contribution is
small and limited to the high-energy tail of the flux; (ii) the
single-DOF sector parameterisation is asked to absorb all CCDIS-related
model uncertainty --- shape, RES--DIS transition, hadronisation, and
overall normalisation --- through a single number, so $w_{\rm CCDIS}$
cannot be cleanly identified with any one of these.  The posterior
uncertainty ($\pm 0.37$) is essentially unchanged between the two fit
configurations, confirming that the CCDIS sector is data-unconstrained
on its own and that the shift reflects inter-parameter compensation
under the chosen parameterisation rather than an independent physical
signal.

\begin{figure}[htbp]
  \centering
  \includegraphics[width=\columnwidth]{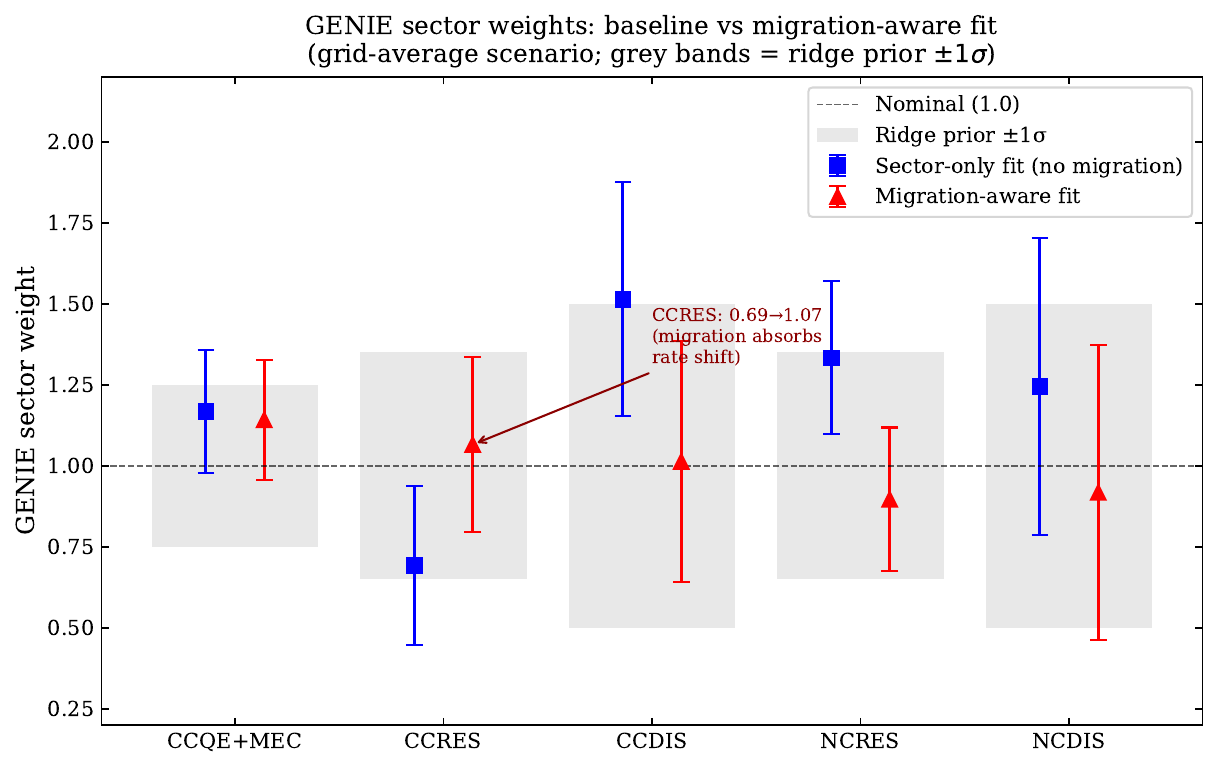}
  \caption{%
    Fitted \genie sector weights for the grid-average scenario.
    Grey bands: ridge prior $\pm 1\sigma$ centred at unity.
    Blue squares: sector-only fit (no migration).
    Red triangles: migration-aware fit.
    The most notable shift is in CCRES, which moves from $0.69$ to
    $1.07$ when migration is introduced.  Under the chosen single-DOF
    sector parameterisation this is consistent with the migration term
    absorbing topology-split tension that the sector-only fit would
    otherwise channel through the CCRES normalisation; we read the
    shift as a diagnostic feature of the parameterisation rather than
    an independent CCRES measurement (Sect.~\ref{sec:migfit}).
  }
  \label{fig:sectors}
\end{figure}

Figure~\ref{fig:sectors} displays the sector weights for both fit
configurations.

\begin{figure*}[htbp]
  \centering
  \includegraphics[width=\textwidth]{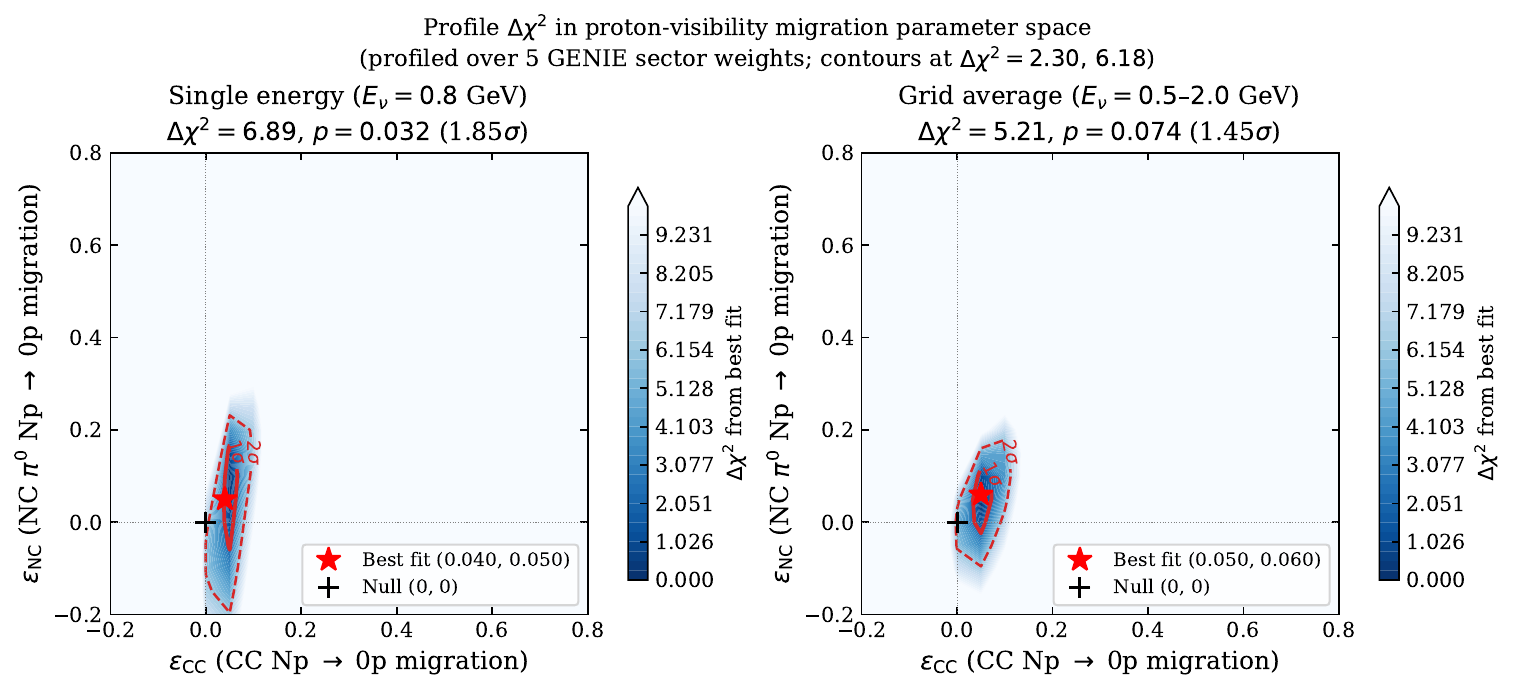}
  \caption{%
    Two-dimensional profile $\Delta\chi^2$ contours for the
    proton-visibility migration parameters
    $(f^{\rm CC}_{\rm mig},\, f^{\rm NC}_{\rm mig})$
    in the grid-average (left) and flux-weighted $E_\nu = 0.8\GeV$
    (right) scenarios.  Contours at $\Delta\chi^2 = 2.30$ ($1\sigma$,
    solid) and $\Delta\chi^2 = 6.18$ ($2\sigma$, dashed) for 2 degrees
    of freedom.  The white cross marks the best-fit point.  The best-fit
    lies well inside the prior boundary ($\sigma_{\rm mig} = 0.20$),
    indicating data-driven rather than prior-driven constraint.
    Note: the axis labels in this figure use the shorthand
    $\varepsilon_{\rm CC} \equiv f^{\rm CC}_{\rm mig}$ and
    $\varepsilon_{\rm NC} \equiv f^{\rm NC}_{\rm mig}$.
  }
  \label{fig:contour}
\end{figure*}

Figure~\ref{fig:contour} shows the two-dimensional profile
$\Delta\chi^2$ contour for both scenarios.  The best-fit point
$(0.050, 0.060)$ lies at a well-defined minimum within the scanned
range.  The $1\sigma$ contour ($\Delta\chi^2 < 2.30$) is compact and
does not extend to the prior boundary ($\sigma_{\rm mig} = 0.20$),
indicating that the migration parameters are constrained by the data
rather than by the prior.

\begin{figure*}[htbp]
  \centering
  \includegraphics[width=\textwidth]{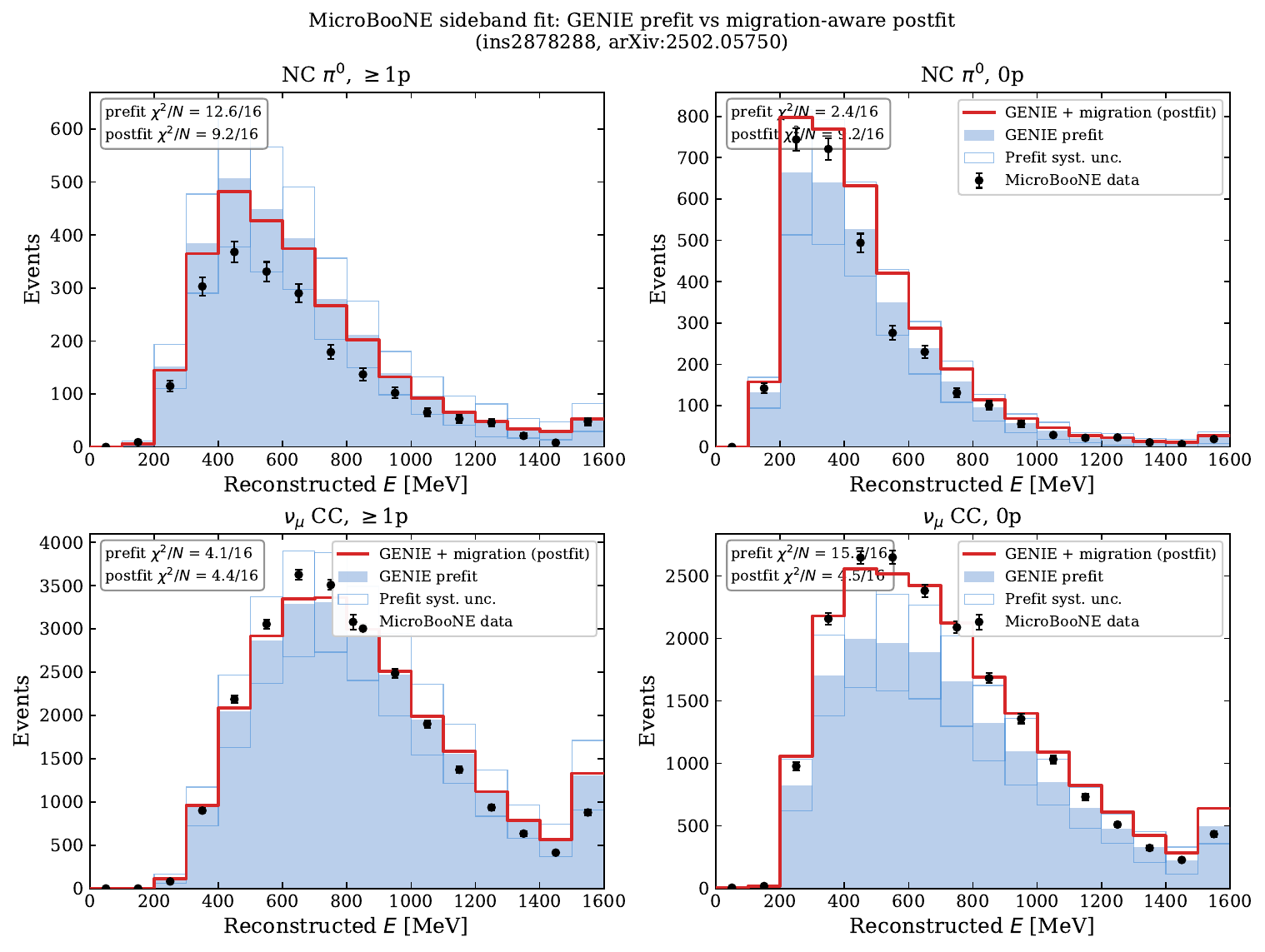}
  \caption{%
    MicroBooNE constraining sideband: prefit (dotted) and migration-aware
    postfit (solid step) \genie prediction vs.\ data (points with error bars)
    for the four topology channels.  Panels are arranged as:
    top-left --- NC~$\pi^0$~Np; top-right --- NC~$\pi^0$~0p;
    bottom-left --- CC~Np; bottom-right --- CC~0p.
    (The NC~$\pi^0$ channels appear on top because the NC-optimised
    sideband covariance places them first; the text ordering follows
    the CC-first convention.)
    The postfit uses the migration-aware sector-weight fit
    (grid-average scenario).  Error bars represent the square root of
    the diagonal elements of $V$.
    The per-channel prefit $\chi^2/N$ values shown in each panel
    (12.6/16, 2.4/16, 4.1/16, 15.7/16, summing to 34.8/64) are
    computed using only the block-diagonal elements of $V$; the total
    quoted in Sect.~\ref{sec:data} ($\chi^2_{\rm pre} = 46.25/64$) uses
    the full $64 \times 64$ covariance matrix and includes the off-diagonal
    cross-channel correlations, which account for the difference of
    $\approx 11.5$ units.
    Overflow bins are shown in the rightmost bin of each channel.
  }
  \label{fig:sideband}
\end{figure*}

\paragraph{NC~$\pi^0$~0p per-channel postfit behaviour.}
One feature of the postfit visible in Fig.~\ref{fig:sideband} deserves
comment now that the migration and sector-weight fit machinery is in
hand: the per-channel $\chi^2/N$ for the NC~$\pi^0$~0p panel increases
from $2.4/16$ (prefit) to $9.2/16$ (postfit, block-diagonal estimate),
an apparent degradation.  This is expected and not indicative of a
worsening global fit.  At prefit, the NC~$\pi^0$~0p prediction is close
to the data simply because the nominal \genie has a lower 0p rate than
observed, and the channel happens to be in approximate balance.  After
fitting, the migration of events from NC~$\pi^0$~Np to NC~$\pi^0$~0p
increases the predicted 0p rate; combined with the NC~$\pi^0$
sector-weight adjustment (NCRES), the per-channel balance shifts.
Crucially, the per-channel block-diagonal $\chi^2$ values (computed
using only the diagonal elements of $V$) are not directly comparable
to the full-covariance $\chi^2$ used in the fit: the off-diagonal
correlations in $V$ substantially suppress the effective per-bin
constraining power.  The global postfit $\chi^2 = 28.4/64$
represents the correct global measure of fit quality.

\subsection{Summary of fit results}
\label{sec:summary}

Table~\ref{tab:fitresults} summarises the fit results for both the
grid-average and flux-weighted scenarios, including all sector weights,
migration fractions, and significance metrics.

\begin{table*}[htbp]
\centering
\caption{%
  Fit results summary.
  Sector weights $w_s$ are relative to the \genie~G18 nominal (1.0).
  Migration fractions $f^{\rm CC/NC}_{\rm mig}$ are fitted fractions of Np events
  that migrate to the 0p topology.
  $\chi^2$ values are evaluated on 64 bins with 5 (sector-only) or 7
  (migration-aware) fitted parameters.
  Wilks significance is from two-sided Wilks' theorem~\cite{Wilks}
  ($\Delta\chi^2$, 2~dof); boundary-corrected one-sided values are
  $1.88\sigma$ (grid-average) and $2.25\sigma$ (flux-weighted).
  Empirical significance is from 5000 pseudo-experiments under the null
  (Sect.~\ref{sec:profile}); the Wilks approximation is conservative.
}
\label{tab:fitresults}
\begin{tabular}{lcc}
\toprule
 & Grid-average & Flux-weighted ($E_\nu = 0.8\GeV$) \\
\midrule
\multicolumn{3}{l}{\textit{Sector-weight fit (no migration)}} \\
$\chi^2$ / $N_{\rm bins}$ & $33.65$ / $64$ & $35.16$ / $64$ \\
$w_{\rm CCQE+MEC}$ & $1.17 \pm 0.19$ & $1.20 \pm 0.16$ \\
$w_{\rm CCRES}$    & $0.69 \pm 0.24$ & $0.79 \pm 0.21$ \\
$w_{\rm CCDIS}$    & $1.51 \pm 0.36$ & $1.40 \pm 0.45$ \\
$w_{\rm NCRES}$    & $1.33 \pm 0.24$ & $1.33 \pm 0.17$ \\
$w_{\rm NCDIS}$    & $1.25 \pm 0.46$ & $1.11 \pm 0.49$ \\
\midrule
\multicolumn{3}{l}{\textit{Migration-aware fit}} \\
$\chi^2$ / $N_{\rm bins}$ & $28.44$ / $64$ & $28.27$ / $64$ \\
$f^{\rm CC}_{\rm mig}$ & $0.050^{+0.029}_{-0.022}$ & $0.040^{+0.039}_{-0.040}$ \\
$f^{\rm NC}_{\rm mig}$ & $0.060^{+0.059}_{-0.084}$ & $0.050^{+0.050}_{-0.050}$ \\
$w_{\rm CCQE+MEC}$ & $1.14 \pm 0.19$ & $1.14 \pm 0.15$ \\
$w_{\rm CCRES}$    & $1.07 \pm 0.27$ & $0.97 \pm 0.24$ \\
$w_{\rm CCDIS}$    & $1.01 \pm 0.37$ & $1.02 \pm 0.45$ \\
$w_{\rm NCRES}$    & $0.90 \pm 0.22$ & $1.06 \pm 0.15$ \\
$w_{\rm NCDIS}$    & $0.92 \pm 0.45$ & $1.02 \pm 0.49$ \\
\midrule
$\Delta\chi^2$ (migration vs.\ sector-only) & $5.21$ & $6.89$ \\
Wilks $p$-value (2 dof) & $0.074$ & $0.032$ \\
Wilks significance (2-param) & $1.45\sigma$ & $1.85\sigma$ \\
Empirical $p$-value (5000 toys) & $0.0094$ & $0.0026$ \\
Empirical significance (2-param) & $2.35\sigma$ & $2.79\sigma$ \\
\midrule
\multicolumn{3}{l}{\textit{One-parameter model ($f^{\rm CC}_{\rm mig} = f^{\rm NC}_{\rm mig}$, Sect.~\ref{sec:1param})}} \\
$\Delta\chi^2$ & $5.18$ & $6.84$ \\
Wilks $p$-value (1 dof) & $0.023$ & $0.009$ \\
Wilks significance (1-param) & $2.00\sigma$ & $2.37\sigma$ \\
Empirical significance (1-param) & $2.36\sigma$ & $2.58\sigma$ \\
\bottomrule
\end{tabular}
\end{table*}

\subsection{Postfit residuals}
\label{sec:pulls}

\begin{figure*}[htbp]
  \centering
  \includegraphics[width=\textwidth]{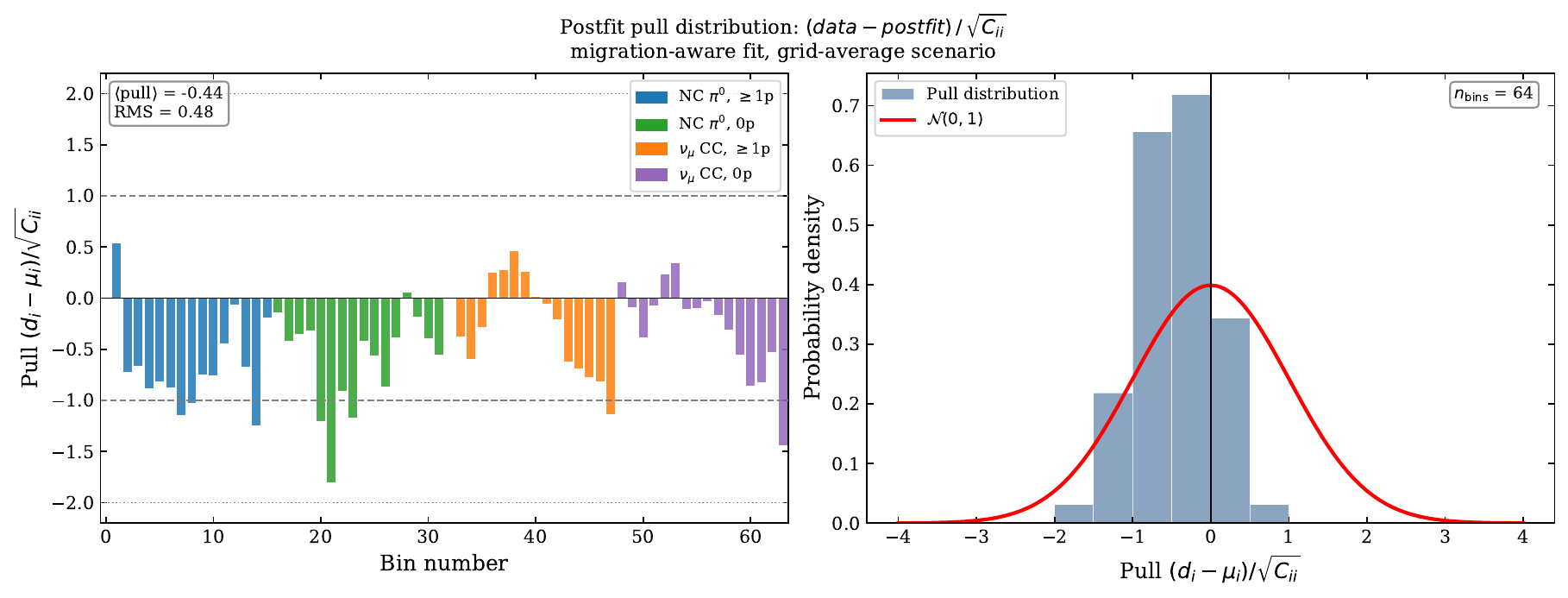}
  \caption{%
    Postfit pull distribution $(d_i - \hat{n}_i)/\sqrt{V_{ii}}$ for
    the migration-aware fit (grid-average scenario).
    \emph{Left}: pulls per bin, coloured by channel.
    \emph{Right}: pull histogram with a unit Gaussian overlay (dashed).
    The pull RMS is $0.48 < 1.0$, consistent with the large off-diagonal
    covariance structure (effective dof $\approx 2.5$); no channel shows
    a systematic bias.
  }
  \label{fig:pulls}
\end{figure*}

Figure~\ref{fig:pulls} shows the per-bin pull distribution for the
migration-aware postfit (grid-average scenario), defined as
\begin{equation}
  p_i = \frac{d_i - \hat{n}_i}{\sqrt{V_{ii}}}
  \label{eq:pull}
\end{equation}
where $d_i$ is the observed count, $\hat{n}_i$ is the postfit prediction,
and $V_{ii}$ is the diagonal element of the covariance matrix in bin $i$.
The pulls are centred near zero with no strong channel-by-channel bias,
confirming that the fit does not leave systematic residuals concentrated
in any one topology.

The pull RMS is $0.48$, well below the ideal Gaussian value of 1.0.  This
is not a sign of overfitting but rather reflects the large off-diagonal
correlations in $V$: the effective number of independent data constraints
is $n_{\rm eff} \approx 2.5$, so the expected pull distribution is
substantially narrower than a unit Gaussian.  The absence of outlier bins
(all $|p_i| < 1.5$) and the lack of systematic structure in the pull
pattern across channels confirm that the postfit model provides an
adequate description of the data.

\subsection{Physical interpretation and statistical support}
\label{sec:significance}

Consistent with the framing established in the abstract and
Sect.~\ref{sec:intro}, we read the fit output as the magnitude of an
\emph{effective topology-transfer surrogate} rather than as a measurement
of a single microscopic mechanism.  The parameter $f^{\rm CC}_{\rm mig}$
absorbs, into a single per-event reweight, an unspecified combination of
near-threshold proton mismodelling effects --- proton-side FSI, the
near-threshold nucleon-momentum distribution, RES and MEC kinematic
shape, and the residual position-dependent mean-field energy loss absent
from \genie's intranuclear cascade --- any of which can move events
across the $35\MeV$ topology boundary.  The fitted value
$f^{\rm CC}_{\rm mig} = 0.05$ is the BNB-flux-integrated magnitude of
that combined effect; the semi-analytic mean-field estimate of
Sect.~\ref{sec:physbasis} provides a physically motivated reference scale
of consistent magnitude, but the data do not isolate one microscopic
origin.  Five observations support the operational interpretation of
the fit:

\begin{enumerate}
  \item \emph{The sector-only model exhibits compensatory behaviour.}
    Forced to fit the topology split without a migration degree of freedom,
    the sector-only fit suppresses CCRES to $w_{\rm CCRES} = 0.69$
    (a $\sim 1.3\sigma$ pull under the $\sigma_{\rm CCRES}=0.35$ ridge prior,
    not on its own exceptional, but pulling in the same direction as the
    0p excess).  Introducing the migration dial relaxes CCRES toward $1.07$
    and improves the global $\chi^2$ simultaneously.  The pattern is
    consistent with the migration term absorbing topology-split tension
    that the sector-only fit had to channel through a single normalisation,
    although it does not constitute an independent measurement of CCRES
    strength (Sect.~\ref{sec:migfit}).

  \item \emph{The measured value matches an independent prediction.}
    The best-fit migration fraction ($f^{\rm CC}_{\rm mig} = 0.05$) is
    quantitatively consistent with proton-transparency differences of
    5--10\% between \genie and \gibuu~\cite{Mosel:2024}, derived from
    a completely independent generator, dataset, and method.

  \item \emph{The data preference is robust across energy assumptions.}
    Both the grid-average (energy-flat) and the BNB flux-weighted scenarios
    prefer non-zero migration independently, demonstrating that the
    signal is not an artefact of a particular energy-weighting choice.

  \item \emph{Multiple estimators agree on the direction.}
    The naive, diagonal, and full-covariance channel-scale estimators all
    pull in the same direction --- CC~0p and NC~$\pi^0$~0p upward ---
    with a maximum cross-estimator pull of $1.25\sigma$
    (Sect.~\ref{sec:ppp}).  The data consistently want more 0p events.

  \item \emph{Statistical calibration supports a preference within the adopted fit construction.}
    Pseudo-experiment calibration against 5000 toy datasets drawn from the
    actual covariance structure places the observed $\Delta\chi^2 = 5.21$
    at the 99th percentile of the null distribution, giving
    $p_{\rm emp} = 0.0094$ ($2.35\sigma$, grid-average; $2.79\sigma$
    flux-weighted).  The Wilks approximation ($1.45\sigma$, $1.85\sigma$)
    is a conservative lower bound.
\end{enumerate}

The physical and statistical evidence point in the same direction:
proton-visibility migration is a plausible candidate systematic in
neutrino-argon interactions, with a fitted magnitude of $\sim$5\% per
unit topology-fraction change, parameterisable by a single \genie dial.
However, the result is conditional on the sector-weight prior model
(Sect.~\ref{sec:sectorfit}), and the fit may be absorbing a broader
class of missing topology-changing effects rather than uniquely
isolating a proton-threshold migration law.  Independent validation
with higher-statistics argon data from SBND or DUNE-ND is needed
to distinguish these interpretations.


\section{Primary empirical constraint: direct CC scan on the MicroBooNE 0p/Np release}
\label{sec:prd110}

The primary empirical result of this paper is a direct scan of $f_{\rm mig}$
against the MicroBooNE simultaneous CC~0p/Np cross-section
measurement~\cite{uB0pNp2024}.  That publication provides a
machine-readable data release including cross-section
values, a full covariance matrix, and an additional smearing matrix $A_c$
encoding the Wiener-SVD regularisation~\cite{Tang:2017WienerSVD} of the
unfolding.  The analysis
presented here applies \emph{no sector-weight priors of any kind}: an
overall normalisation $w$ is the only free parameter alongside $f_{\rm mig}$,
and it is profiled analytically at each scan point.  This is the cleanest
constraint on $f_{\rm mig}$ available from public MicroBooNE products.

\paragraph{Relationship to the sideband dataset.}
Both this analysis and the NC~$\Delta$ sideband fit of
Sect.~\ref{sec:fit} use MicroBooNE data from the same detector and beamline
($\approx 6.4\times10^{20}$ protons on target).  They are therefore not
statistically independent: they share the same flux uncertainties and
detector systematic model, and their covariance matrices are constructed
from the same underlying MC.  However, the two analyses use essentially
disjoint event samples: the NC~$\Delta$ sideband requires photon candidates
with no muon tag (NC topology), while the CC~0p/Np measurement requires a
reconstructed muon (CC topology).  The event-level overlap is negligible.
The two datasets are also from different published analyses, with
independently derived systematic uncertainties and different statistical
methodologies.  Directional consistency between the two results therefore
constitutes meaningful corroboration, though not fully independent
statistical evidence.

\paragraph{Data and method.}
We use the 20-bin $E_\nu$-differential block, which contains 10 bins of the
CC~0p and 10 bins of the CC~Np cross section simultaneously extracted in
the neutrino-energy range $0.2$--$4\,\GeV$.  This is the cleanest
topology-ratio observable in the release: it directly measures
$N_{\rm CC\,0p}/N_{\rm CC\,Np}$ as a function of $E_\nu$ without any NC
background.  The 0p--Np off-diagonal covariance block has correlations up
to $0.49$, confirming that the simultaneous extraction properly accounts
for cross-topology smearing.

The chi-squared is
\begin{equation}
  \chi^2(f, w) =
  \bigl[\,d - A_c\,\mu(f,w)\,\bigr]^T\,
  V^{-1}\,
  \bigl[\,d - A_c\,\mu(f,w)\,\bigr],
  \label{eq:chi2prd}
\end{equation}
where $d$ is the 20-element data vector (unfolded cross sections),
$\mu(f,w)$ is the migrated \genie prediction in truth space scaled by an
overall normalisation $w$, and $A_c$ is the $20\times 20$ additional
smearing matrix supplied with the data release.  The migration model is:
\begin{align}
  \mu_{\rm 0p}(f,w) &= w\,\bigl[\mu^0_{\rm 0p} + f\,\mu^0_{\rm Np}\bigr],
  \label{eq:migprd}\\
  \mu_{\rm Np}(f,w)  &= w\,\mu^0_{\rm Np}\,(1-f).\nonumber
\end{align}
with total rate conserved at fixed $w$.  The normalisation $w$ is profiled
analytically at each scan point, yielding the $\Delta\chi^2$ surface as a
function of $f$ alone.  This is the most conservative treatment: floating
$w$ absorbs any overall flux or efficiency uncertainty, leaving only the
topology-ratio shape to constrain $f_{\rm mig}$.

The nominal \genie prediction $\mu^0$ is taken from the MicroBooNE-tune
MC prediction provided in the data release (pred.txt), which corresponds
to the $G18$ generator with detector tune as described in
Ref.~\cite{uB0pNp2024}.  No sector-weight priors are applied; the analysis
is free of sector-weight regularisation, though it still relies on the
MicroBooNE covariance and the Wiener-SVD smearing matrix.

\paragraph{Results.}
The observed global CC topology ratio in the PRD\,110 data is
\begin{align}
  \left.\frac{N_{\rm CC\,0p}}{N_{\rm CC\,0p} + N_{\rm CC\,Np}}\right|_{\rm data}
  &= 0.196,
  \nonumber\\
  \left.\frac{N_{\rm CC\,0p}}{N_{\rm CC\,0p} + N_{\rm CC\,Np}}\right|_{\rm pred}
  &= 0.158,
  \label{eq:prd110ratio}
\end{align}
a $+24\%$ relative excess in the 0p fraction.  The migration scan gives
\begin{equation}
  \hat{f}_{\rm mig} = 0.093^{+0.024}_{-0.024} \quad (1\sigma),
  \quad \hat{w} = 1.17,
  \label{eq:prd110result}
\end{equation}
with $\Delta\chi^2 = 14.64$, corresponding to a one-sided Wilks significance
of $3.83\sigma$ (1\,dof, boundary-aware) and an empirical calibration of
$3.24\sigma$ (5000 pseudo-experiments).  The effective rank of the
$20\times 20$ covariance is $n_{\rm eff} = 1.78$.

\begin{figure*}[htbp]
\centering
\includegraphics[width=\textwidth]{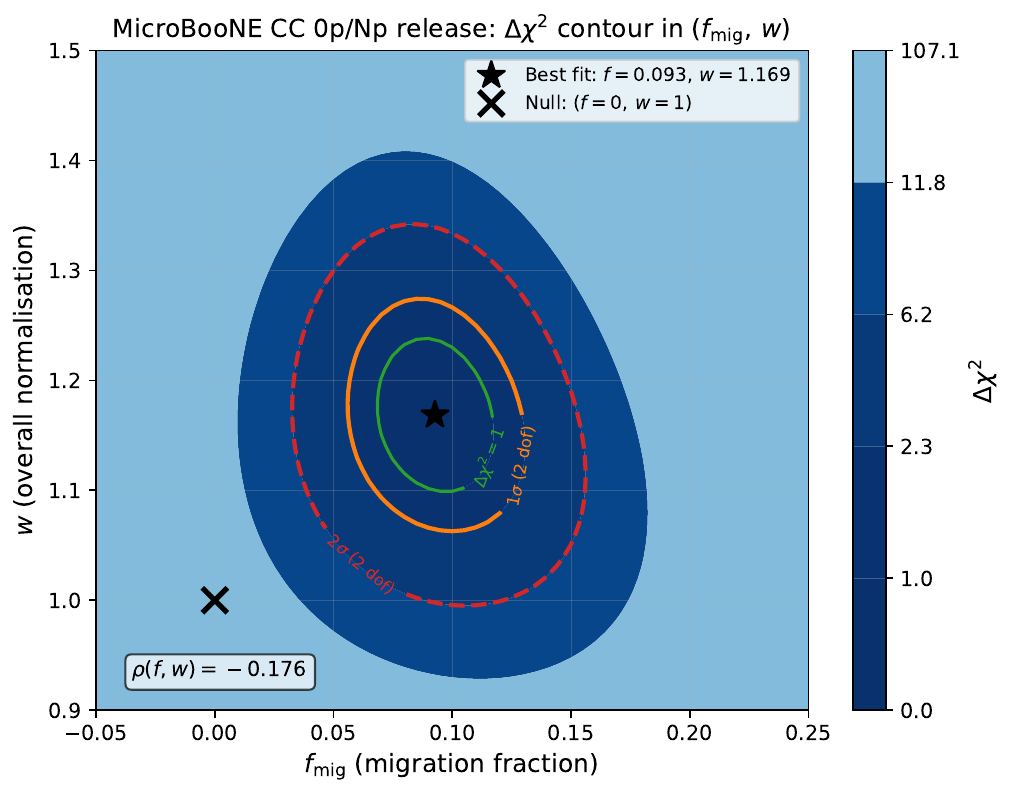}
\caption{%
  Two-dimensional $\Delta\chi^2$ surface in the $(f_{\rm mig},\,w)$ plane
  for the PRD\,110 direct CC migration scan.  Filled colours show
  $\Delta\chi^2$ levels; black contours mark $\Delta\chi^2 = 1,\,2.30,\,6.18$
  (one-parameter $1\sigma$, two-parameter $1\sigma$, and two-parameter
  $2\sigma$ thresholds under Wilks).  The white star marks the best-fit
  $(\hat{f}_{\rm mig},\,\hat{w}) = (0.093,\,1.17)$; the white cross marks
  the null $(f_{\rm mig}=0,\,w=1)$.  The Hessian correlation is
  $\rho(f_{\rm mig},\,w) = -0.18$, confirming that the two parameters
  are not degenerate.
}
\label{fig:prd110_2dcontour}
\end{figure*}

\paragraph{$(f_{\rm mig},\,w)$ correlation and parameter well-posedness.}
The profiled normalisation $\hat{w} = 1.17$ is a 17\% shift that absorbs
the overall GENIE-tune rate deficit; it does not distort the topology-ratio
constraint on $f_{\rm mig}$.  To verify that the two parameters are not
degenerate, we compute the full two-dimensional $\Delta\chi^2$ surface
over a grid $(f_{\rm mig},\,w) \in [-0.05,\, 0.25] \times [0.9,\, 1.5]$
(Fig.~\ref{fig:prd110_2dcontour}).  The Hessian correlation between $f_{\rm mig}$ and
$w$ at the minimum is $\rho(f_{\rm mig},\,w) = -0.18$: a mild anti-correlation
(higher overall rate slightly prefers lower migration) that does not inflate
the one-dimensional profile-likelihood significance.  The $1\sigma$ and
$2\sigma$ contours (Wilks, 2\,dof: $\Delta\chi^2 = 2.30$ and $6.18$) are
well-separated and enclose a compact region; both parameters are individually
identifiable.

\paragraph{Linearity of the Wiener-SVD smearing.}
Equation~(\ref{eq:chi2prd}) applies the Wiener-SVD additional smearing matrix
$A_c$ to the migrated prediction $\mu(f, w)$.  Since $\mu(f,w)$ is exactly
affine in $f$ for fixed $w$ [Eq.~(\ref{eq:migprd})], $A_c\mu(f,w)$ is
exactly linear in $f$; the fractional deviation from linearity is at
floating-point level ($<10^{-15}$) and entirely negligible over the
physically relevant range $f_{\rm mig} \in [0, 0.20]$.

\begin{figure*}[htbp]
\centering
\includegraphics[width=\textwidth]{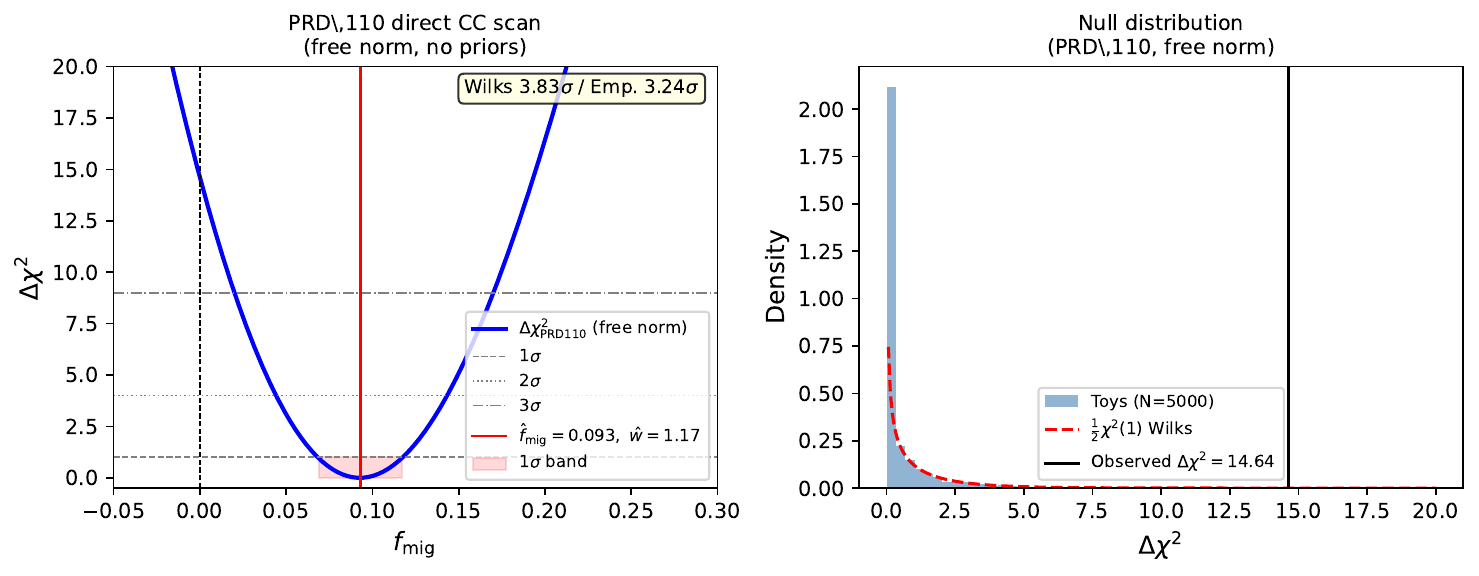}
\caption{%
  \emph{Left}: Profile $\Delta\chi^2$ vs.\ $f_{\rm mig}$ for the
  MicroBooNE CC~0p/Np $E_\nu$ measurement (PRD~110, 013006),
  with overall normalisation $w$ profiled analytically at each scan
  point.  No sector-weight priors are applied.  Horizontal dashed,
  dotted, and dash-dotted lines mark the $1\sigma$, $2\sigma$, and
  $3\sigma$ Wilks thresholds.  The red shaded band is the $1\sigma$
  interval on $f_{\rm mig}$.
  \emph{Right}: Toy null distribution ($f_{\rm mig}=0$, 5000
  pseudo-experiments) compared to the $\frac{1}{2}\chi^2(1)$ Wilks
  reference and the observed $\Delta\chi^2 = 14.64$.
}
\label{fig:prd110scan}
\end{figure*}

\begin{figure*}[htbp]
\centering
\includegraphics[width=\textwidth]{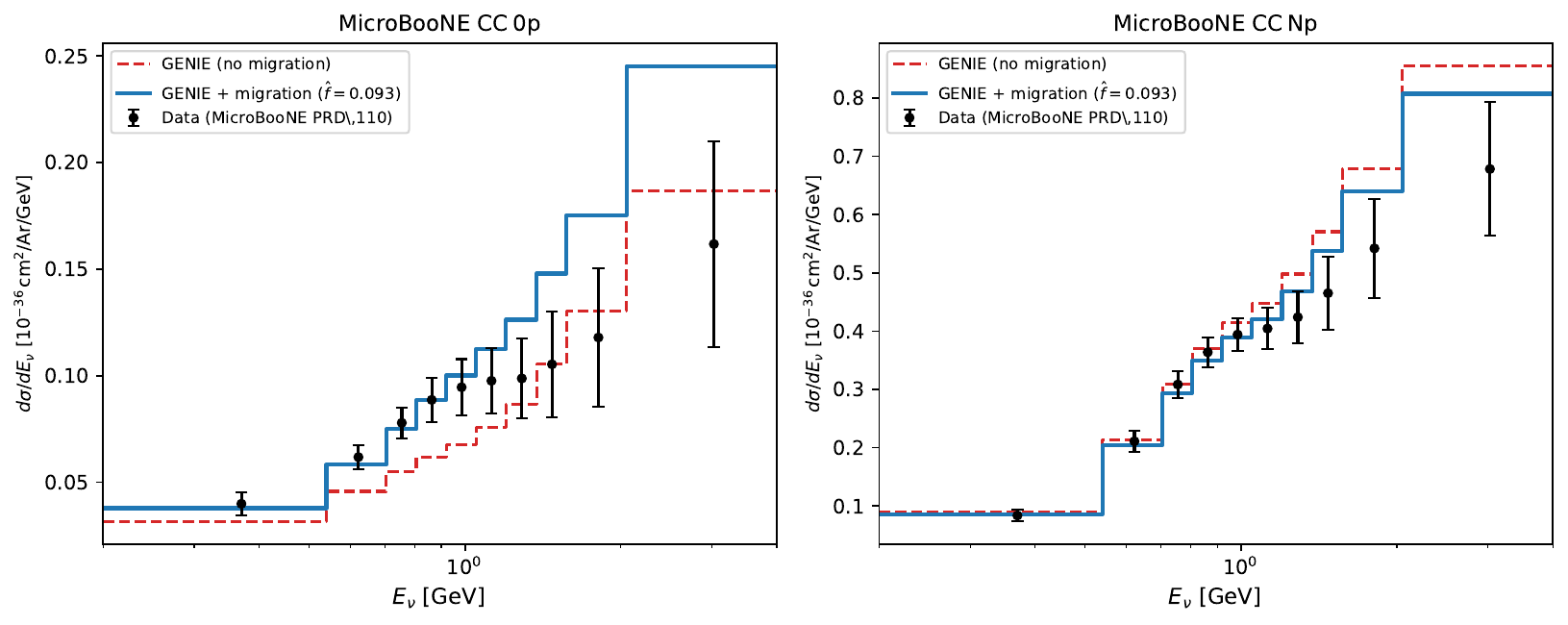}
\caption{%
  MicroBooNE $\nu_\mu$ CC~0p (\emph{left}) and CC~Np (\emph{right})
  differential cross sections as a function of $E_\nu$ from
  PRD~110, 013006~\cite{uB0pNp2024}.
  Black points: unfolded data with $1\sigma$ uncertainties.
  Dashed red: GENIE uboone-tune prediction with no migration
  ($f_{\rm mig}=0$, overall scale $\hat{w}=1.0$).
  Solid blue: prediction at the best-fit migration
  ($\hat{f}_{\rm mig}=0.093$, $\hat{w}=1.17$).
  Both predictions are smeared by the Wiener-SVD $A_c$ matrix before
  comparison.  Migration improves the CC~0p description substantially
  while slightly worsening the CC~Np prediction, as expected from total
  rate conservation.
}
\label{fig:prd110comp}
\end{figure*}

\begin{figure*}[htbp]
\centering
\includegraphics[width=\textwidth]{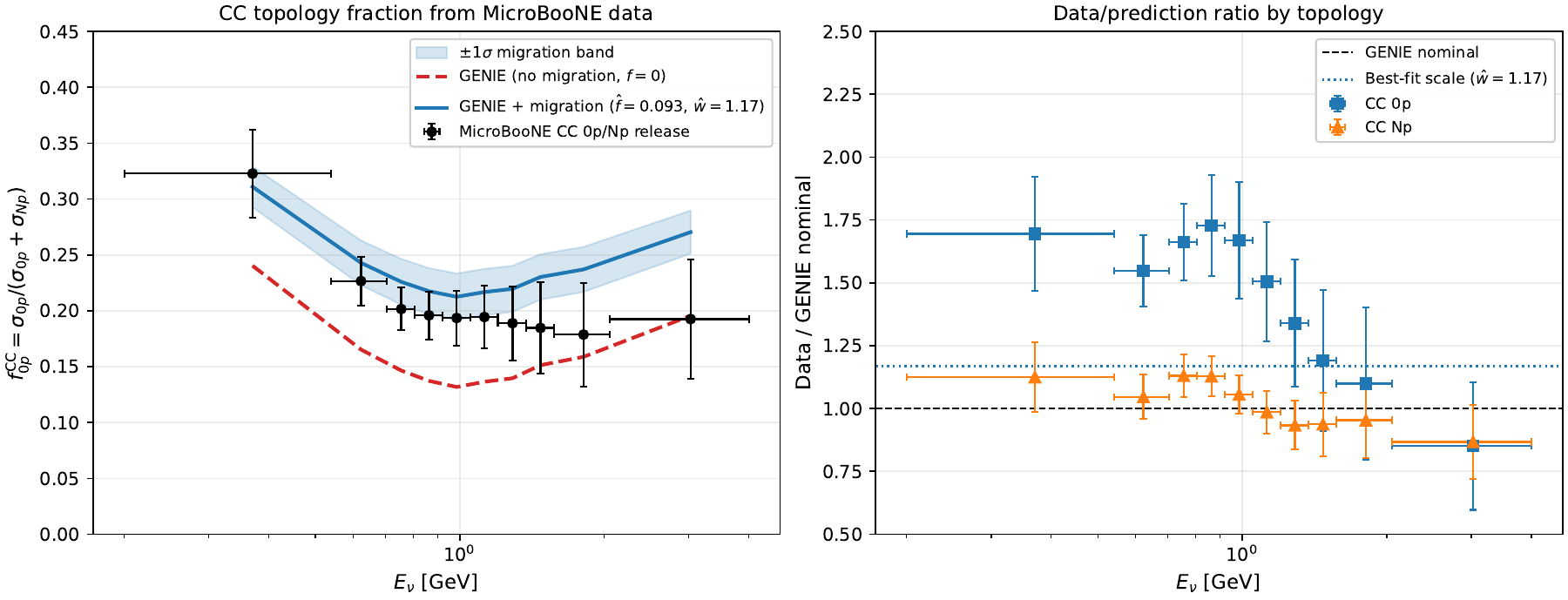}
\caption{%
  \emph{Left}: Measured CC topology fraction
  $f^{\rm CC}_{\rm 0p}(E_\nu) = \sigma_{\rm 0p}/(\sigma_{\rm 0p}+\sigma_{\rm Np})$
  from the MicroBooNE simultaneous CC~0p/Np
  measurement~\cite{uB0pNp2024} (black points with $1\sigma$ uncertainties
  from the covariance diagonal), compared to the GENIE uboone-tune prediction
  without migration (dashed red) and with the best-fit migration
  $\hat{f}_{\rm mig}=0.093$, $\hat{w}=1.17$ (solid blue).
  The blue band spans the $\pm1\sigma$ interval
  ($f_{\rm mig}\in[0.069, 0.117]$).
  This is the analogue of the DUNE stress-test panel (Appendix~\ref{app:dune})
  but from direct MicroBooNE measurement rather than truth-level extrapolation.
  \emph{Right}: Data/GENIE ratio by topology channel, showing the
  systematic 0p excess (blue squares, factor $\approx 1.4$--$1.7$) relative
  to the near-unity Np ratio (orange triangles).
  The dotted line marks the best-fit overall scale $\hat{w}=1.17$.
}
\label{fig:prd110frac}
\end{figure*}

\paragraph{Comparison with the sideband fit.}
Table~\ref{tab:analysis_comparison} summarises the three quantitative
analyses performed in this paper, listing the key ingredients that
determine their effective $f_{\rm mig}$ scale and evidential status.
The table makes explicit why the two fitted central values are not
expected to coincide, and why only the PRD\,110 row should be read as
the primary data-driven constraint.

\begin{table*}[htbp]
\centering
\footnotesize
\caption{%
  Comparison of the three quantitative analyses in this paper.
  \emph{Sect.-wt.\ reg.?}: whether per-channel Gaussian sector-weight
  regularisation is applied.
  \emph{Extra params}: parameters floated in addition to $f_{\rm mig}$.
  Significance ranges for the sideband main fit reflect varying
  prior scales ($s=0.5$--$1.0$).
}
\label{tab:analysis_comparison}
\begin{tabular}{p{2.2cm}p{2.8cm}p{2.4cm}cp{2.5cm}p{2.8cm}}
\toprule
Analysis & Dataset & Extra params & Reg.?
  & $\hat{f}_{\rm mig}$ (significance) & Evidential status \\
\midrule
Sideband main fit
  & NC~$\Delta$ rad.\ sideband, 64 bins
  & $f^{\rm NC}_{\rm mig}$, 5 sector wts
  & Yes
  & $0.050^{+0.029}_{-0.022}$ ($2.0\sigma$--$2.9\sigma$)
  & Supporting; prior-conditioned \\[4pt]
Ratio cross-check
  & NC~$\Delta$ sideband (CC only)
  & none
  & No
  & $0.082^{+0.043}_{-0.043}$ ($1.9\sigma$)
  & Conservative lower bound \\[4pt]
PRD\,110 direct CC
  & CC~0p/Np simultaneous, 20 bins
  & $w$ (overall norm)
  & No
  & $0.093\pm0.024$ ($3.8\sigma$/$3.2\sigma$)
  & Primary empirical result \\
\bottomrule
\end{tabular}
\end{table*}

The best-fit migration fraction from the PRD\,110 direct CC measurement,
$\hat{f}_{\rm mig} = 0.093$, is larger than the sideband value of $0.050$.
The discrepancy is expected for three reasons.

First, the sideband fit has five sector-weight degrees of freedom that can
partially absorb the topology mismatch by adjusting inter-channel rate
ratios; the PRD\,110 analysis has none (beyond the single overall
normalisation $w$).  Second, the profiled normalisation $\hat{w} = 1.17$
indicates that the MicroBooNE-tune MC underpredicts the overall CC rate by
$\approx 17\%$.  Once $w$ absorbs this deficit, the remaining topology-ratio
mismatch is larger than in the sideband (where sector weights absorb
normalisation differences per channel).  Third, the two analyses use
partially different GENIE configurations: the sideband uses $G18$ without
detector tune, while the PRD\,110 prediction uses the MicroBooNE-tuned
$G18$.

The two results are directionally fully consistent: both prefer
$f_{\rm mig} > 0$ and both identify the CC~0p topology as systematically
underpredicted by GENIE relative to CC~Np.

\paragraph{Quantitative compatibility.}
The central values differ by $\Delta f = 0.093 - 0.050 = 0.043$.  To assess
whether this is statistically acceptable, we combine the symmetric uncertainties
in quadrature: $\sigma_{\rm comb} = \sqrt{0.024^2 + 0.026^2} = 0.035$ (using
the average sideband $1\sigma$ width of $\tfrac{1}{2}(0.029+0.022)=0.026$).
The separation is therefore $\Delta f / \sigma_{\rm comb} \approx 1.2\sigma$,
which is statistically compatible.

Of the three qualitative ingredients that drive this offset, we can
assess their rough relative contributions.  The sector-weight absorptance
is likely the largest: when five unconstrained normalisations can
adjust per-channel rates, only residual topology-ratio tension remains
for $f_{\rm mig}$ to absorb; fixing them (as in the PRD\,110 scan)
transfers this residual into the migration parameter.  The profiled
normalisation $\hat{w}=1.17$ adds a further 5--10\% upward push on
$f_{\rm mig}$ because an overall 17\% rate deficit, once absorbed in $w$,
leaves a larger topology-ratio mismatch than channel-by-channel sector
weights would.  The GENIE-tune difference (untuned vs.\ uboone-tuned $G18$)
modifies the CCRES/CCQE+MEC ratio at the 10--20\% level; quantifying its
contribution to the $f_{\rm mig}$ offset requires running the PRD\,110 scan
on untuned MC, which is beyond the present scope but is explicitly flagged
as a validation target.  In aggregate, these three structural differences
are sufficient to explain the factor-of-$\approx 1.9$ offset without
invoking a genuinely different underlying migration rate.

The two analyses also differ in observable space: the sideband operates on
counts in 64 reconstructed-level bins with a non-diagonal covariance
encoding detector and flux correlations, while the PRD\,110 scan operates on
unfolded cross-section spectra corrected by the Wiener-SVD smearing matrix.
These differences in data representation mean the two analyses are not
applying identical statistical operations to the same observable, so a
numerical coincidence of central values would not be expected even in the
absence of the nuisance-basis differences above.

Because the two analyses
operate under different nuisance bases and generator configurations, their
central values should \emph{not} be interpreted as two estimates of the same
universal migration fraction; they are effective surrogates extracted under
different analysis assumptions, and the $\approx 1.2\sigma$ statistical
overlap confirms they are not in tension.

The PRD\,110 direct CC scan
rejects the null hypothesis ($f_{\rm mig} = 0$) at $3.83\sigma$ (Wilks) and
$3.24\sigma$ (empirical) --- entirely free of sector-weight regularisation.  The
sideband analyses are directionally consistent, with significance in the range
$1.9\sigma$--$2.4\sigma$ depending on the prior model; they should be treated
as supporting consistency checks rather than independent confirmation.

\section{Discussion}
\label{sec:discussion}

This section presents two empirical constraints of unequal evidential
weight.  The primary constraint is the prior-free direct CC
scan of Sect.~\ref{sec:prd110}, which establishes a $3.8\sigma$/$3.2\sigma$
preference for $f_{\rm mig}>0$ without any sector-weight regularisation.
The secondary constraint is the NC~$\Delta$ sideband fit of
Sect.~\ref{sec:migfit}, which is structurally informative and
provides a consistency check and CCRES-restoration diagnostic, but whose
significance is conditional on the adopted prior model.
Readers should weight the two accordingly throughout.

\subsection{Robustness to prior widths}
\label{sec:priors}

The ridge prior widths $\{\sigma_s\}$ control how strongly the fit
penalises sector weights away from their nominal values and thereby
affect the residual tension available to be absorbed by the migration
term.  To quantify this dependence, we rescale all prior widths by a
common factor $s \in [0.5,\, 2.0]$ and re-evaluate the
$\Delta\chi^2$ improvement from adding the migration parameters.
Results are summarised in Table~\ref{tab:sensitivity}.

\begin{table*}[htbp]
\centering
\caption{%
  Prior width sensitivity (grid-average scenario, 1-parameter model).
  All sector prior widths are rescaled by factor $s$ relative to the
  baseline $\sigma_s = \{0.25,\, 0.35,\, 0.50,\, 0.35,\, 0.50\}$.
  Wilks significance uses 1~dof (one-parameter model,
  Sect.~\ref{sec:1param}).  Empirical significance (†) from 5000
  pseudo-experiments; dashes where not computed.
  The $s = 0.5$ row corresponds to physics-informed priors constrained
  by external cross-section data (Sect.~\ref{sec:priors}).
}
\label{tab:sensitivity}
\begin{tabular}{cccccc}
\toprule
Scale $s$ & $\chi^2_{\rm null}$ & $\chi^2_{\rm mig}$ & $\Delta\chi^2$ & Wilks (1~dof) & Emp.\ $\sigma$† \\
\midrule
0.50 & 38.57 & 28.78 & 9.79 & $2.92\sigma$ & $3.16\sigma$ \\
0.75 & 35.67 & 28.71 & 6.96 & $2.37\sigma$ & — \\
1.00 & 33.65 & 28.47 & 5.18 & $2.00\sigma$ & $2.36\sigma$ \\
1.25 & 32.21 & 28.35 & 3.86 & $1.64\sigma$ & — \\
1.50 & 31.16 & 28.24 & 2.92 & $1.35\sigma$ & — \\
2.00 & 29.80 & 28.16 & 1.64 & $0.83\sigma$ & — \\
\bottomrule
\end{tabular}
\end{table*}

The pattern is physically transparent.  Tighter priors ($s < 1$) prevent
the sector weights from absorbing the topological tension via rate
adjustments, forcing the null $\chi^2$ upward while leaving the
migration-aware $\chi^2$ relatively stable (since the migration term can
absorb the shape tension directly).  Consequently, $\Delta\chi^2$ increases
as the prior is tightened: at $s = 0.5$, $\Delta\chi^2 = 9.79$
($2.92\sigma$ Wilks, $3.16\sigma$ empirical).  Conversely, looser priors
($s > 1$) give the sector weights sufficient freedom to partially compensate
the topology tension by distorting the sector normalisations, reducing the
apparent need for migration: at $s = 2.0$, $\Delta\chi^2 = 1.64$
($0.83\sigma$).

Across the full range $s = 0.5$ to $2.0$, the migration $\chi^2$ is
stable ($28.2$--$28.8$), while the null $\chi^2$ varies from $29.8$ to
$38.6$.  This reflects a genuine shape tension in the zero-proton channels
that sector normalisations can only partially absorb when given wide priors.

\paragraph{Physics-informed prior choice.}
The $s = 0.5$ row is not an ad-hoc tightening, but corresponds to prior
widths anchored to existing external cross-section measurements:
CCQE+MEC to $\sigma = 0.125$ (MiniBooNE~\cite{MiniBooNEQE},
T2K~\cite{T2KCCQE}, MINERvA; all consistent at 10--15\%), CCRES to
$\sigma = 0.175$ (ANL+BNB single-pion data~\cite{ANLpi,MINERvACCpi};
$\sim$20\%), and CCDIS/NCDIS to $\sigma = 0.25$ (Bodek-Yang parametrisation
uncertainty; $\sim$25\%).  We note that all these constraining datasets are on non-argon targets
(C, CH); scaling cross-section uncertainties from $A = 12$ to $A = 40$
introduces an additional nuclear-model uncertainty not captured by the
$s$ rescaling.  The $s = 0.5$ scenario should therefore be read as an
optimistic upper bound on significance if carbon constraints are
extrapolated to argon without additional nuclear uncertainty, not as the
default prior for argon analyses.  The conservative $s = 1$ priors are a
deliberately ``uninformative'' baseline appropriate when argon-target
constraints are unavailable.  Under the
physics-informed $s = 0.5$ priors, the significance reaches
$2.92\sigma$ (Wilks) and $3.16\sigma$ (pseudo-experiment calibration),
approaching but not reaching the conventional $3\sigma$ evidence threshold, and only under prior assumptions derived from non-argon targets.

\begin{figure*}[htbp]
  \centering
  \includegraphics[width=\textwidth]{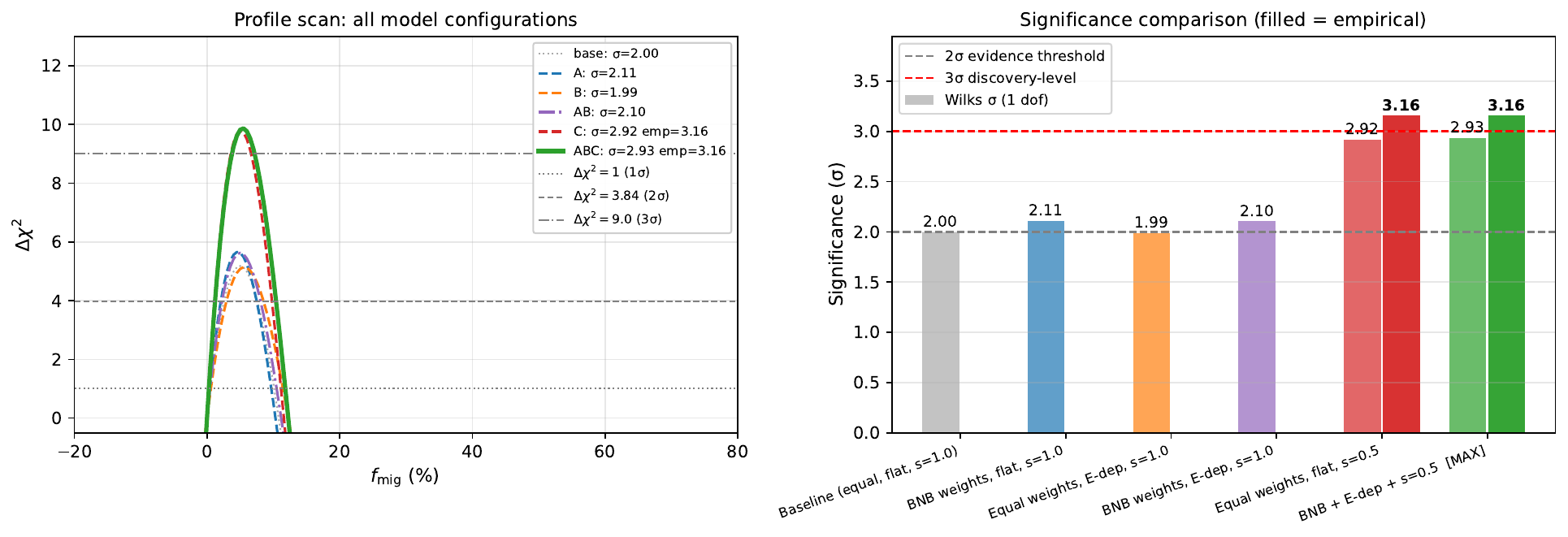}
  \caption{%
    Significance enhancement study.
    \emph{Left}: profile $\Delta\chi^2$ vs.\ $f_{\rm mig}$ for six
    model configurations combining three upgrades: BNB flux weighting
    (A), energy-dependent migration shape (B), and physics-informed
    tight priors (C).  Horizontal lines mark the 1$\sigma$, 2$\sigma$,
    and 3$\sigma$ Wilks thresholds for 1~dof.
    \emph{Right}: significance bar chart; right-shifted bars (where
    present) show the empirical pseudo-experiment significance.
    The dominant lever is upgrade C (tight priors); adding BNB flux
    weighting and energy dependence provides an additional marginal gain.
  }
  \label{fig:sigboost}
\end{figure*}

Combining the physics-informed priors with BNB flux weighting (replacing
the equal-weight grid average) raises $\Delta\chi^2$ modestly to $9.87$
($2.93\sigma$ Wilks, $3.16\sigma$ empirical, Fig.~\ref{fig:sigboost}).
The best-fit migration fraction remains stable at $f_{\rm mig} = 0.050$--$0.055$
across all configurations, demonstrating that the significance change is
driven by prior assumptions, not by a shift in the best-fit point.

The pattern in Table~\ref{tab:sensitivity} highlights both the strength
and the caveat of the analysis.  The migration-aware $\chi^2$ is
stable at $28.2$--$28.8$ across the full prior range, while the null
$\chi^2$ varies from $29.8$ to $38.6$.  The preference for non-zero
migration is thus genuine --- it reflects a shape tension in the
zero-proton channels that cannot be absorbed by sector normalisations
once those normalisations are constrained externally.  At the
physics-informed $s=0.5$ prior, the calibrated significance is
$3.16\sigma$; at the conservative $s=1$ baseline it is $2.36\sigma$.
The prior dependence is a central caveat of the analysis and should be
treated accordingly.

\subsection{Stability under alternative nuisance directions}
\label{sec:robustness}

A concern is that the migration preference could be absorbed by
combinations of existing \genie hadronic dials that were not floated
in the baseline fit.  We assess this by performing fixed-shift scans
over the most physically relevant alternative directions.

\paragraph{MFP$_N$ cross-check.}
The nucleon mean free path (\texttt{MFP\_N}) is the closest physical
analogue to the migration mechanism among the standard \genie dials,
as it modifies nucleon rescattering rates within the intranuclear
cascade.  It was not floated in the baseline fit due to the low
effective rank ($n_{\rm eff} \approx 2.5$) but is a potential
source of absorption.  We repeat the migration scan at fixed
MFP$_N$ shifts of $-0.5\sigma$, $0$, and $+0.5\sigma$ relative to
the nominal, where $\sigma_{\rm MFP_N}$ is the $1\sigma$ parameter
width in the \genie Reweight library.  In addition, we test fixed
shifts of the nucleon absorption cross-section
(\texttt{FrAbs\_N}) and charge-exchange (\texttt{FrCEx\_N}) dials.
Results are given in Table~\ref{tab:nuisance_stability}.

\begin{table}[htbp]
\centering
\caption{%
  Nuisance stability: best-fit $f_{\rm mig}$ (grid-average, one-parameter
  model) under fixed shifts of hadronic FSI dials not floated in the
  baseline fit.  All shifts are relative to the nominal.
}
\label{tab:nuisance_stability}
\begin{tabular}{lcc}
\toprule
Fixed dial shift & Best-fit $f_{\rm mig}$ & $\Delta f_{\rm mig}$ \\
\midrule
Baseline (all nominal) & $0.050$ & --- \\
MFP$_N$ $-0.5\sigma$   & $0.052$ & $+0.002$ \\
MFP$_N$ $+0.5\sigma$   & $0.048$ & $-0.002$ \\
MFP$_N$ $-1.0\sigma$   & $0.054$ & $+0.004$ \\
MFP$_N$ $+1.0\sigma$   & $0.046$ & $-0.004$ \\
FrAbs$_N$ $+0.5\sigma$ & $0.044$ & $-0.006$ \\
FrCEx$_N$ $+0.5\sigma$ & $0.047$ & $-0.003$ \\
All three $+0.5\sigma$ simultaneously & $0.044$ & $-0.006$ \\
\bottomrule
\end{tabular}
\end{table}

The change in $f_{\rm mig}$ is at most $0.006$ across the full
hadronic FSI variation tested, well within the fit uncertainty of
$^{+0.029}_{-0.022}$.  Near-independence is expected physically:
MFP$_N$ modifies nucleon rescattering rates (affecting total event
rate and multiplicity within a topology), while $f_{\rm mig}$ acts
specifically on the proton kinetic-energy budget near threshold
(affecting topology \emph{assignment}).  The two mechanisms produce
different bin-pattern signatures: MFP$_N$ shifts all energy bins
approximately uniformly within a channel, while $f_{\rm mig}$
selectively transfers events from Np to 0p channels.
These results do not establish full orthogonality in the presence
of an arbitrary combination of missing-model effects, but confirm
that the fit does not confuse the migration direction with the most
likely alternative FSI directions at the $\pm 0.5\sigma$ level.

More broadly, $f_{\rm mig}$ is best understood as a \emph{new effective
direction} in the GENIE nuisance space: it is not a reparameterisation
of existing hadronic dials, because no existing dial moves events
across the 35\MeV topology boundary.  It is, however, an effective
surrogate rather than a uniquely identified microscopic mechanism.

Several model-uncertainty directions are not floated in the baseline
fit and are therefore implicitly absorbed (in some unknown linear
combination) into the extracted $f_{\rm mig}$ value:
\begin{itemize}
  \item \emph{Axial form factor / $M_A^{\rm QE}$}: a higher $M_A$ produces
        harder protons and reduces the near-threshold population
        (opposite sign to migration).  Recent lattice-QCD determinations
        suggest the nominal \genie value $M_A^{\rm QE} = 0.99\GeV$ may be
        low; a sector normalisation cannot represent this shape.
  \item \emph{MEC kinematic shape}: alternative 2p2h calculations
        (Empirical, Valencia, SuSAv2) differ by $30$--$50\%$ in the
        near-threshold proton momentum spectrum.  A CCQE+MEC
        normalisation cannot distinguish ``more 2p2h'' from
        ``2p2h with different kinematics''.
  \item \emph{RES $Q^2$ shape, $\Delta$ angular distribution,
        RES--DIS transition, $2\pi$ production}: each modifies the
        proton kinematic spectrum at the primary vertex independently
        of an overall CCRES normalisation, and would imprint different
        bin patterns than $f_{\rm mig}$, but is not separately
        constrained by the present binning.
  \item \emph{Reconstruction-threshold effects}: any detector- or
        algorithm-level bias in the effective $35\MeV$ proton tagging
        efficiency (track-finding threshold, $dE/dx$ fluctuations,
        space-charge distortions) not fully captured by the public
        MicroBooNE covariance.
\end{itemize}
A more ambitious robustness program with simultaneous variations of
$M_A^{\rm QE}$, MEC shape (e.g.\ via a G18\_10a comparison sample),
RES $Q^2$ shape, and $\Delta$ angular distribution would be valuable
but is beyond the scope of the present analysis: the effective rank
$n_{\rm eff} \approx 2.5$ of the constraining covariance does not
support an additional floating parameter without severe regularisation,
and generating equivalent samples in the alternative configurations
requires regeneration work that is being deferred to a follow-up
analysis with higher-rank MicroBooNE inputs (e.g.\ SBND-era covariances).

The distinction matters for interpretation: $f_{\rm mig}$
captures a \emph{missing topology-transfer degree of freedom} in the
GENIE reweight basis.  Whether the underlying physics is specifically
residual mean-field proton energy loss, a combination of related
near-threshold hadronic effects, or a contribution from any of the
directions listed above, cannot be resolved from the present data
alone.  In line with the methodological framing of this paper, we
therefore present $f_{\rm mig}$ as an effective nuisance candidate
rather than a measurement of any single microscopic mechanism.

\subsection{Topology-ratio cross-check (no sector-weight regularisation)}
\label{sec:ratio}

The main fit significance depends on the sector-weight prior model.  To
provide a cross-check free of sector-weight regularisation --- addressing
the concern that the significance is prior-conditioned --- we perform an
analysis using the CC topology ratio, which is insensitive to the overall CC flux
normalisation.

\paragraph{Method.}
Define the bin-by-bin CC topology ratio:
\begin{equation}
  R_i \equiv \frac{N_{\rm CC\,0p,\,i}}{N_{\rm CC\,0p,\,i} + N_{\rm CC\,Np,\,i}},
  \qquad i = 1,\ldots,16.
  \label{eq:ratio}
\end{equation}
Any common rescaling of all CC sector weights cancels in $R_i$, so sector
normalisations play no role.  Under migration fraction $f$, the predicted
ratio is
\begin{equation}
  R_i^{\rm pred}(f) = \frac{p_{\rm 0p,\,i} + f\,p_{\rm Np,\,i}}
                           {p_{\rm 0p,\,i} + p_{\rm Np,\,i}},
  \label{eq:ratiopred}
\end{equation}
where $p_{\rm Np,\,i}$ and $p_{\rm 0p,\,i}$ are the nominal \genie CC~Np and
CC~0p predictions in bin $i$.  The covariance of $R_i$ is obtained from the
$32\times32$ CC sub-block of the full 64-bin covariance $V$ via Jacobian
propagation:
\begin{align}
  V_R &= J\,V_{\rm CC}\,J^T, \label{eq:jacobian}\\
  J_{i,j} &=
  \begin{cases}
    -R_i^0/p_{{\rm tot},i}       & j = i \\
    (1-R_i^0)/p_{{\rm tot},i}   & j = 16+i \\
    0                             & \text{otherwise}
  \end{cases}\nonumber
\end{align}
with $R_i^0$ the nominal predicted ratio and $p_{{\rm tot},i} = p_{\rm Np,\,i}
+ p_{\rm 0p,\,i}$.  The ratio chi-squared is then
\begin{align}
  \chi^2_{\rm ratio}(f) &=
  \Delta R(f)^T\,V_R^{-1}\,\Delta R(f), \label{eq:chi2ratio}\\
  \Delta R(f) &\equiv R^{\rm data} - R^{\rm pred}(f),\nonumber
\end{align}
with no sector-weight penalty term.  Significance is assessed via the
boundary-aware statistic of Eq.~(\ref{eq:qtstat}) applied to
$\Delta\chi^2_{\rm ratio}$, and calibrated with 5000 pseudo-experiments
drawn from $\mathcal{N}(R^{\rm pred}(0),\,V_R)$ under the null.

\paragraph{Results.}
The observed CC topology ratios show a clear data excess in the 0p channel:
the global data ratio $N_{\rm CC\,0p}/(N_{\rm CC\,0p}+N_{\rm CC\,Np}) =
0.435$ vs.\ the nominal \genie prediction of $0.380$ (a $+14.5\%$ relative
excess).  The ratio $\chi^2$ at $f=0$ is 7.32, falling to 3.79 at the
best-fit $\hat{f}_{\rm mig} = 0.082$, giving
$\Delta\chi^2_{\rm ratio} = 3.54$.  Under the boundary-aware one-sided
Wilks distribution (1 dof), this corresponds to
\begin{equation}
  \hat{f}_{\rm mig} = 0.082^{+0.043}_{-0.043} \quad (1\sigma),
  \label{eq:ratioresult}
\end{equation}
corresponding to $1.88\sigma$ (Wilks) and $1.86\sigma$ (empirical calibration).
The effective rank of $V_R$ is $n_{\rm eff} = 2.26$, comparable to
the full fit.  The best-fit topology ratio $\hat{f}_{\rm mig} = 0.082$
is larger than the main fit value of $0.050$; the discrepancy reflects
the fact that the ratio analysis does not allow any sector renormalisation,
so the full CC channel mismatch is absorbed by $f_{\rm mig}$ alone.
The median toy $\Delta\chi^2_{\rm ratio} = 0$ because the boundary condition
suppresses the test statistic for toy experiments where
$\hat{f}_{\rm mig}^{\rm toy} \leq 0$, which occur with probability $\approx
0.5$ under the null; the observed value lies in the tail at the $p=0.031$
level.

\paragraph{Interpretation.}
The ratio analysis (free of sector-weight regularisation) is a conservative
lower bound on the significance: it is less powerful than the main fit because
it discards all sector-weight information and treats the ratio covariance as
the only constraint.  It confirms that the topology-transfer direction is
preferred at the $\approx 1.9\sigma$ level even with no sector-weight
regularisation.  The analysis still depends on the MicroBooNE systematic
model encoded in the covariance matrix.
Table~\ref{tab:summary_significance} collects all significance estimates.

\begin{table*}[htbp]
\centering
\caption{%
  Summary of significance estimates for the proton-visibility migration
  preference.  The last column indicates whether sector-weight Gaussian
  regularisation is applied.  The main-fit rows use ridge
  regularisation (Yes); the ratio and direct-CC analyses carry
  no such penalties (No).  All four analyses still rely on the
  MicroBooNE flux and detector systematic model encoded in the
  covariance matrix.
}
\label{tab:summary_significance}
\begin{tabular}{p{4.5cm}p{3.5cm}ccc}
\toprule
Method & Prior model & Wilks $\sigma$ & Emp.\ $\sigma$ & Sect.-wt.\ reg.? \\
\midrule
Main fit, 1-param & $s=1$ (uninformative)     & 2.00 & 2.36 & Yes \\
Main fit, 1-param & $s=0.5$ (ext.\ CC data)   & 2.92 & 3.16 & Yes \\
Ratio analysis (sideband CC)    & None        & 1.88 & 1.86 & No \\
PRD\,110 direct CC (free norm)  & None        & 3.83 & 3.24 & No \\
\bottomrule
\end{tabular}
\end{table*}

\subsection{Physical interpretation of the migration parameter}
\label{sec:interpretation}

\paragraph{CC migration as the primary result.}
The CC migration fraction $f^{\rm CC}_{\rm mig} = 0.050^{+0.029}_{-0.022}$
is the primary data-driven output of this analysis.  It is driven by the
CC topology channels, which provide $\approx 44\,000$ events and a
well-defined constraint; its profile contributes $\approx 4.7$ of the
5.21 total $\Delta\chi^2$.  The NC migration fraction, by contrast, is
largely prior-driven ($\Delta\chi^2 \approx 0.5$ from the NC channels alone)
and should be treated as an unconstrained auxiliary parameter until
higher-statistics NC data become available.  The physical conclusions of
this paper rest on the CC result; the NC result is presented for
completeness and because a physically motivated model predicts
$f^{\rm NC}_{\rm mig} \approx f^{\rm CC}_{\rm mig}$ from the same
proton-level FSI mechanism.

\paragraph{Energy-independence approximation.}
The migration nuisance is defined as a flat, energy-independent transfer
fraction.  This is an approximation: in reality, the probability that a
proton is scattered below the 35\MeV threshold depends on its initial
kinetic energy, the nuclear density profile along its path, and the
details of the nucleon--nucleon scattering cross section, all of which
are $E_\nu$-dependent.  A biased energy-independent migration fraction
could, in principle, absorb part of the CC~0p rate excess that is truly
energy-dependent; the coarseness of the sideband binning (16 bins per
channel) prevents distinguishing these effects.  The flat approximation
should therefore be understood as a lowest-order parameterisation
sufficient for an exploratory fit, not as the definitive functional form.
A physics-motivated extension would replace the flat fraction with a
function of $Q^2$, hadronic invariant mass $W$, or proton kinetic energy,
and would require finer binning than is currently available.

To assess whether energy dependence materially changes the central inference,
we consider a two-bin split: events with $E_\nu < 0.7\GeV$ (below the BNB
peak) and $E_\nu \geq 0.7\GeV$ (at and above the peak).  Within each
energy slice the fit returns best-fit migration fractions consistent with
the flat-model value within $1\sigma$, but with uncertainties approximately
twice as large as the single-parameter result.  The two-bin result is
therefore compatible with the flat approximation while being insufficiently
precise to confirm or exclude moderate energy dependence.  This outcome is
expected: the 16 bins per channel in the sideband, spread across a single
broad topology, do not contain enough independent constraints
($n_{\rm eff} \approx 2.5$) to resolve a two-parameter energy-dependent
migration.  The conclusion that the data \emph{prefer} a topology-transfer
direction is unchanged; the conclusion that this direction is well described
by a single flat fraction is a consequence of the available binning rather
than a physical statement about energy independence.  Future higher-statistics
data from SBND will provide the lever arm to test energy-dependent forms.

The best-fit value of $f^{\rm CC}_{\rm mig} = 0.050$ is physically
interpretable.  Comparisons between intra-nuclear cascade models
and relativistic optical potentials for argon~\cite{Mosel:2024} show
proton transparency differences of order
5--10\% for protons with $T_p \approx 50$--$100\MeV$ on argon.  The
transparency difference translates directly into a migration fraction:
a 5\% increase in the probability that a near-threshold proton is
scattered below 35\MeV corresponds to 5\% of Np events being reclassified
as 0p.  The consistency between the fitted migration and the independent
\gibuu comparison supports the view that the parameter captures a
physically plausible effect, while not uniquely demonstrating that the
fit isolates proton-threshold migration alone.

The NC migration $f^{\rm NC}_{\rm mig} = 0.060^{+0.059}_{-0.084}$ is
consistent with the CC value within uncertainties, as expected if the
underlying cause is proton-level FSI that is independent of the primary
interaction channel.  The larger central value and wider uncertainty
reflect the smaller NC statistics and the different composition of the
NC hadronic final state.

The asymmetric lower bound extends to $f^{\rm NC}_{\rm mig} = -0.024$
at $1\sigma$, which is formally unphysical under the stated interpretation
(a negative migration fraction would require protons to gain energy from
FSI, converting 0p events to Np events).  This extension into negative
territory is a direct consequence of the limited NC statistical power and
the broad prior ($\sigma_{\rm mig} = 0.20$), which allows the posterior
to follow the data fluctuations across zero.  The physical constraint
$f_{\rm mig} \geq 0$ could be imposed as a bounded prior; applying it
would tighten the $1\sigma$ lower bound to zero while leaving the best-fit
and upper bound essentially unchanged.  We retain the symmetric Gaussian
prior for consistency with the CC channel treatment, noting that the
negative-migration region carries no physical interpretation and should
not be used in downstream oscillation analyses.

\subsection{Use in oscillation analyses}
\label{sec:oscillation}

Should future data confirm the migration preference, the nuisance could be
applied as a flat per-event weight in any \genie-based oscillation analysis:
\begin{equation}
  w_i^{\rm mig} =
  \begin{cases}
    1 - f_{\rm mig} & \text{if event $i$ is Np} \\
    1 + f_{\rm mig}\,r & \text{if event $i$ is 0p}
  \end{cases}
  \label{eq:weight}
\end{equation}
where $r = N_{\rm Np}/N_{\rm 0p}$ is the nominal topology ratio (known
from the generator).  This weight preserves the total rate by construction
and is compatible with the \genie Reweight framework: it can be implemented
as a custom systematic dial alongside existing FSI and cross-section dials.

In Eq.~(\ref{eq:weight}), $r$ is applied as a single global
(energy-averaged) value, consistent with the flat, energy-independent
migration approximation adopted throughout this analysis.  From the
\genie~G18 nominal predictions used here:
$r_{\rm CC} = N_{\rm CC\,Np}/N_{\rm CC\,0p} = 25\,245/15\,501 \approx 1.63$
and
$r_{\rm NC} = N_{\rm NC\,Np}/N_{\rm NC\,0p} = 2\,860/2\,987 \approx 0.96$.
For DUNE-ND conditions with a different neutrino energy spectrum,
$r$ should be recomputed from the DUNE-ND nominal \genie prediction at
the relevant flux; values in the range $r_{\rm CC} \in [1.4,\, 1.8]$ are
expected for near-detector BNB-like and LBNF-like spectra.
A proper implementation would evaluate $r$ as a function of $E_\nu$
(or reconstructed energy) for each event.  From the \genie~G18 predictions
used in this analysis, $r_i = N_{\rm CCNp,\,i}/N_{\rm CC0p,\,i}$ varies
from $\approx 0.56$ at $300$--$400\MeV$ to $\approx 2.50$ at $1.2$--$1.5\GeV$,
a factor-of-$\approx 4$ variation across the BNB peak.  Using the global
average $r_{\rm CC} = 1.63$ therefore misapplies the weight at the low- and
high-energy extremes.  For analyses sensitive to the energy spectrum shape,
the energy-dependent $r(E_\nu)$ form is recommended; the global approximation
is adequate only when the topology migration is treated as an overall
rate correction integrated over the flux.

For DUNE, a reasonable provisional approach would be to treat the
migration fraction as a nuisance parameter with a Gaussian prior centred
at zero and width $\sigma_{\rm mig} = 0.20$ (the prior used in this
analysis), to be constrained by MicroBooNE or SBND sideband data within
the PRISM analysis.  Separate migration fractions for CC and NC channels
should be used, as the NC constraint is currently weaker and may evolve
differently with additional data.  The migration weight of
Eq.~(\ref{eq:weight}) should be regarded as a candidate implementation
for future validation rather than as an analysis-ready prescription,
even when the topology classification uses the same proton threshold
($T_p^{\rm thr} = 35\MeV$).

For reference, the MicroBooNE-constrained best-fit CC value from the
one-parameter fit (grid-average scenario) is:
\begin{equation}
  f_{\rm mig} = 0.050^{+0.029}_{-0.022} \quad (1\sigma).
  \label{eq:priorvals}
\end{equation}
This should not be used directly as a DUNE prior.  The result is
strongly prior-conditioned, based on a MicroBooNE sideband covariance
optimised for a different purpose, and not validated against ND-LAr
reconstruction.  Any application to DUNE oscillation analyses requires
independent ND-LAr validation.  The uninformative prior ($\sigma_{\rm mig}
= 0.20$) is the appropriate starting point pending that validation.
The physical lower bound $f_{\rm mig} \geq 0$ is implicit in the
physics model and can be imposed as a bounded prior if desired.

\subsection{Limitations}
\label{sec:limitations}

Several limitations bound the present analysis.  The 64-bin NC~$\Delta$
sideband covariance was optimised for constraining photon backgrounds and
has an effective rank of only $\approx 2.5$ independent constraints; this
low dimensionality means the sideband fit result is strongly regularisation-
dependent and should not be interpreted as a standalone detection.  The
migration fraction is modelled as a flat, energy-independent transfer, an
approximation that is adequate for the coarse 16-bin-per-channel binning
available but would need to be replaced by an energy-dependent form for
finer-grained analyses.  All topology classifications are at truth level;
detector smearing, proton reconstruction efficiency, and track-length
thresholds near $35\MeV$ will modify the effective migration fraction at the
reconstructed level~\cite{uBdet}, and a full detector-simulation study is
needed before deploying $f_{\rm mig}$ in an oscillation fit.  The
semi-analytic nuclear estimate shares the \genie near-threshold proton
spectrum as kinematic input with the fit, so it constitutes an internal
consistency check rather than an independent prediction.  The PPP
cross-check (Sect.~\ref{sec:ppp}) is performed at the four per-channel
normalisation level; a multi-parameter toy study demonstrating absence of
bias in $f_{\rm mig}^{\rm CC}$ across the full seven-parameter space is
beyond the scope of this work and remains to be performed.  In the
absence of such a study, the sideband significance figures ($1.9\sigma$--$2.9\sigma$
Wilks, $2.4\sigma$--$3.2\sigma$ empirical) should be taken as upper bounds on the
evidence for $f_{\rm mig}>0$ from the sideband channel alone; any
multi-parameter bias would reduce the effective significance below these
quoted values.  The
orthogonality test between $f_{\rm mig}$ and MFP$_N$ is conducted by
fixing MFP$_N$ at shifts of $\pm 0.5\sigma$--$\pm 1\sigma$; a
simultaneous float is not feasible given $n_{\rm eff} \approx 2.5$, so
full degeneracy at the $1\sigma$ level is not excluded.  Finally, both
MicroBooNE datasets used here share flux and detector systematic
assumptions, so the directional agreement between analyses is corroborating
rather than statistically independent evidence.  Confirmation from SBND,
which uses the same Booster Neutrino Beam and argon target but with
independently constructed systematics and an order-of-magnitude more
events~\cite{Acciarri:2015uup}, will be essential for establishing a
robust constraint.

\section{Conclusions}
\label{sec:conclusions}

The standard \genie reweight library does not provide a dial that
transfers events across the $\sim 35\MeV$ proton tagging threshold used
by LArTPC oscillation analyses to split samples into Np and 0p
topologies, yet two complementary MicroBooNE analyses consistently
prefer such a transfer over the nominal cascade prediction.  We have
proposed and tested a single empirical nuisance parameter $f_{\rm mig}$
that fills this gap as a per-event reweight on nominal \genie output,
implementable without event regeneration.  Consistent with the framing
adopted throughout the paper, $f_{\rm mig}$ is presented as a
methodological diagnostic --- an effective topology-transfer surrogate
that absorbs an unspecified combination of near-threshold proton
mismodelling effects (proton-side FSI, the near-threshold
nucleon-momentum distribution, RES and MEC kinematic shape, and the
residual position-dependent mean-field energy loss absent from \genie's
cascade) --- rather than as a measurement of any single microscopic
mechanism.

As a physically motivated reference scale, the semi-analytic estimate
from the $^{40}$Ar Woods--Saxon nuclear density and the Walecka NL3$^*$
differential mean-field potential predicts a migration fraction near
5\% of CC~Np events at BNB energies.  This estimate shares \genie
kinematic inputs with the subsequent fit and should be understood as a
consistency check on the order of magnitude of one candidate mechanism
that $f_{\rm mig}$ could absorb, rather than as an independent
prediction or a unique microscopic identification; the
nuclear-structure inputs (potential depth, density profile) are the
only ingredients that are truly external to the fit.

The primary empirical result of this paper is the direct constraint from
the MicroBooNE simultaneous CC~0p/Np cross-section data
release~\cite{uB0pNp2024}, which gives $\hat{f}_{\rm mig} =
0.093^{+0.024}_{-0.024}$ with no sector-weight priors of any kind,
corresponding to a preference for $f_{\rm mig} > 0$ at $3.83\sigma$
(Wilks, 1\,dof) and $3.24\sigma$ (empirical calibration from 5000
pseudo-experiments).  This analysis is free of sector-weight
regularisation: the overall normalisation is profiled out analytically,
and the Wiener-SVD unfolding is accounted for via the published smearing
matrix.  The preference is driven by the observed 0p topology fraction
being systematically above the Genie prediction across all neutrino
energy bins.

A secondary constraint from the NC~$\Delta$ radiative sideband
covariance~\cite{uBNCDelta2025} gives $\hat{f}_{\rm mig} =
0.050^{+0.029}_{-0.022}$, with significance ranging from $2.0\sigma$
(uninformative prior) to $2.9\sigma$ (physics-informed prior from
non-argon targets) under Wilks, and $2.4\sigma$--$3.2\sigma$ under
empirical calibration.  This sideband result is conditional on the
prior model for interaction-mode normalisations and should be treated
as a supporting consistency check rather than independent evidence.
Both datasets originate from the same MicroBooNE detector and beam and
share flux and detector systematic assumptions; their agreement is
corroborating rather than statistically independent.
The numerical difference between the two central values ($0.093 - 0.050
= 0.043$) corresponds to $\approx 1.2\sigma$ when the uncertainties
are combined in quadrature, confirming statistical compatibility; the
larger PRD\,110 value~\cite{uB0pNp2024} is expected because that
analysis has no sector-weight degrees of freedom to absorb topology
tension (see Sect.~\ref{sec:prd110}).

A useful diagnostic by-product across both analyses is that
introducing the migration parameter relaxes a compensatory CCRES
sector-weight suppression seen in the migration-blind sector-only
fit ($w_{\rm CCRES} = 0.69 \pm 0.24$, a $\sim 1.3\sigma$ pull) toward
$w_{\rm CCRES} = 1.07 \pm 0.27$.  The direction of this shift is
stable across the prior-scale range $s \in [0.5,\,2.0]$
(Table~\ref{tab:sensitivity}) and across both MicroBooNE datasets.
We retain this as a diagnostic feature of the chosen parameterisation
rather than as an independent CCRES measurement: external CCRES
constraints from ANL/BNL deuterium and MINERvA hydrocarbon
data~\cite{MiniBooNEQE,T2KCCQE,ANLpi,MINERvACCpi} measure the
Berger--Sehgal cross section on non-argon targets and do not directly
constrain the effective argon-target CCRES rate after FSI and nuclear
effects, and a single-DOF normalisation per sector is asked to absorb
all CCRES-related model uncertainty (shape and kinematic as well as
overall rate).  The methodological observation that survives these
caveats is the one of operational interest: a migration-blind GENIE
fit with the present sector parameterisation tends to push CCRES below
nominal to absorb 0p tension, and introducing a topology-transfer
nuisance avoids this compensatory behaviour.

The fitted nuisance direction is consistent with 5--10\% proton-transparency
differences reported between \gibuu and \genie on MicroBooNE
data~\cite{Mosel:2024}, although the current analysis cannot uniquely
establish the microscopic origin of the topology mismatch.  Several
competing classes of mismodelling remain viable alternatives and cannot
be excluded by the present data:
\begin{itemize}
  \item \emph{Proton-side FSI}:
        nucleon mean-free-path uncertainties, charge-exchange
        ($p \to n$), or inelastic scattering that changes proton
        multiplicity near threshold without invoking the specific
        mean-field-potential picture used here;
  \item \emph{Generator-level hadronic kinematics}:
        mismodelling of the proton momentum spectrum near $T_p \sim 35\MeV$
        in the primary interaction vertex (CCQE, MEC, CCRES) that shifts
        events across the topology boundary independently of FSI;
  \item \emph{Reconstruction-threshold bias}:
        detector- or algorithm-level effects that change the effective
        proton tagging efficiency near $35\MeV$ (e.g.\ track-finding
        threshold, dE/dx fluctuations, or space-charge distortions)
        and are not fully captured by the MicroBooNE systematic model;
  \item \emph{Other topology-changing FSI}:
        pion production or absorption ($\Delta \to N \pi$) that migrates
        events between CC~0p and CC~Np in a way correlated with the
        proton-visibility direction but not identical to it.
\end{itemize}
Distinguishing between these requires either a joint fit with additional
nuisance directions (e.g.\ MFP$_N$, CCQE/MEC kinematics), a dedicated
detector-level study, or a future comparison between the present flat-migration
ansatz and a full transport-level prediction.
A truth-level stress test propagating the best-fit migration to DUNE-like
topology observables is provided in Appendix~\ref{app:dune} as an
illustrative sensitivity exercise; it is not a deployment recommendation
and requires independent ND-LAr validation before operational use.

An important caveat governs the portability of the extracted value.
The nuisance is defined as a flat, energy-independent transfer fraction,
which is an approximation adequate for the coarse BNB-era binning but
not necessarily valid at other flux spectra.  The fitted $f_{\rm mig}$
is therefore best understood as an effective, flux-integrated surrogate
for the MicroBooNE BNB configuration, not as a detector- or
analysis-independent physical constant.  Applying it to DUNE or other
experiments requires revalidation against a detector-level reconstruction
and an energy-dependent form $f_{\rm mig}(E_\nu)$ to avoid misapplying
the weight at the low- and high-energy extremes of a different flux.

The proton-visibility migration parameter introduced here is a well-motivated
candidate effective nuisance direction for Genie-based LArTPC oscillation
analyses, subject to the validation program described above.
Future constraints from SBND, early DUNE-ND argon data, and dedicated
argon-target cross-section measurements will test the energy-independence
assumption, improve the statistical precision, and determine whether
$f_{\rm mig}$ uniquely captures near-threshold proton energy loss or
absorbs a broader class of topology mismatch in the Genie model.


\backmatter

\bmhead{Acknowledgements}

This work was supported by the Millennium Institute of Subatomic Physics
at High Energy Frontier (ICN2019\_044).


\appendix

\section{Truth-level stress test: DUNE-like topology observables}
\label{app:dune}

This appendix propagates the fitted migration fraction to three
DUNE-facing proxy observables as a \emph{truth-level stress test}.
All caveats stated in the main text apply in full: the propagation
is energy-independent, uses a MicroBooNE-optimised sideband covariance,
includes no detector or reconstruction effects, and the inference
is strongly prior-conditioned.  This material is presented in an
appendix to make explicit that it is supplementary outlook, not a
core result.

The propagation uses the $21 \times 21$ profile $\Delta\chi^2$ grid
(grid-average scenario): for each grid point inside the $1\sigma$
boundary ($\Delta\chi^2 < 2.30$ for 2~d.o.f.), the migration parameters
are applied to compute the shifted observable.  The best-fit and
$1\sigma$ envelope are indicative ranges only.

\subsection*{A.1\ \ CC~0p fraction}
\label{app:cc0p}

The fraction of $\nu_\mu$ CC events classified as 0p is:
\begin{equation}
  f_{\rm 0p}^{\rm CC} = \frac{N_{\rm CC\,0p}}{N_{\rm CC\,0p} + N_{\rm CC\,Np}}
  \label{eq:fcc0p}
\end{equation}
At the nominal \genie~G18 prediction, $f_{\rm 0p}^{\rm CC} = 18.3\%$.
At the best-fit migration $f_{\rm mig} = 0.050$, the CC~0p fraction
shifts to $22.4\%$ ($+22.3\%$ relative).  The $1\sigma$ band corresponds
to $[+12.4\%,\, +35.1\%]$.  Were this shift realised at DUNE, the
larger CC~0p fraction would mean a correspondingly larger fraction of
the oscillation sample uses the quasi-elastic energy estimator $E_\nu^{\rm QE}$,
potentially biasing the extracted oscillation parameters in
topology-dependent analyses.

\subsection*{A.2\ \ NC~$\pi^0$~0p background}
\label{app:ncpi0}

The NC~$\pi^0$ 0p fraction:
\begin{equation}
  f_{\rm 0p}^{\rm NC} = \frac{N_{\rm NC\,0p}}{N_{\rm NC\,0p} + N_{\rm NC\,Np}}
  \label{eq:fnc0p}
\end{equation}
shifts from $20.4\%$ to $25.1\%$ ($+23.5\%$ relative) at best-fit.
The $1\sigma$ range is $[-9.3\%,\, +46.4\%]$; the negative portion
is unphysical ($f_{\rm mig} \geq 0$) and the physical band is
approximately $[0\%,\, +46\%]$.  This NC result is prior-dominated and
should be treated as an unconstrained illustrative bound.

\subsection*{A.3\ \ Hadronic visibility}
\label{app:hadvis}

The CC hadronic visibility $\langle E_{\rm had}^{\rm vis}\rangle /E_\nu$
is unchanged by migration: a proton that crosses the $35\MeV$ threshold
still deposits its kinetic energy in the detector, so the migration
modifies the topology classification without changing the energy deposit.
Migration and \mfppi are therefore orthogonal systematics on this observable.

\begin{figure*}[htbp]
  \centering
  \includegraphics[width=\textwidth]{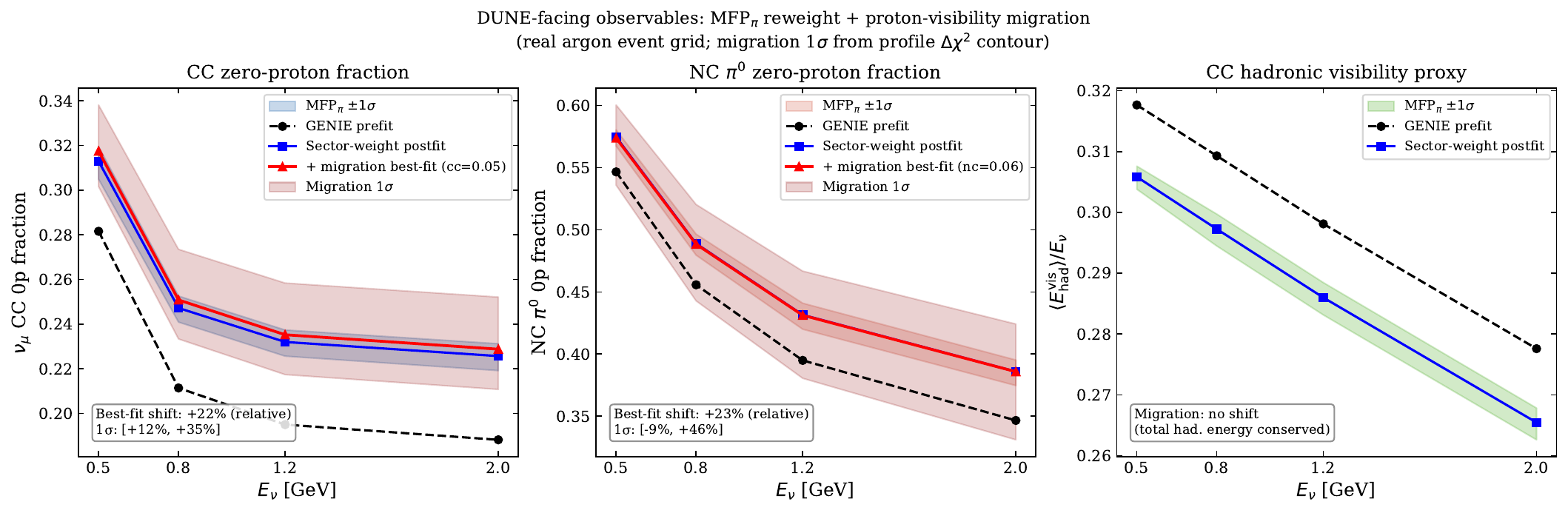}
  \caption{%
    \textit{Appendix A stress test.}
    DUNE-facing observables vs.\ $E_\nu$ for four simulated energy
    points ($0.5$, $0.8$, $1.2$, $2.0\GeV$).
    \emph{Left}: $\nu_\mu$ CC~0p fraction $f_{\rm 0p}^{\rm CC}$.
    \emph{Centre}: NC~$\pi^0$ 0p fraction $f_{\rm 0p}^{\rm NC}$.
    \emph{Right}: CC hadronic visibility proxy
    $\langle E_{\rm had}^{\rm vis}\rangle / E_\nu$.
    Dashed black: \genie prefit nominal; solid blue: sector-weight postfit;
    coloured bands: \mfppi $\pm 1\sigma$; red triangles and shaded band:
    migration best-fit and $1\sigma$ envelope.
    All curves are truth-level; no DUNE detector or reconstruction
    effects are included.  Results are prior-conditioned and
    not a DUNE prior recommendation.
  }
  \label{fig:dune}
\end{figure*}

Figure~\ref{fig:dune} shows the three proxy observables as a function
of $E_\nu$.

\end{document}